\documentclass[longauth]{aa}  
\usepackage{graphicx}
\usepackage{txfonts,textcomp}
\usepackage{gensymb} 
\usepackage{xspace}
\usepackage{tabularx}
\usepackage[dvipsnames]{xcolor}
\usepackage{placeins}
\usepackage{threeparttable}
\usepackage{ulem}
\usepackage{natbib,twoopt}
\usepackage[breaklinks=true]{hyperref} 
\bibpunct{(}{)}{;}{a}{}{,} 
\makeatletter
\newcommandtwoopt{\citeads}[3][][]{\href{http://adsabs.harvard.edu/abs/#3}%
{\def\hyper@linkstart##1##2{}%
\let\hyper@linkend\@empty\citealp[#1][#2]{#3}}}
\newcommandtwoopt{\citepads}[3][][]{\href{http://adsabs.harvard.edu/abs/#3}%
{\def\hyper@linkstart##1##2{}
\let\hyper@linkend\@empty\citep[#1][#2]{#3}}}
\newcommandtwoopt{\citetads}[3][][]{\href{http://adsabs.harvard.edu/abs/#3}%
{\def\hyper@linkstart##1##2{}
\let\hyper@linkend\@empty\citet[#1][#2]{#3}}}
\newcommandtwoopt{\citeyearads}[3][][]%
{\href{http://adsabs.harvard.edu/abs/#3}
{\def\hyper@linkstart##1##2{}%
\let\hyper@linkend\@empty\citeyear[#1][#2]{#3}}}
\makeatother
\def\ms{\hbox{m\,s$^{-1}$}}         
\def\m2s2{\hbox{\,m$^{2}$\,s$^{-2}$}} 
\def\Msun{$M_{\odot}$\xspace}             
\def\Rsun{$R_{\odot}$\xspace}
\def\Mjup{\hbox{$\mathrm{M}_{\rm J}$}\xspace}

\def\ten[#1]{$\;\times 10^{#1}$}
\def\teff{$T_{\rm eff}$}
\def\logg{$\log g$}

\newcommand{\e}[1]{{\times10^{#1}}}

\newcommand{\Rnom}{\hbox{$\mathcal{R}^{\rm N}_{\odot}$}} 

\newcommand{\GMnom}{\hbox{$\mathcal{(GM)}^{\rm N}_{\odot}$}}

\newcommand{\Renom}{\hbox{$\mathcal{R}^{\rm N}_{e \rm E}$}}

\newcommand{\GMenom}{\hbox{$\mathcal{(GM)}^{\rm N}_{\rm E}$}}

\newcommand{\rebound}{{\sc \tt REBOUND}\xspace}
\newcommand{\whf}{{\sc \tt WHFast}\xspace}
\newcommand{\emcee}{{\sc \tt emcee}\xspace}
\newcommand{\juliet}{{\sc \tt juliet}\xspace}
\newcommand{\batman}{{\sc \tt batman}\xspace}

\newcommand{\celerite}{{\sc \tt celerite}\xspace}

\newcommand{\PARSEC}{{\sc \tt PARSEC}\xspace}

\newcommand{\pipe}{{\sc \tt PIPE}\xspace}
\newcommand{\astropy}{{\sc \tt astropy}\xspace}
\newcommand{\prose}{{\sc \tt prose}\xspace}
\newcommand{\photutils}{{\sc \tt photutils}\xspace}
\newcommand{\kiauhoku}{{\sc \tt kiauhoku}\xspace}
\newcommand{\nuance}{{\sc \tt nuance}\xspace}
\newcommand{\everest}{{\sc \tt EVEREST}\xspace}
\newcommand{\gastli}{{\sc \tt GASTLI}\xspace}

\newcommand{\REarth}{$\mathrm{R_E}$\xspace}
\newcommand{\MEarth}{$\mathrm{M_E}$\xspace}

\newcommand{\red}{\color{red}}
\newcommand{\cyan}{\color{cyan}}
\newcommand{\blue}{\color{blue}}
\newcommand{\lightblue}{\color{SkyBlue}}
\newcommand{\grey}{\color{lightgray}}
\newcommand{\green}{\color{Green}}
\newcommand{\purple}{\color{Orchid}}

\newcommand{\lightpink}{\color{CarnationPink}}
\newcommand{\lightgreen}{\color{YellowGreen}}
\newcommand{\orange}{\color{Orange}}

\newcommand{\be}{\begin{equation}}
\newcommand{\ee}{\end{equation}}
\newcommand{\hand}{\hspace{0.5cm}{\rm and}\hspace{0.5cm}}
\newcommand{\rn}[1]{(\ref{#1})}
\newcommand{\mv}{\mathrm{v}}
\newcommand{\bea}{\begin{eqnarray}}
\newcommand{\eea}{\end{eqnarray}}

\def\logg{$\log g$}
\def\Msun{$M_{\odot}$\xspace}            
\def\Rsun{$R_{\odot}$\xspace}
\newcommand{\orcid}[1]{\protect\href{https://orcid.org/#1}{\protect\includegraphics[width=8pt]{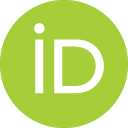}}}
\hypersetup{colorlinks=true, citecolor=blue, linkcolor=blue}

\begin{document} 

   \title{A decade of transit photometry for K2-19: Revised system architecture\thanks{This study uses CHEOPS data observed as part of the guest observer programmes PR240003 (PI Jiang) and PR240013 (PI Almenara).}}
    

   \author{
        J.M.~Almenara\orcid{0000-0003-3208-9815}\inst{\ref{Geneva},\ref{Grenoble}}
        \and R.~Mardling\orcid{0000-0001-7362-3311}\inst{\ref{Australia},\ref{Geneva}}
        \and A.~Leleu\orcid{0000-0003-2051-7974}\inst{\ref{Geneva}}
        \and R.F.~D\'{i}az\orcid{0000-0001-9289-5160}\inst{\ref{ITBA},\ref{BA}}
        \and X.~Bonfils\orcid{0000-0001-9003-8894}\inst{\ref{Grenoble}}
        \and Ing-Guey~Jiang\orcid{0000-0001-7359-3300}\inst{\ref{Taiwan_phys}} 
        \and Li-Chin~Yeh\orcid{0000-0001-8677-0521}\inst{\ref{Taiwan_mx}} 
        \and Ming~Yang\orcid{0000-0002-6926-2872}\inst{\ref{Shanghai}} 
        \and Keivan~G.~Stassun\orcid{0000-0002-3481-9052}\inst{\ref{Nashville}} 
        \and Napaporn~A-thano\orcid{0000-0001-7234-7167}\inst{\ref{Thailand}} 
        \and Billy~Edwards\orcid{0000-0002-5494-3237}\inst{\ref{Netherlands}} 
        \and F.~Bouchy\orcid{0000-0002-7613-393X}\inst{\ref{Geneva}} 
        \and V.~Bourrier\orcid{0000-0002-9148-034X}\inst{\ref{Geneva}} 
        \and A.~Deline\inst{\ref{Geneva}} 
        \and D.~Ehrenreich\orcid{0000-0001-9704-5405}\inst{\ref{Geneva}} 
        \and E.~Fontanet\orcid{0000-0002-0215-4551}\inst{\ref{Geneva}}
        \and T.~Forveille\orcid{0000-0003-0536-4607}\inst{\ref{Grenoble}}
        \and J.M.~Jenkins\orcid{0000-0002-4715-9460}\inst{\ref{AMES}}
        \and L.K.W.~Kwok\orcid{0000-0003-4493-510X}\inst{\ref{Geneva}}
        \and M.~Lendl\orcid{0000-0001-9699-1459}\inst{\ref{Geneva}} 
        \and A.~Psaridi\orcid{0000-0002-4797-2419}\inst{\ref{Barcelona1},\ref{Barcelona2},\ref{Geneva}}
        \and S.~Udry\orcid{0000-0001-7576-6236}\inst{\ref{Geneva}}
        \and J.~Venturini\orcid{0000-0001-9527-2903}\inst{\ref{Geneva}}
        \and J.~Winn\inst{\ref{Princeton}}
        }
   \institute{
        Observatoire de Gen\`eve, Département d’Astronomie, Universit\'e de Gen\`eve, Chemin Pegasi 51b, 1290 Versoix, Switzerland\label{Geneva}
        \and Univ. Grenoble Alpes, CNRS, IPAG, F-38000 Grenoble, France\label{Grenoble}
        \and School of Physics \& Astronomy, Monash University, Victoria, 3800, Australia\label{Australia}
        \and Instituto Tecnol\'ogico de Buenos Aires (ITBA), Iguaz\'u 341, Buenos Aires, CABA C1437, Argentina\label{ITBA}
        \and Instituto de Ciencias F\'isicas (ICIFI; CONICET), ECyT-UNSAM, Campus Miguelete, 25 de Mayo y Francia, (1650) Buenos Aires, Argentina\label{BA}
        \and Department of Physics and Institute of Astronomy, National Tsing-Hua University, Hsinchu 30013, Taiwan\label{Taiwan_phys}
        \and Institute of Computational and Modeling Science, National Tsing-Hua University, Hsinchu 30013, Taiwan\label{Taiwan_mx}
        \and College of Surveying and Geo-Informatics, Tongji University, Shanghai, 200092, People's Republic of China\label{Shanghai}
        \and Department of Physics and Astronomy, Vanderbilt University, Nashville, TN 37235, USA\label{Nashville}
        \and National Astronomical Research Institute of Thailand, Chiang Mai, 50180, Thailand\label{Thailand}
        \and SRON, Space Research Organisation Netherlands, Niels Bohrweg 4, NL-2333 CA, Leiden, The Netherlands\label{Netherlands}
        \and NASA Ames Research Center, Moffett Field, CA 94035, USA\label{AMES}
        \and Institute of Space Sciences (ICE, CSIC), Carrer de Can Magrans S/N, Campus UAB, Cerdanyola del Valles, E-08193, Spain\label{Barcelona1}
        \and Institut d’Estudis Espacials de Catalunya (IEEC), 08860 Castelldefels (Barcelona), Spain\label{Barcelona2}
        \and Department of Astrophysical Sciences, Princeton University, NJ 08544, USA\label{Princeton}
             }

   \date{}

 
  \abstract
  {
The star K2-19 hosts a pair of Neptunian planets deep inside the 3:2 resonance. They induce strong transit-timing variations with two incommensurate frequencies. Previous photodynamical modeling of 3.3 years of transit and radial velocity data produced mass estimates of $32.4\pm 1.7\,M_\oplus$ and $10.8\pm 0.6\,M_\oplus$ for planets b and c, respectively, and corresponding eccentricity estimates of $0.20\pm 0.03$ and $0.21\pm 0.03$. These high eccentricities raise questions about the formation origin of the system, and this motivated us to extend the observing baseline in an attempt to better constrain their values. We present a photodynamical analysis of 10 years of transit data that confirms the previous mass estimates ($30.8\pm 1.3\,M_\oplus$ and $11.1\pm 0.4\,M_\oplus$), but reduces the median eccentricities to $0.04\pm 0.02$ and $0.07\pm 0.02$ for b and c, respectively. These values are more consistent with standard formation 
models, but still involve nonzero free eccentricity. The previously reported high eccentricities appear to be due to a single transit for which measurements taken at twilight mimicked ingress. This resulted in a 12-minute error in the midtransit time. The data that covered 1.3 and 5 so-called super and resonant periods were used to match a Fourier analysis of the transit-timing variation signal with simple analytic expressions for the frequencies and amplitudes to obtain planet mass estimates within 2\% of the median photodynamical values, regardless of the eccentricities. Theoretical details of the analysis are presented in a companion paper. Additionally, we identified a possible planet candidate situated exterior to the b–c pair. Finally, in contrast to a previous study, our internal structure modeling of K2-19\,b yields a metal mass fraction that is consistent with core accretion. 
    }
   

   \maketitle
%

\section{Introduction}\label{section:introduction}

The migration process is a complex mechanism that takes place in protoplanetary disks. It can result in the capture of planets in mean motion resonance (MMR) chains \citep{Cresswell2008}. When the protoplanetary disk dissipates, most of the resonant chains are expected to break \citep{Izidoro2017,Dai2024} because most systems are observed outside the MMR \citep{Fabrycky2014}. It is particularly interesting to study the few remaining resonant chains because it can help us to understand the early life of planetary systems. K2-19 \citep{Foreman-Mackey2015,Montet2015,Armstrong2015} is one of these systems. It is composed of three known transiting planets: an inner Earth-size planet \citep[K2-19\,d, 2.5~days period,][]{Vanderburg2016}, and an outer pair of Neptunian planets (K2-19\,b and c) inside the 3:2 MMR (with periods 7.9 and 11.9~days) that present significant transit-timing variations \citep[TTVs;][]{Agol2005,Holman2005}. These TTVs allowed the first measurements of the masses of K2-19\,b and c \citep{Barros2015,Narita2015}. The latest characterization of the system by \citet{Petigura2020}, who  used a mix of TTVs and a photodynamical modeling \citep{Carter2011}, found eccentricities of $0.20 \pm 0.03$ and $0.21 \pm 0.03$ for K2-19\,b and c, respectively. Disk migration does not favor significant orbital eccentricity excitation, however, because planet-disk interactions tend to damp eccentricities \citep{Bitsch2013}. \citet{Petit2020} proposed secular angular momentum exchange with the interior planet d at 2.5~days, as well as an additional planet in the system. When this additional planet is included, the eccentricities of K2-19\,b and c are reduced.

The precision of the radial velocities in the K2-19 system is moderate because  the target is faint, and significant variability is caused by stellar activity \citep{Dai2016,Nespral2017,Petigura2020}. Therefore, transit photometry (notably through TTVs) remains the main technique for investigating this system, and it alone currently has the sensitivity to accurately measure orbital eccentricities and mutual inclinations. These measurements might help us to unveil the formation history of the system.

Motivated by the work of \citet{Petit2020}, we obtained additional transit observations of K2-19\,b and c with several facilities from the ground and space. They extend the observing baseline of \citet{Petigura2020} from 3 to 10 years and cover more than a complete superperiod \citep{Lithwick2012}.
With 18 and 9 additional transits of planets b and c, respectively, we find that the system is considerably closer to the stationary state of the 3:2 resonance (see Fig.~\ref{figure:poincare}). Our posterior included nonaligned systems with very little free eccentricity (consistent with the standard model of disk migration) as well as aligned systems with a free eccentricity no higher than 0.1. The forced eccentricities of planets b and c are about 0.015 and 0.045 for all samples in the posterior (see Sect.~\ref{section:TTVs}).

The article is organized as follows: We present the new transit photometry of K2-19\,b and c in Sect.~\ref{section:observations}. We revise the stellar parameters in Sect.~\ref{section:stellar_parameters}. We describe our analysis in Sect.~\ref{section:analysis}, and we present our results in Sect.~\ref{section:results}. Finally, we discuss our results in Sect.~\ref{section:discussion}.

\section{Observations}\label{section:observations}

\begin{table*}
\tiny
  \setlength{\tabcolsep}{5pt} 
\renewcommand{\arraystretch}{0.95}
\caption{Log of the transit observations of K2-19 b and c.}             
\label{table:log}      
\centering                          
\begin{tabular}{c c c c c c c c c}        
\hline\hline                 
Epoch & Epoch & Mid-transit date (UT) & Telescope & Band & Exposure time & Coverage & Data & Source \\    
b & c & YYYY-MM-DD & Instrument & & (s) & & \\
\hline                        
  0 to 9 & 0 to 6    & 2014-06-04 to 2014-08-19 & K2 & Kepler & 1765 & Full & lc & This work \\
 30 &     & 2015-01-28 & FLWO 1.2~m & i' & 30 & Partial & tt & \citet{Narita2015} \\
 34 &     & 2015-03-01 & TRAPPIST-South & Exo & 10 & Full & tt & \citet{Narita2015} \\
 34 &     & 2015-03-01 & NITES & no filter & 20 & Partial & lc & \citet{Armstrong2015} \\
 35 &     & 2015-03-09 & 1-m C2PU Omicron & Johnson-R & 60 & Full & lc & \citet{Barros2015} \\
 36 &     & 2015-03-17 & Belesta 82-cm & Johnson-R & 120 & Full & lc & \citet{Barros2015} \\
 41 &     & 2015-04-25 & OAO 1.88~m / MuSCAT & g', r', z$_{\rm S}$ & 60, 30, 60 & Partial & tt & \citet{Narita2015}\\
133 &     & 2017-04-23 & Spitzer & $4.5~\mu$m & 2 & Full & tt & \citet{Petigura2020} \\ 
    &  87 & 2017-04-08 & Spitzer & $4.5~\mu$m & 2 & Full & tt & \citet{Petigura2020} \\
141 &     & 2017-06-25 & LCOGT-SAAO 1~m & i' & 120 & Partial & lc & This work \\ 
150 &     & 2017-09-05 & Spitzer & $4.5~\mu$m & 2 & Full & tt & \citet{Petigura2020} \\
    & 102 & 2017-10-04 & Spitzer & $4.5~\mu$m & 2 & Full & tt & \citet{Petigura2020} \\
311 &     & 2021-03-03 & ExTrA T23 & ExTrA & 60 & Full & lc & This work\\  

343, 345 & 228, 229 & 2021-11-11 to 2021-11-27 & TESS s45 & TESS & 120 & Full & lc & This work \\

346 to 349 & 230, 232 & 2021-12-05 to 2021-12-29 & TESS s46 & TESS & 120 & Full & lc & This work \\

361 &     & 2022-04-03 & ExTrA T23 & ExTrA & 60 & Partial & lc & This work\\ 
362 &     & 2022-04-11 & ExTrA T23 & ExTrA & 60 & Partial & lc & This work\\ 
398 &     & 2023-01-21 & ExTrA T23 & ExTrA & 60 & Partial & lc & This work\\ 
399 &     & 2023-01-29 & ExTrA T123 & ExTrA & 60 & Full & lc & This work\\      
400 &     & 2023-02-06 & ExTrA T123 & ExTrA & 60 & Partial & lc & This work\\  
    & 272 & 2023-04-19 & ExTrA T23 & ExTrA & 60 & Full & lc & This work\\  

436, 438 & 290, 291    & 2023-11-18 to 2023-12-04& TESS s72 & TESS & 120 & Full & lc & This work \\

449 &     & 2024-02-29 & CHEOPS & CHEOPS & 60 & Intermittent & lc & This work\\
449 &     & 2024-02-29 & ExTrA T12 & ExTrA & 60 & Full & lc & This work\\ 
450 &     & 2024-03-08 & Euler / ECAM & NGTS & 60 & Full & lc & This work\\ 
450 &     & 2024-03-08 & ExTrA T12 & ExTrA & 60 & Full & lc & This work\\ 
451 &     & 2024-03-16 & Euler / ECAM & NGTS & [30, 35] & Full & lc & This work\\ 
451 &     & 2024-03-16 & ExTrA T12 & ExTrA & 60 & Partial & lc & This work\\ 
    & 301 & 2024-03-29 & CHEOPS & CHEOPS & 60 & Intermittent & lc & This work\\
    & 304 & 2024-05-04 & Euler / ECAM & NGTS & 75 & Full & lc & This work\\ 
    & 304 & 2024-05-04 & ExTrA T23 & ExTrA & 60 & Full & lc & This work\\ 
458 &     & 2024-05-10 & CHEOPS & CHEOPS & 60 & Intermittent & lc & This work\\

\hline                                   
\end{tabular}
\tablefoot{The epoch is the number of orbital periods since the first observed transit. For the TESS observations, the sector number is specified after the s. For the ExTrA observations, T stands for telescope, and the numbers following it represent the specific telescopes that observed the transit. For example, T123 means that telescopes 1, 2, and 3 all observed the transit. The Data column indicates how each transit was incorporated into the analysis of Sect.~\ref{section:photodynamical}. The term lc is used when the transit light curve was used, and tt is employed when only the transit timing determined in the literature was used.}
\end{table*}

Table~\ref{table:log} summarizes all transit observations we used. We detail the new observations by telescope below together with the revision of two datasets from the literature.

\subsection{K2}

K2-19 was discovered during Campaign~1 of the K2 mission of the {\it Kepler} space telescope \citep{Borucki2010}. The light curve presents significant systematics because the fine-pointing capability of the spacecraft was lost. We used \everest \citep{Luger2016,Luger2018} to correct for these systematics by masking the transits of the known transiting planets (including candidate K2-19\,e; Sect.~\ref{section:nuance}) to minimize changes in the transits during the detrending process. 

\subsection{LCOGT}

\citet{Petigura2020} presented one partial transit of K2-19\,b observed with the Las Cumbres Observatory Global Telescope \citep[LCOGT;][]{Brown2013} on June 25, 2017\footnote{The date for this transit reported in Sect.~2.3 of \citet{Petigura2020} do not correspond with the quoted timing in their Table~1.}, using a 1~m telescope at the South African Astronomical Observatory. We reanalyzed these transit data because we found that the transit timing reported by \citet{Petigura2020} deviated from the model posterior (Sect.~\ref{section:photodynamical}) by $3.5\,\sigma$ (equivalent to $12.4\pm3.6$~minutes\footnote{This offset suggests that the LCOGT model of \citet{Petigura2020} underestimates the transit duration by approximately 25 minutes.}). We note that the observations they interpreted as the transit ingress were acquired during twilight. We performed photometry with \prose \citep{Garcia2022} on images calibrated with the standard LCOGT pipeline \citep[BANZAI;][]{McCully2018}.

\subsection{ExTrA}

We started our follow-up of K2-19\,b and c in 2021 with the facility called Exoplanets in Transits and their Atmospheres \citep[ExTrA;][]{Bonfils2015}, which is located at La Silla Observatory in Chile. ExTrA is a low-resolution near-infrared (0.85 to 1.55~$\mu$m) multi-object spectrograph fed by three 60~cm telescopes. We used 8$\arcsec$ aperture fibers and the lowest-resolution mode ($R$$\sim$20) of the spectrograph with an exposure time of 60~seconds. During the 2021 and 2022 seasons, we targeted several transit windows of K2-19\,b and c using the transit forecast from \citet{Petigura2020}, which had uncertainties between 35 and 65 minutes for K2-19\,b and between 2 and 4 hours for K2-19\,c. We were able to observe one full transit and two partial transits of K2-19\,b and no transit of K2-19\,c. The observed transits of K2-19\,b differed between $2.5\,\sigma$ to $4.5\,\sigma$ from the predictions by \citet{Petigura2020}. From season 2023 onward, we used a TTV modeling of the available timings to forecast transits for ExTrA and other facilities. As K2-19\,b and c have detected gravitational interactions, the observations of K2-19\,b also improved the transit forecast of K2-19\,c. This allowed us to finally observe a transit of K2-19\,c on April 19, 2023 with ExTrA, which to our knowledge is the first time that the transit of this 6-hour TTV planet has been observed from the ground.

\subsection{TESS}

The Transiting Exoplanet Survey Satellite \citep[TESS;][]{Ricker2015} observed K2-19 in sectors 45, 46, and 72 with a cadence of 2 minutes. For the analysis, we used the presearch data-conditioning simple aperture photometry (PDCSAP; \citealt{Smith2012}; \citealt{Stumpe2012,Stumpe2014}) light curve of K2-19 that was produced by the TESS Science Processing Operations Center \citep[SPOC;][]{Jenkins2016}.

\subsection{CHEOPS}

The CHaracterising ExOPlanet Satellite \citep[CHEOPS;][]{Benz2021} observed two transits of K2-19\,b on February 29, 2024 and May 10, 2024, and one transit of K2-19\,c on March 29, 2024 (Table~\ref{table:cheops}). The raw data were automatically processed by the CHEOPS data reduction pipeline \citep[DRP, version 14.1.3;][]{Hoyer2020}. We performed point-spread function photometry with the \pipe package\footnote{\url{https://github.com/alphapsa/PIPE}} \citep{Brandeker2024}. Unfortunately, even with an observation efficiency of 93.7\%\footnote{Efficiency refers to the proportion of the observational time during which usable data are acquired.}, no first contact of the transit of K2-19\,c was observed.

\begin{table*}
\small
\renewcommand{\arraystretch}{1.0}
\caption{Log of CHEOPS observations.}             
\label{table:cheops}      
\centering                          
\begin{tabular}{c c c c c c}        
\hline\hline                 
File key & OBS ID & UTC start & Visit duration & Exposure time &  Efficiency \\    
\hline                        
\verb|CH_PR240003_TG000101_V0300| & 2361633 & 2024-02-29T02:44:43 & 47002 s & 1 x 60.0 s & 66.9\% \\
\verb|CH_PR240013_TG000101_V0300| & 2377170 & 2024-03-29T01:48:43 & 41059 s & 1 x 60.0 s & 93.7\% \\
\verb|CH_PR240003_TG000102_V0300| & 2410093 & 2024-05-10T07:38:43 & 47002 s & 1 x 60.0 s & 54.9\% \\
\hline                                   
\end{tabular}
\end{table*}

\subsection{EulerCam}

We observed two transits of K2-19\,b on March 8 and 16, 2024, and one transit of K2-19\,c on May 4, 2024 with the EulerCam (ECAM) at the Swiss 1.2~m Euler telescope located at La Silla Observatory. The image reduction was carried out following \citet{Lendl2012} and \citet{Lendl2014}. Aperture photometry was performed using \prose \citep{Garcia2022}, which relies on \astropy \citep{astropy2013,astropy2018,astropy2022} and \photutils \citep{Bradley2023}. The optimal differential photometry followed \citet{Broeg2005}.

\section{Stellar parameters}\label{section:stellar_parameters}

To determine the stellar parameters of K2-19, we fit its spectral energy distribution (SED) using stellar atmosphere and evolution models, along with the parallax determination from Gaia \citep{gaia, gaiaDR3}. We constructed the SED of K2-19 using the magnitudes from Gaia data release 3 \citep{Riello2021}, the 2-Micron All-Sky Survey \citep[2MASS;][]{2mass,Cutri2003}, and the Wide-field Infrared Survey Explorer \citep[WISE;][]{wise,Cutri2013}. The measurements are listed in Table~\ref{table:stellar_params}. 
We used the PHOENIX/BT-Settl \citep{Allard2012} stellar atmosphere model and the two stellar evolution models Dartmouth \citep{Dotter2008} and \PARSEC \citep{Chen2014}. We modeled the SED using the procedure described by \citet{Diaz2014}, with informative priors for the effective temperature ($T_{\mathrm{eff}}$), surface gravity (\logg), and metallicity ($[\rm{Fe/H}]_\star$) from \citet{Sousa2021}, and for the distance from the Gaia zeropoint-corrected parallax \citep{Lindegren2021}. We added an uncertainty of 2\% quadratically to the $T_{\mathrm{eff}}$ following \citet{Tayar2022}. We used uninformative priors for the remaining parameters. We used a jitter \citep{Gregory2005} for each set of photometric bands (Gaia, 2MASS, and WISE). The offset in the posterior median between the two tested stellar evolution models is 0.008~\Rsun and 0.017~\Msun. We merged the results from the two different stellar evolution models assuming an equal probability for each. The parameters priors and posteriors are listed in Table~\ref{table:sed}. The data with the maximum a posteriori (MAP) stellar atmosphere model are shown in Fig.~\ref{figure:sed}. 
For the stellar radius, we used a systematic uncertainty floor of 4\% following \citet{Tayar2022} (we added an uncertainty of 4\% quadratically to the radius estimated with the SED analysis). We confirmed the mass values with the four stellar evolution models implemented by \kiauhoku \citep{Claytor2020} and determined a mean value of 0.882~\Msun, a standard deviation between models of 0.012~\Msun, and a maximum offset of 0.026~\Msun. Our mass value and uncertainty already encompass the differences between the models in \kiauhoku. We compare the stellar parameters derived in this section (Table~\ref{table:stellar_params}) with literature values listed in Table~\ref{table:literature}. The stellar mass and radius were used as priors in the photodynamical modeling (Sect.~\ref{section:photodynamical}).

We used the stellar mass and the stellar rotation period determined in Sect.~\ref{section:photodynamical} ($P_{\rm rot} = 18.36 \pm 0.51$~days) to derive a gyrochronological age, neglecting the influence of the planets. Based on the formulation of \citet{Barnes2010} and \citet{BarnesKim2010} with initial periods P$_0$ between 0.12 and 3.4~days, the age is $1.70 \pm 0.25$~Gyr \citep[we added a 10\% systematic error to the statistical error based on][]{Meibom2015}.

\begin{table}[!ht]
\centering
\small
  \setlength{\tabcolsep}{4pt}
      \caption[]{Stellar parameters for K2-19 (EPIC~201505350, TIC~281885301, TOI-5145, Gaia~DR3~3798833775141351552, and 2MASS~J11395048+0036129).}
         \label{table:stellar_params}
         \begin{tabular}{lcc}
            \hline
            \noalign{\smallskip}
            Parameter & \text{Value} & \text{Refs} \\
            \noalign{\smallskip}
            \hline
            \noalign{\smallskip}
            \textit{Astrometry} \\
            Right ascension (J2016), $\alpha$ & 11$^{\rm h}$39$^{\rm m}$50.46$^{\rm s}$ & 1 \\
            Declination (J2016), $\delta$ & 00$^{\rm o}$36'12.95'' & 1 \\
            Parallax, $\pi$ (mas) & $3.368 \pm 0.020$ & 1, 2 \\
            Distance, d (pc) & $296.9 \pm 1.7$ & 3 \\
            Proper motion $\alpha$ (mas/year) & $-18.673 \pm 0.022$ & 1 \\
            Proper motion $\delta$ (mas/year) & $4.571 \pm 0.015$ & 1 \\
            \noalign{\smallskip}
            \textit{Photometry} \\
            Kepler magnitude (mag) & 12.806 & 4 \\
            TESS magnitude (mag) & $12.3049 \pm 0.0068$ & 5 \\
            Gaia-BP (mag) & $13.2080 \pm 0.0030$  & 1 \\
            Gaia-G (mag) & $12.8060 \pm 0.0028$ & 1\\
            Gaia-RP (mag) & $12.2387 \pm 0.0029$  & 1 \\
            2MASS J (mag) & $11.596 \pm 0.024$ & 6 \\
            2MASS H (mag) & $11.208 \pm 0.022$ & 6 \\
            2MASS $K_s$ (mag) & $11.161 \pm 0.026$ & 6 \\
            WISE W1 (mag) & $11.105 \pm 0.023$  & 7 \\
            WISE W2 (mag) & $11.152 \pm 0.021$  & 7 \\
            WISE W3 (mag) & $10.92 \pm 0.12$  & 7 \\
            \noalign{\smallskip}
            \textit{Stellar parameters} \\
            Spectral type & G9 & 8 \\
            Prior stellar mass, $m_{\star}$ (\Msun) & $0.885 \pm 0.034$ & 3 \\
            Prior stellar radius, $R_{\star}$ (\Rsun) & $0.845 \pm 0.039$ & 3 \\
            Effective temperature, $T_{\rm eff}$ (K) & $5350 \pm 110$ & 9, 3 \\
            Surface gravity, log g (cgs) &  $4.531 \pm 0.044$ & $R_\star$, $m_\star$ \\
            Metallicity, [Fe/H]$_\star$ (dex) & $0.03 \pm 0.02$ & 9 \\ 
            Age (Gyr) & $1.70 \pm 0.25$ & 3 \\
            \noalign{\smallskip}
            \hline
        \end{tabular}
        \begin{tablenotes}
        \small
        \item References : 1)~\cite{gaiaDR3}, 2)~\cite{Lindegren2021}, 3)~This work, 4)~\cite{Huber2016}, 5)~\cite{Stassun2019}, 6)~\cite{Cutri2003}, 7)~\cite{Cutri2013}, 8)~\cite{Pecaut2013}, 9)~\cite{Sousa2021}.
        \end{tablenotes}
\end{table}

\section{Analysis}\label{section:analysis}

\subsection{Search for additional transiting planets}\label{section:nuance}

\begin{figure*}[t]
  \centering
  \includegraphics[width=0.99\textwidth]{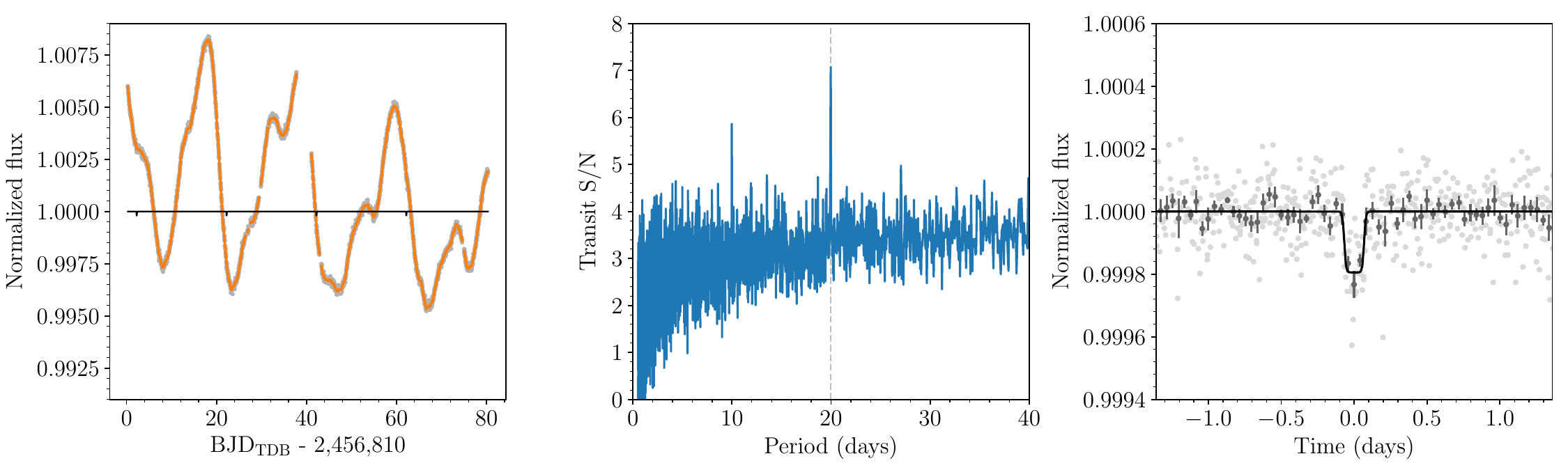}
  \caption{Detection of the candidate planet~e. {\it Left}: Gray data points represent the K2 data without the transits of planets b, c, and d. The orange data points show the mean GP model. The black light curve indicates the four transits we found. {\it Center}: Periodogram of the \nuance algorithm. {\it Right}: Phased light curve without the noise model (gray points), binned (dark gray), and transit model (black line).} \label{figure:nuance}
\end{figure*}

The K2 light curve of K2-19 is dominated by rotational modulation, with an amplitude variability (peak to peak) of about 1.2\% and a periodicity of $18.36 \pm 0.51$~days (Sect.~\ref{section:photodynamical}). We searched for additional transiting planets in the K2 data using \nuance \citep{Garcia2024}, a transit search algorithm that accounts for correlated noise (stellar variability in this case), which we modeled with the rotation kernel Gaussian process (GP).

We found the already known planets with an S/N of 54 (K2-19\,b), 33 (K2-19\,c), and 13 (K2-19\,d). Then, we detected an additional signal, the candidate planet~e, at 19.97~days with an S/N of 7.1 (Fig.~\ref{figure:nuance}). We fit this planet candidate with \juliet \citep{Espinoza2019,Speagle2020} using \batman \citep{Kreidberg2015} for the transit model and the quasi-periodic kernel GP included in \celerite \citep{Foreman-Mackey2017} to model the stellar variability. We modeled the K2 light curve without the transits of K2-19\,b, c, and d. We used a normal prior for the stellar density and uninformative priors for the rest of the parameters. We obtained an orbital period of $19.9657_{-0.0067}^{+0.0047}$~days, a transit duration of $4.28 \pm 0.31$~hours, and a planet size of $1.26 \pm 0.17$~\REarth. We repeated the analysis after removing the transits of the candidate planet~e from the model. A model comparison strongly favors \citep{Kass1995} the model with planet~e over the model without the planet (log-Bayes factor $\mathcal{Z}_{\rm planet~e}/\mathcal{Z}_{\rm no~planet~e}$ of $5.60 \pm 0.37$). We repeated the analysis with \juliet with the timing of each transit as a free parameter. The individual timings have a precision of about 20 minutes and are compatible with a linear ephemeris. Interestingly, the candidate planet~e is close to the 5:3 MMR with K2-19\,c. We included this planet candidate (K2-19\,e) in the analysis of the system.

Subsequent searches yielded low S/N candidates that are not favored by the model comparison. One interesting candidate that emerged is a $0.591 \pm 0.025$~\REarth planet with a period of 1.89~days (close to the 4:3 MMR with K2-19\,d), however. When this candidate was added to the analysis described in Sect.~\ref{section:photodynamical}, its mass was $0.089 \pm 0.035$~\MEarth.

\subsection{Photodynamical modeling}\label{section:photodynamical}

\begin{table*}
  \tiny
\renewcommand{\arraystretch}{1.22}
\centering
\caption{Inferred system parameters.}\label{table:results}
\begin{tabular}{lccccc}
\hline
Parameter & Units & Prior & Median and 68.3\% CI & Prior & Median and 68.3\% CI \\
\hline
\emph{\bf Star} \\
Stellar mass, $m_\star$              & (\Msun)     & $N(0.885, 0.034)$   & $0.892 \pm 0.034$ \\
Stellar radius, $R_\star$            & (\Rnom)     & $N(0.845, 0.039)$   & $0.824 \pm 0.024$ \\
Stellar mean density, $\rho_{\star}$ & ($\mathrm{g\;cm^{-3}}$) &         & $1.59 \pm 0.12$ \\
Surface gravity, \logg\              & (cgs)           &                 & $4.556 \pm 0.024$ \\

K2 GP $P_{\rm rot}$ & (d) & $U(0.1, 100)$ & $18.36 \pm 0.51$ \\

\emph{\bf Planets} & & \emph{\bf Planet~b} & \emph{\bf Planet~b} & \emph{\bf Planet~c} & \emph{\bf Planet~c} \\
Semi-major axis, $a$                      & (au)                   &              & $0.07487 \pm 0.00093$         &              & $0.0982 \pm 0.0012$ \\
Eccentricity, $e$                         &                        &              & $0.043 \pm 0.024$             &              & $0.067 \pm 0.017$ \\
Argument of pericenter, $\omega$          & (\degree)              &              & $197 \pm 26$                  &              & $252^{+18}_{-11}$ \\
Inclination, $i$                          & (\degree)              & $S(0, 180)$  & $89.37 \pm 0.26$              & $S(0, 90)$   & $89.37 \pm 0.15$ \\
Longitude of the ascending node, $\Omega$ & (\degree)              &              & 180 (fixed at $t_{\mathrm{ref}}$)&  $U(90, 270)$ & $180.84 \pm 0.45$ \\
Mean anomaly, $M_0$                       & (\degree)              &              & $257 \pm 26$                  &              & $202^{+13}_{-20}$\\

Mass ratio, $m_{\mathrm{p}}/m_\star$      &                        & $U(0, 1)$  & $(10.357 \pm 0.085)\e{-5}$      & $U(0, 1)$  & $(3.746 \pm 0.056)\e{-5}$\\
Radius ratio, $R_{\mathrm{p}}/R_\star$    &                        & $U(0, 1)$  & $0.07332 \pm 0.00054$           & $U(0, 1)$  & $0.04321 \pm 0.00044$ \\
Scaled semi-major axis, $a/R_{\star}$     &                        &            & $19.53 \pm 0.51$               &            & $25.61 \pm 0.66$ \\
Impact parameter, $b$                     &                        &            & $0.217 \pm 0.089$               &            & $0.299 \pm 0.065$ \\
$T_0'$\;-\;2\;450\;000                    & (BJD$_{\mathrm{TDB}}$) & $U(5829, 7829)$ & $6829.22231 \pm 0.00016$   & $U(5829, 7829)$ & $6829.18490 \pm 0.00054$ \\
$P'$                                      & (d)                    & $U(5, 10)$  & $7.922806 \pm 0.000090$        & $U(10, 15)$ & $11.89770 \pm 0.00046$ \\
$K'$                                      & (\ms)                  &             & $10.66 \pm 0.17$               &             & $3.368 \pm 0.063$ \\

Planet mass, $m_{\mathrm{p}}$             &(\MEarth)               &             & $30.8 \pm 1.3$                 &             & $11.12 \pm 0.44$ \\
Planet radius, $R_{\mathrm{p}}$           &(\Renom)                &             & $6.59 \pm 0.22$                &             & $3.88 \pm 0.14$\\
Planet mean density, $\rho_{\mathrm{p}}$  &($\mathrm{g\;cm^{-3}}$) &             & $0.590 \pm 0.053$              &             & $1.046 \pm 0.097$  \\
Planet surface gravity, $\log$\,$g_{\mathrm{p}}$ &(cgs)            &             & $2.841 \pm 0.028$              &             & $2.859 \pm 0.028$\\
Equilibrium temperature, T$_{\rm eq}$     & (K)                    &             & $857 \pm 21$                   &             & $749 \pm 19$\\

Mutual inclination, $I_{b,c}$                   & (\degree)            &   &  $0.85 \pm 0.44$  \\
$e_{\mathrm{c}} \cos \omega_{\mathrm{c}} - \frac{a_{\mathrm{b}}}{a_{\mathrm{c}}} e_{\mathrm{b}} \cos \omega_{\mathrm{b}} \,\,$ &  & $U(-1, 1)$ &  $0.00918 \pm 0.00020$ \\
$e_{\mathrm{c}} \cos \omega_{\mathrm{c}} + \frac{a_{\mathrm{b}}}{a_{\mathrm{c}}} e_{\mathrm{b}} \cos \omega_{\mathrm{b}} \,\,$ &  & $U(-1, 1)$ &  $-0.050 \pm 0.040$ \\
$e_{\mathrm{c}} \sin \omega_{\mathrm{c}} - \frac{a_{\mathrm{b}}}{a_{\mathrm{c}}} e_{\mathrm{b}} \sin \omega_{\mathrm{b}} \,\,$ &  & $U(-1, 1)$ &  $-0.05386 \pm 0.00082$ \\
$e_{\mathrm{c}} \sin \omega_{\mathrm{c}} + \frac{a_{\mathrm{b}}}{a_{\mathrm{c}}} e_{\mathrm{b}} \sin \omega_{\mathrm{b}} \,\,$ &  & $U(-1, 1)$ &  $-0.071 \pm 0.027$ \\

\emph{\bf Planets} & & \emph{\bf Planet~d} & \emph{\bf Planet~d} & \emph{\bf Candidate planet~e} & \emph{\bf Candidate planet~e} \\
Semi-major axis, $a$                      & (au)                   &              & $0.03478 \pm 0.00043$   &              & $0.1387 \pm 0.0017$ \\
Eccentricity, $e$                         &                        &              & < 0.44$^\dagger$        &              & < 0.15$^\dagger$ \\
Argument of pericenter, $\omega$          & (\degree)              &              &  $248^{+72}_{-110}$     &              &  $245 \pm 37$ \\
Inclination, $i$                          & (\degree)              & $S(0, 180)$  &  $89.7 \pm 1.6$         & $S(0, 180)$  &  $89.20^{+0.43}_{-0.17}$\\
Longitude of the ascending node, $\Omega$ & (\degree)              & $U(90, 270)$ & $179.3 \pm 7.3$         & $U(90, 270)$ & $179 \pm 10$ \\
Mean anomaly, $M_0$                       & (\degree)              &              & $200 \pm 100$           &              & $152 \pm 41$ \\
$\sqrt{e}\cos{\omega}$                    &                        & $U(-1, 1)$   & $0.02 \pm 0.27$         & $U(-1, 1)$   & $-0.08 \pm 0.12$ \\
$\sqrt{e}\sin{\omega}$                    &                        & $U(-1, 1)$   & $-0.17^{+0.22}_{-0.15}$ & $U(-1, 1)$   & $-0.174 \pm 0.068$ \\

Mass ratio, $m_{\mathrm{p}}/m_\star$      &                        & $U(0, 1)$    & < 2.1$\e{-5}^\dagger$   & $U(0, 1)$    & < 0.67$\e{-5}^\dagger$ \\
Radius ratio, $R_{\mathrm{p}}/R_\star$    &                        & $U(0, 1)$    & $0.01165 \pm 0.00066$   & $U(0, 1)$    & $0.0135 \pm 0.0016$ \\
Scaled semi-major axis, $a/R_{\star}$     &                        &              & $9.07 \pm 0.24$         &              & $36.17 \pm 0.94$ \\
Impact parameter, $b$                     &                        &              & $0.17^{+0.18}_{-0.12}$  &              & $0.544^{+0.099}_{-0.15}$ \\
$T_0'$\;-\;2\;450\;000                    & (BJD$_{\mathrm{TDB}}$) & $U(5829, 7829)$ & $6828.9909 \pm 0.0035$ & $U(5829, 7829)$ & $6832.2143 \pm 0.0066$ \\
$P'$                                      & (d)                    & $U(0, 5)$   & $2.50827^{+0.00027}_{-0.00014}$        & $U(15, 25)$ & $19.9724 \pm 0.0054$ \\
$K'$                                      & (\ms)                  &             & < 3.1$^\dagger$          &             & < 0.51$^\dagger$\\

Planet mass, $m_{\mathrm{p}}$             &(\MEarth)               &             & < 6.2$^\dagger$          &             & < 2.0$^\dagger$ \\
Planet radius, $R_{\mathrm{p}}$           &(\Renom)                &             & $1.044 \pm 0.064$        &             & $1.22 \pm 0.15$ \\
Planet mean density, $\rho_{\mathrm{p}}$  &($\mathrm{g\;cm^{-3}}$) &             & < 31$^\dagger$           &             & < 8.4$^\dagger$ \\
Planet surface gravity, $\log$\,$g_{\mathrm{p}}$ &(cgs)            &             & $2.95 \pm 0.63$          &             & $2.46 \pm 0.40$ \\
Equilibrium temperature, T$_{\rm eq}$     & (K)                    &             & $1258 \pm 31$            &             & $630 \pm 16$ \\

\hline
\end{tabular}
\tablefoot{\tiny The table lists the prior, the posterior median, and the 68.3\% CI for the photodynamical analysis (Sect.~\ref{section:photodynamical}). The Jacobi orbital elements are given for the reference time $t_{\mathrm{ref}}=2\,456\,829.222$~BJD$_{\mathrm{TDB}}$. $^\dagger$ Upper limit at 99\% confidence. The planetary equilibrium temperature is computed for zero albedo and full day-night heat redistribution. $P'$ and $T_0'$ should not be confused with the period or the time of conjunction, respectively, and they were only used to reduce the correlations between jump parameters, replacing the semimajor axis and the mean anomaly at $t_{\mathrm{ref}}$. $T'_0 \equiv t_{\mathrm{ref}} - \frac{P'}{2\pi}\left(M_0-E+e\sin{E}\right)$ with $E=2\arctan{\left\{\sqrt{\frac{1-e}{1+e}}\tan{\left[\frac{1}{2}\left(\frac{\pi}{2}-\omega\right)\right]}\right\}}$, $P' \equiv \sqrt{\frac{4\pi^2a^{3}}{\mathcal G m_{\star}}}$, $K' \equiv \frac{m_p \sin{i}}{m_\star^{2/3}\sqrt{1-e^2}}\left(\frac{2 \pi \mathcal G}{P'}\right)^{1/3}$. CODATA 2018: $\mathcal G = 6.674\,30$\ten[-11]$\rm{m^3\,kg^{-1}\,s^{-2}}$. IAU 2012: au = $149\,597\,870\,700$~m$\,$. IAU 2015: \Rnom = 6.957\ten[8]~m, \GMnom = 1.327$\,$124$\,$4\ten[20]~$\rm{m^3\,s^{-2}}$, \Renom~=~6.378$\,$1\ten[6]~m, \GMenom = 3.986$\,$004\ten[14]~$\rm{m^3\,s^{-2}}$, \Msun$ = \GMnom/\mathcal G$, \MEarth = \GMenom/$\mathcal G$, $k^2$ = \GMnom$\,(86\,400~\rm{s})^2$/$\rm{au}^3$. $N(\mu, \sigma)$: Normal distribution with mean $\mu$ and standard deviation $\sigma$. $U(a, b)$: A uniform distribution defined between a lower $a$ and upper $b$ limit. $S(a, b)$: A sinusoidal distribution defined between a lower $a$ and upper $b$ limit. Transits occur when the sum of the true anomaly and the argument of pericenter equals -90\degree.}
\end{table*}

\begin{figure*}[!ht]
  \centering
  \includegraphics[width=0.98\textwidth]{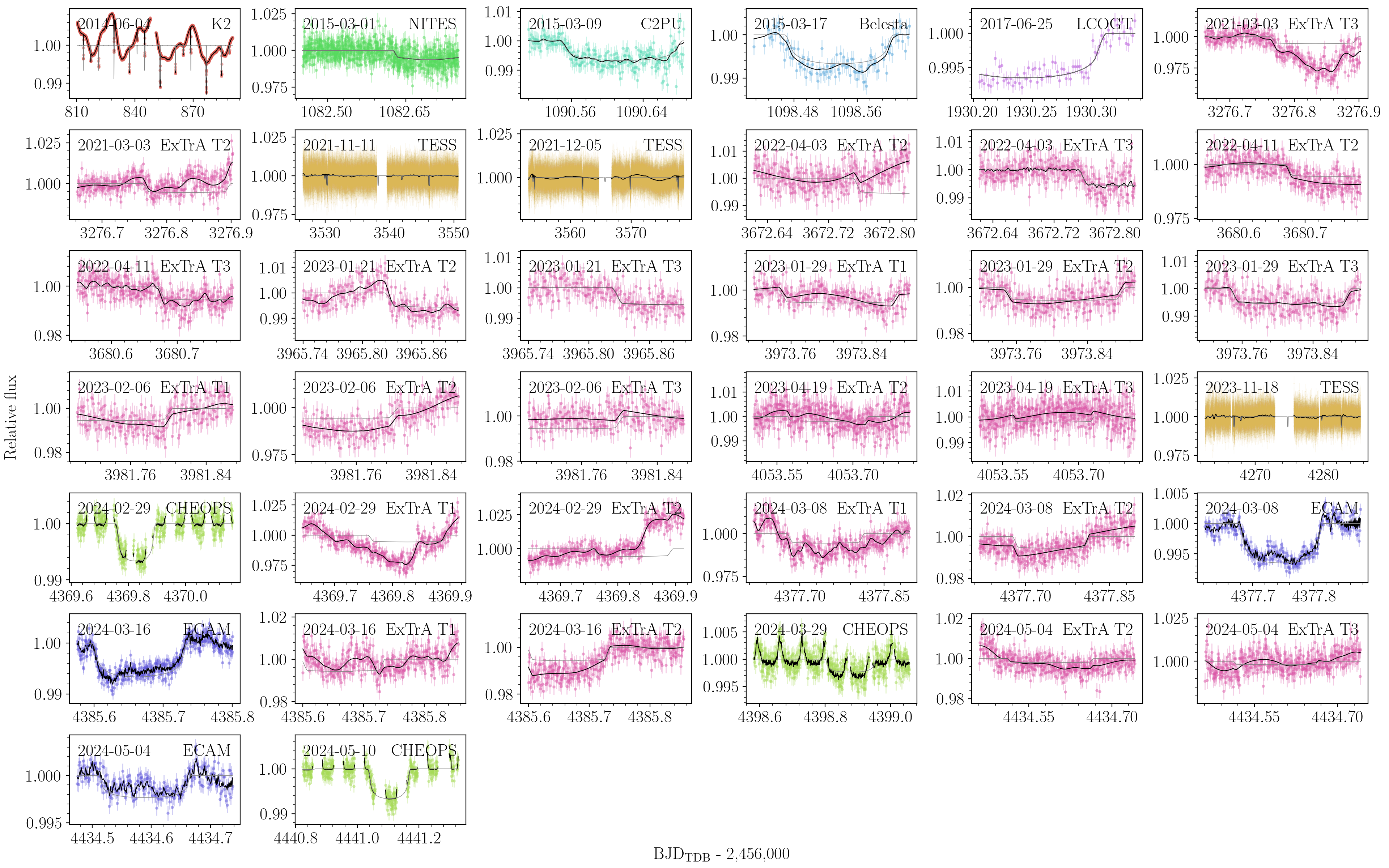}
  \caption{Photodynamical modeling of the transit photometry. Each dataset is shown in a different panel, labeled with the midtransit date (or the start date of the observation for K2 and TESS) and the telescope (or instrument). The error bars, in different colors for each telescope, represent the observations. The black line shows the MAP model that combines transits and noise. The gray line shows the transit model.} \label{figure:phot}
\end{figure*}

We fit the observed photometry accounting for the gravitational interactions between the five bodies assumed in the system using a photodynamical model. Its positions and velocities in time were obtained through an n-body integration. The sky-projected positions were used to compute the light curve \citep{Mandel2002} with \batman \citep{Kreidberg2015} using a quadratic limb-darkening law \citep{Manduca1977}, which we parameterized following \citet{Kipping2013}. To account for the integration time of 30 minutes of the K2 light curve, the model was oversampled by a factor of 30, and then binned back to match the cadence of the data points \citep{Kipping2010}. We used the n-body code \rebound \citep{Rein2012} with the \whf integrator \citep{Rein2015} and an integration step of 0.01~days, which resulted in a maximum error \citep{almenara2018} of $\sim$5~ppm for the photometric model. The light-time effect \citep{Irwin1952} was not included; with an expected amplitude of $\sim$6~ms in the transit timing, it is a negligible effect for this system. We included some transit timings from the literature, assuming that the reported time is the calculated time of minimum sky-projected planet–star separation. Table~\ref{table:log} specifies whether we used the time-series photometry or only the reported timing for each observation. The model was parameterized using the stellar mass and radius, limb-darkening coefficients, planet-to-star mass and radius ratios, and Jacobi orbital elements (Table~\ref{table:results}) at the reference time, $t_{\mathrm{ref}} = 2456829.222$~BJD$_{\mathrm{TDB}}$, during the K2 observations. Because the problem is symmetrical, we fixed the longitude of the ascending node of K2-19\,b $\Omega_{\mathrm{b}}$ at $t_{\mathrm{ref}}$ to 180\degree, and we limited the inclination of K2-19\,c to $i_c<90$\degree. In order to reduce the correlation between the jump parameters, we used the \citet{Huber2013} parameterization for the eccentricity and the argument of pericenter of K2-19\,b and c.

We used a GP regression \citep[\celerite;][]{Foreman-Mackey2017} for the model of the error terms of the transit light curves. We used the rotational kernel for K2 and an approximate Matern kernel for the rest of the datasets. For CHEOPS observations, we used the satellite roll angle as a regressor instead of the time for the rest, and we added a linear model with $\log$-background (for the transit on May 10, 2024, we also added a quadratic term). For the ECAM transits, we added linear models with the point spread function centroid shifts, full width at half maximum, and peak flux value, and a quadratic term for the airmass.

In total, the model had 149 free parameters. We used normal priors for the stellar mass and radius from Sect.~\ref{section:stellar_parameters} and uninformative prior distributions for the rest of the parameters. The joint posterior distribution was sampled using the \emcee algorithm \citep{Goodman2010, emcee}.

Our parameterization of the eccentricity and argument of pericenter for planets~b and c leads to a nonuniform prior in eccentricity. To account for this, we applied the algorithm called sampling importance resampling \citep[SIR;][]{Rubin1987} as an a posteriori correction.

\section{Results}\label{section:results}

In Table~\ref{table:results} we list the median and the 68\% credible interval (CI) of the marginal distribution of the inferred system parameters. The one- and two-dimensional projections of the posterior sample of selected system parameters are shown in Fig.~\ref{figure:pyramid}.
The MAP model of the transit photometry is plotted in Fig.~\ref{figure:phot}. The noise-model-corrected transits for each planet are plotted in Figs.~\ref{figure:transit1} to \ref{figure:transit4}. Figure~\ref{figure:orbits} shows the posterior of the planet orbits. 
The value of $P_{\rm rot}$ derived from the K2 data with the \celerite rotational kernel is $18.36 \pm 0.51$~days, which is lower ($3.7\,\sigma$) than the $20.54 \pm 0.30$~days of \citep{Nespral2017} that were obtained by applying the autocorrelation function \citep{McQuillan2013} to the K2 light curve.

The current transit-timing $1\,\sigma$ uncertainty for K2-19\,d and e is about 7 and 33~hours, respectively. With a transit depth of about 190 and 270 ppm for K2-19\,d and e, respectively, these planets are difficult to follow-up (or confirm). Notably, their shallow transit depths place them below the detection threshold of TESS, which otherwise would have been well suited for recovering planets with currently poorly constrained ephemerides. In addition, K2-19\,e might not always transit in the long term (Sect.~\ref{section:long-term}). If the candidate K2-19\,e is not real, the data provide an upper limit of 2.0~\MEarth at a confidence of 99\% for a body in 5:3 MMR with K2-19 c. We verified the results for planets b and c did not change significantly when planet e was excluded from the model.
In the subsequent sections, we concentrate on K2-19\,b and c because we have achieved a more detailed characterization of these planets.

\subsection{Evolution of the orbital parameters during the observations}\label{section:short-term}

\begin{figure}
  \centering
  \includegraphics[width=0.48\textwidth]{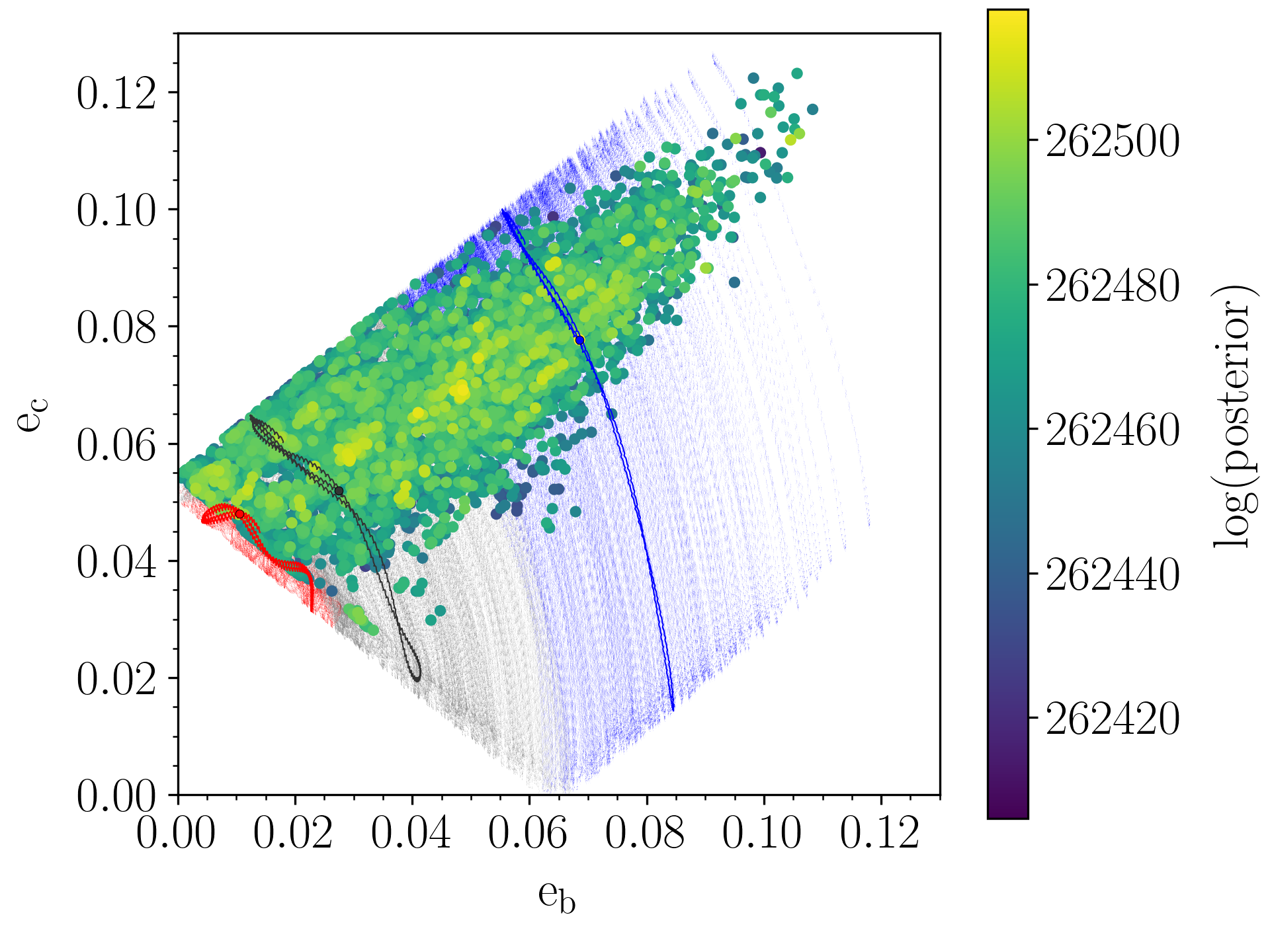}
  \caption{Correlation of the eccentricities of K2-19\,b and K2-19\,c. The dots represent the posterior samples at $t_{\mathrm{ref}}$, and the color scale shows their $\log$-posterior value. The evolution during the observations for 1000 random draws from the posterior distribution is shown in light gray (black for a random sample), light red (red for a random sample), and light blue (blue for the MAP model). The color coding depends on whether $\varpi_b-\varpi_c$ circulates, librates around $\pi$, or librates around zero (the same scheme as in Fig.~\ref{figure:ecc_longperi}). The blue dot represents the MAP model values at $t_{\mathrm{ref}}$ (and from there, it starts to move upward), and similarly for the two random samples in red (left) and black (upward). Section~\ref{section:TTVs_ecc} discusses the shape of the domain and resonance angle behavior.} \label{figure:ecc}
\end{figure}

\begin{figure}
  \centering
    \includegraphics[width=0.49\textwidth]{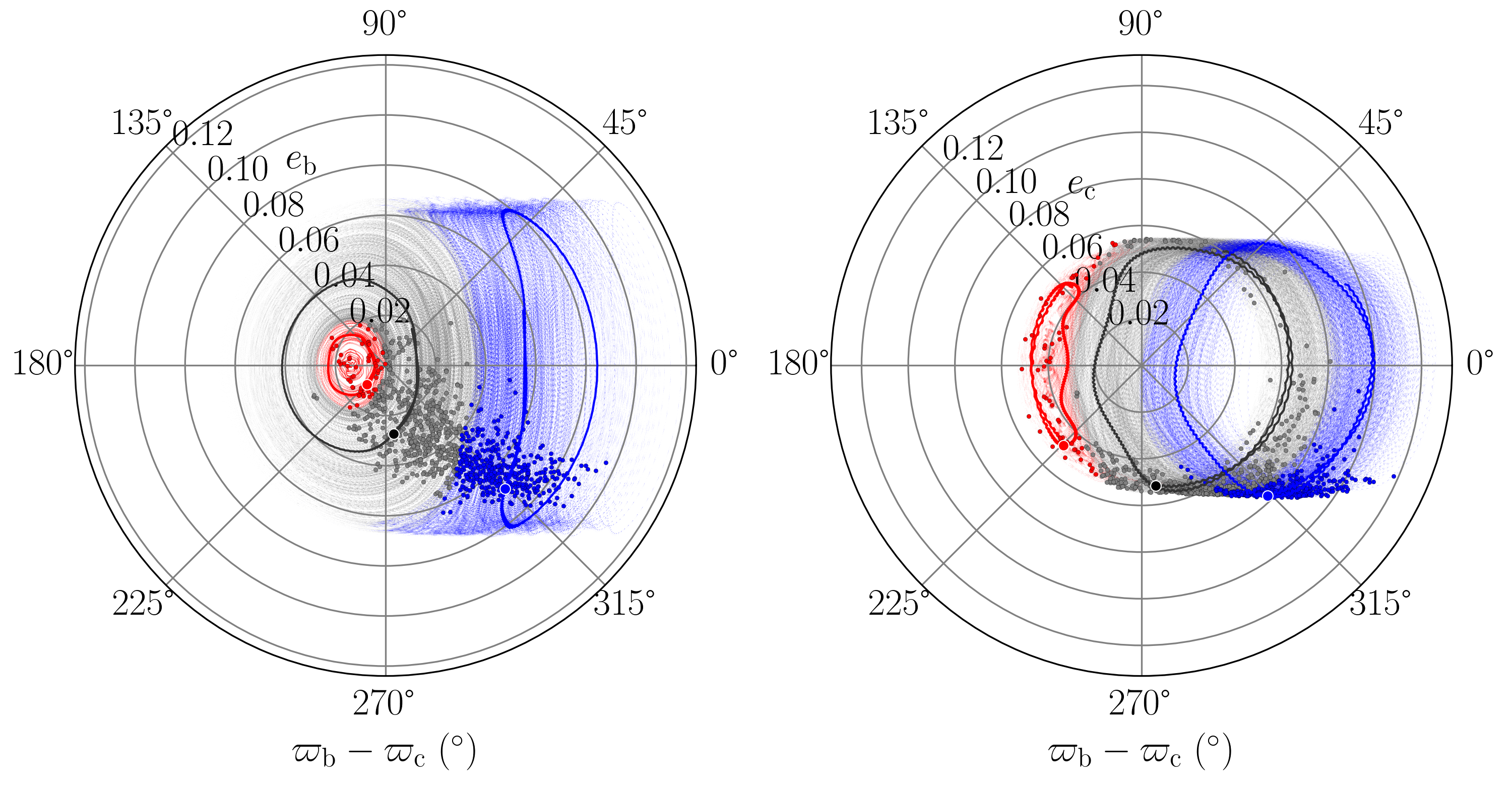}
  \caption{Correlation of the eccentricity of K2-19\,b (left panel) and K2-19\,c (right panel) with the difference in the longitudes of periastron over the time span covered by the observations for 1000 random draws from the posterior distribution. Samples that circulate are shown in light gray (one random sample is shown in black), those that librate around $\varpi_{\mathrm{b}}-\varpi_{\mathrm{c}} = 0\degree$ are shown in light blue, and those that librate around $\varpi_{\mathrm{b}}-\varpi_{\mathrm{c}} = 180\degree$ are shown in light red (one random sample is shown in red). The MAP model, which librates around 0, is shown in blue. The dots, with the same color code, represent the positions at $t_{\mathrm{ref}}$.} \label{figure:ecc_longperi}
\end{figure}

We studied the changes in the orbital parameters of the planets during the period covered by the observations (10~years) from 1000 random draws from the posterior distribution. 
The eccentricities of K2-19\,b and c vary significantly during the observations for most posterior samples and are correlated. The periodicity is generally dominated by the superperiod (Fig.~\ref{figure:ecc}).
The correlation between the eccentricities and difference in the longitudes of periastron exhibits three types of behavior (see Fig.~\ref{figure:ecc_longperi}): Libration around $\varpi_{\mathrm{b}}-\varpi_{\mathrm{c}} = 0\degree$ (53\% of the posterior samples), circulation (41\%), and libration around $\varpi_{\mathrm{b}}-\varpi_{\mathrm{c}} = 180\degree$ (6\%). This behavior is discussed in detail in the companion paper (hereafter C2).

The period ratio librates around its mean value $3/2+\Delta\sigma$ with a small amplitude $A_\sigma\Delta\sigma$ and with a period equal to the resonant period (see Sect.~\ref{section:TTVs}), each determined by conditions at the time of formation, which generally result in two-planet systems having period ratios slightly larger than exact commensurability \citep{Fabrycky2014}. The posterior median and 68.3\% CI values for $\Delta\sigma$ and $A_\sigma$ are $0.00291^{+0.00001}_{-0.00002}$ and $0.40\pm 0.02$, respectively, suggesting that the system is resonant and close to its equilibrium state (Sect.~\ref{section:TTVs}).

\subsection{Secular behavior}\label{section:long-term}

Over longer timescales, the precession of the orbital planes exhibits a period of approximately 165 years (Fig.~\ref{figure:LongTermEvolution}), which results in oscillations in both the difference in orbital inclinations and the relative longitudes of the ascending nodes. During certain epochs, this configuration permits mutual eclipses between planets b and c. 

\subsection{Transit-timing variations: Mass estimates}\label{section:TTVs}

\begin{figure}
  \centering
  \includegraphics[width=0.47\textwidth]{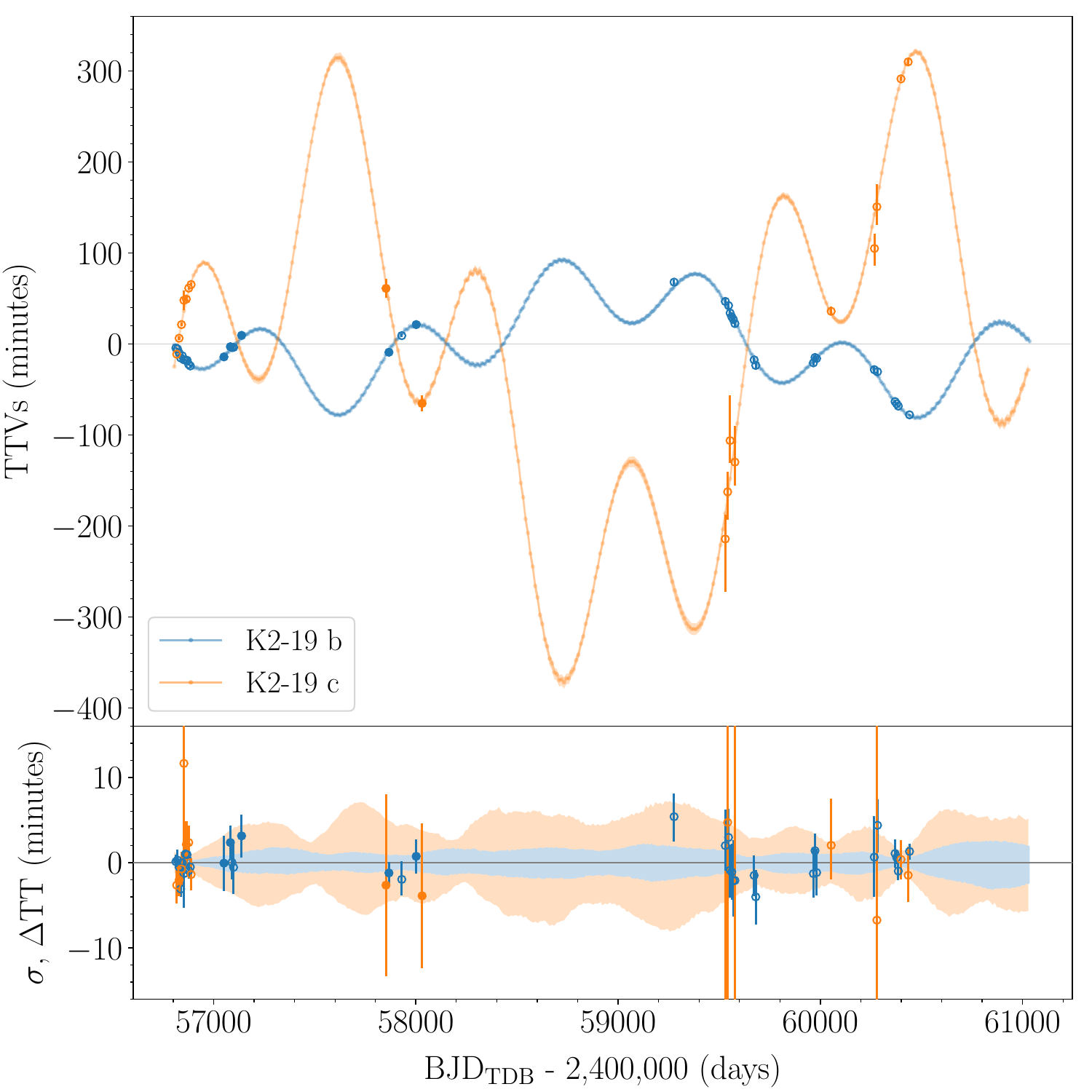}
  \caption{Posterior TTV predictions of K2-19\,b (blue band) and K2-19\,c (orange band) computed relative to a linear ephemeris (2456813.386969 + 7.9209002\;$\times$\;epoch [BJD$_{\rm TDB}$], 2456817.280436 + 11.8983248\;$\times$\;epoch [BJD$_{\rm TDB}$] for planet b and c, respectively). We used 1000 random draws from the posterior distribution to estimate the TTV median value and its uncertainty (68.3\% CI). In the upper panel, the posterior TTV values are shown and compared with individual transit-time determinations (Sect.~\ref{section:transit_times}, open and filled error bars). The filled error bars were used as data in the modeling. In the lower panel, the posterior median transit-timing value was subtracted to visualize the uncertainty of the distribution. The posterior median transit time was also subtracted from each observed epoch for the individual transit-time determinations to allow a better comparison with the posterior of the photodynamical modeling.} \label{figure:TTVs}
\end{figure}

Figure~\ref{figure:TTVs} 
shows the posterior TTVs of K2-19\,b and c, obtained from the times of minimum projected separation of the star and planet.
Two well-separated periodicities are clearly evident, with 
the ratio of the associated super and (shorter, noncommensurate) resonant periods, $P_S$ and $P_R$, respectively,
which immediately indicates that the system is formally resonant and in the librating state, even before any analysis is performed. 
Moreover, the magnitudes of the amplitudes of each component
are similar, which reveals (again without a detailed analysis) 
that the system is close to (but not at) the equilibrium state of the resonance 
(cyan star in Fig.~\ref{figure:poincare}):
This is the primary fixed point of the phase space in which the system resides. 
An algorithm for drawing these conclusions by simple inspection is given in C2.

\begin{figure}
  \centering
\includegraphics[width=0.49\textwidth]{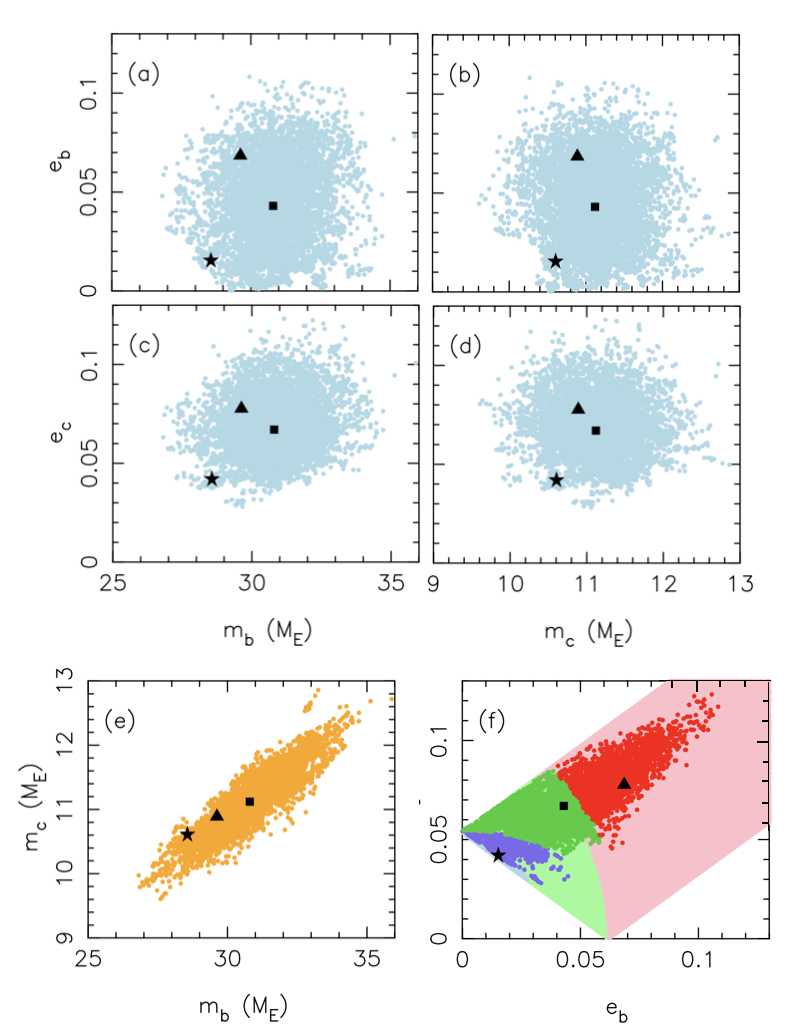}
  \caption{Posterior projections of the mass and eccentricity distributions. They demonstrate that the eccentricities and planet masses are hardly correlated (panels a-d), while the planet masses themselves are correlated (panel e),
  as are the eccentricities (panel f). The color scheme in (f) is {\purple purple} for when both resonance angles librate such that
  $\varpi_b-\varpi_c\equiv\Delta\varpi$ librates around $\pi$;
  {\green green} for when $\phi_1$ circulates and $\phi_2$ librates so that $\Delta\varpi$ circulates; 
 {\red red} for when both resonance angles circulate such that $\Delta\varpi$ librates around zero; and
 {\lightpink pink} and {\lightgreen light green} for the possible values of the eccentricities given the MAP value of $u$ 
 and arbitrary values of $\upsilon$.
   The true resonance angle $\psi$ librates with a small amplitude around zero 
  for all posterior solutions (Sect.~\ref{section:TTVs_ecc} and Fig.~\ref{figure:poincare}). The stars indicate the posterior
  sample closest to the fixed-point eccentricities, 
  and the squares and triangles indicate the median and MAP values, respectively.
    Fig.~\ref{figure:pyramid} shows the projections of the other parameters.
  } 
  \label{figure:eccs-masses}
\end{figure}

When a system is in this state and the quality of the timing data is high enough, the Bayesian analysis process is expected to find
little correlation between the masses and eccentricities, in contrast to
the mass-eccentricity degeneracy that is understood to plague many systems exhibiting TTVs.
Figure~\ref{figure:eccs-masses}
%
shows selected posterior projections (see also Figure~\ref{figure:pyramid}), demonstrating that this is indeed the case
for the K2-19 system, while the planet masses themselves are correlated, as are the two eccentricities 
(panels (e) and (f), respectively).
A full analysis of the nature of these correlations is presented in C2 \citep[see also][]{Nesvorny2016}, but we note that the masses can be determined independently when the resonant and superfrequencies are distinct and their associated Fourier contributions to the TTV signal can be accurately separated
and modeled. Moreover,
while in general the interference between the resonant and super harmonics is nonlinear,
for a system close to the primary fixed point such as K2-19\,b,c there is little interference, and
the masses can be directly determined from the amplitudes of the supercomponent together with 
the ratio of the superperiod and resonant period. This result is summarized below, and a more general treatment for systems that are not necessarily close to the primary fixed point (using the information in the resonant harmonic)
is presented in C2.

The analysis in C2 is based on a new formalism for the study of the dynamics of systems near first-order resonances
(Mardling, in prep.). A key result we used here is that for a system that is close to the primary fixed point of the 
forced part of the associated phase space,\footnote{We refer to the forced and free eccentricity
and defer their definitions to C2.}
a quantity $R$ called the {\it resonance parameter} relates the ratio of the superperiod to the resonant 
period to the planet-to-star mass ratios such that
\bea
R
&=&
(P_S/P_R)^2-1\nonumber\\
&=&
3\left(\frac{\sigma}{\Delta\sigma}\right)^2\left(\overline m_b+ \alpha\, \overline m_c\right)
\left(\mv_1(\alpha)\,e_b^{(+)}+\mv_2(\alpha)\,e_c^{(+)}\right),
\label{PR}
\eea
for a system close to the $n+1:n$ commensurability. Here, $\sigma=(n+1)/n$,
$\alpha=\sigma^{-2/3}$ is the ratio of the semimajor axes at exact commensurability, $\Delta\sigma$
is the distance of the primary fixed point to exact commensurability,\footnote{Note that for a long enough baseline,
$\Delta\sigma\simeq \Delta S\equiv\bar P_c/\bar P_b-(n+1)/n$ for systems close to the fixed point.}
$\overline m_{b,c}=m_{b,c}/m_*$ are the planet-to-star mass ratios,
$\mv_1(\alpha)$ and $\mv_2(\alpha)$ are constants related to Laplace coefficients, such that $\mv_1=-2.025$ and 
$\mv_2=2.484$ for $\sigma=3/2$, and
\be
e_b^{(+)}=\frac{\alpha^{-1/2} |\mv_1|\,\overline m_c}{n\Delta\sigma}
\hand
e_c^{(+)}=\frac{\mv_2\,\overline m_b}{n\Delta\sigma}
\label{ebc}
\ee
are the values of the forced eccentricities at the primary fixed point. Equation~\rn{PR}
represents one constraint on the three quantities $\overline m_b$, $\overline m_c$, and $\Delta\sigma$.
The other two come from the amplitudes of the supercomponents of the TTV signals, as described below.

\begin{figure}
  \centering
\includegraphics[width=0.45\textwidth]{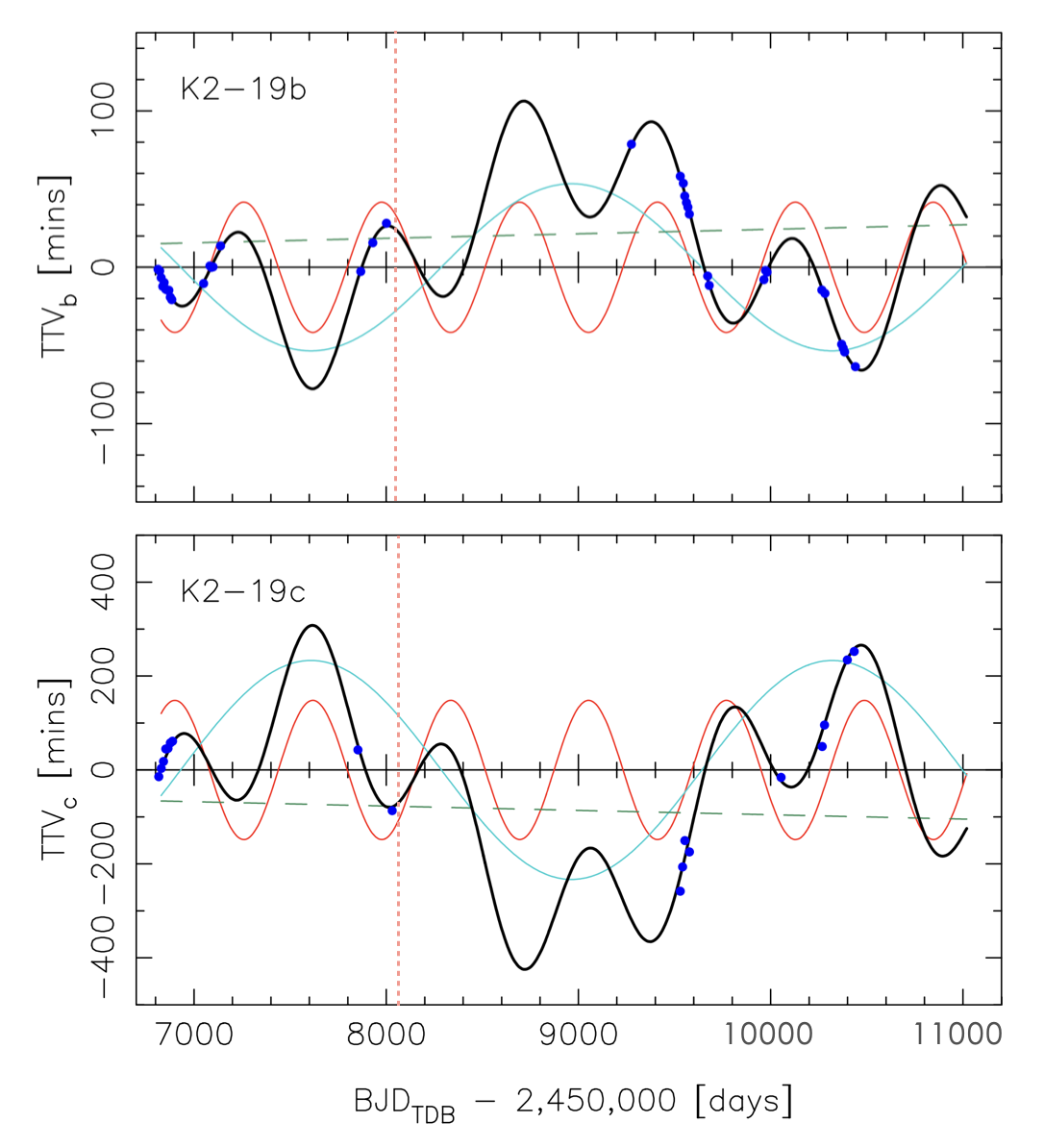}
  \caption{Least-squares fits of the TTV data shown in Fig.~\ref{figure:TTVs} (blue dots).
  The red and cyan curves show the individual resonant and superharmonics, while the dashed green lines show
  the linear offsets. The black curves show the sum of all three, and the vertical dotted lines indicate the observing baseline
  of \citet{Petigura2020}.
  The two panels use different $y$-axis scales.  
  } 
  \label{figure:fourier-fits}
\end{figure}

As discussed in C2, an accurate model for the TTVs of a 
system in the librating state is such that the $jth$ TTV of planet b is given by
\bea
{\rm TTV}_b(j)
&=&
A_S^{(b)}\cos[\nu_S (t_j-t_{\rm ref})+\beta_S]\nonumber\\
&&\hspace{-0.2cm}
+A_R^{(b)}\cos[\nu_R (t_j-t_{\rm ref})+\beta_R]+c_1^{(b)}t_j+c_0^{(b)}
\label{TTVb}
\eea
with a similar expression for the $kth$ TTV of planet c.
Here, $t_j$ is the $jth$ midtransit time, $t_{\rm ref}$ is the reference time,
$\bar P_b$ is the long-term average value of the inner orbital period,
which is well approximated by its linear ephemeris value for a long enough baseline,
$A_S^{(b)}$ and $A_R^{(b)}$ are the amplitudes of the supercomponent and resonant component, $\beta_S$ and $\beta_R$
are the associated phases, and $c_0^{(b)}$ and $c_1^{(b)}$ are constants that tend to zero in the case of an 
unlimited observing baseline (or more generally, model any long-term secular trends that are not accounted for in the model).
For our purposes, we have in particular that
\be
A_S^{(b)}\simeq\bar P_b\,e_b^{(+)}/\pi\propto\overline m_c
\hand
A_S^{(c)}\simeq\bar P_c\,e_c^{(+)}/\pi\propto\overline m_b,
\label{ASbc}
\ee
where $e_b^{(+)}$  and $e_c^{(+)}$ are given in \rn{ebc}, 
while $A_R^{(b)}\propto\overline m_c$ and $A_R^{(c)}\propto\overline m_b$ and the phases 
$\beta_S$ and $\beta_R$ depend 
on the forced eccentricities away from their fixed-point values (C2).
If the K2-19\,b,c system were precisely at the fixed point, we would have
$A_R^{(b)}=A_R^{(c)}=0$ and 
\be
\beta_S=n\lambda_b(t_{\rm ref})-(n+1)\lambda_b(t_{\rm ref}),
\ee
the latter being minus the longitude of conjunctions at the reference time,
while if the system were farther from the fixed point, the resonant contribution would dominate the TTV signal.
Figure~\ref{figure:fourier-fits} 
%
plots the best fits to the median TTV values shown in Fig.~\ref{figure:TTVs}, 
and the individual contributing harmonics and residual linear trends, verifying that the 
superamplitude and resonant amplitude are similar, while $P_S/P_R\simeq 3.8$. The values for the Fourier parameters
are listed in  Table~\ref{table:fourier},
%
%
\begin{table*}
  \tiny
\centering
\caption{Fourier and photodynamical mass estimates}\label{table:fourier}
\begin{tabular}{lccccccc}
\hline
Parameter  & Units & Median and 68.3\% CI  & MAP & \phantom{+}$\nu_R/\nu_S$, $A_S$$^{(1)}$ 
&  \phantom{-}Petigura$^{(2)}$ & From Radii$^{(3)}$ \\
&& Photodyn & Photodyn & Fourier & Photodyn\phantom{-}  \\
&& 10 yr & 10 yr & 10 yr    & 3.3 yr \phantom{-} &  \\
\hline
\vspace{-2mm}
&\\
$m_b$  & $M_\oplus$ &  ${\bf 30.8\pm 1.3}$  & \bf{29.62}  & {\bf 30.03}  &  { $\bf 32.4\pm 1.7$}  & 39.6 \\
$m_c$  & $M_\oplus$ &  {$\bf 11.12\pm 0.44$}  & {\bf 10.94} & {\bf 11.08}   &  { $\bf 10.8\pm 0.6$} & 18.0\\
\vspace{1mm}
$\Delta m_{b,c}/m_{b,c}$ &  &    & $-0.04$, $-0.02$  & $-0.02$, $-0.004$      & $+0.05$, $+0.03$   & $+0.29$, $+0.62$           \\
%
%
$\rho_b$ & ${\rm g\,cm}^{-3}$ & 0.59 & 0.57 & 0.58  &  0.52 & 0.76  \\
\vspace{1mm}
$\rho_c$ & ${\rm g\,cm}^{-3}$  & 1.05 & 1.03 & 1.04  &  0.86 & 1.69  \\
$P_S$  ($\overline P_S$) & d &  &  & 2711.2    & & (2751.9) \\
\vspace{1mm}
$P_R$ & d &  & & 717.4     \\
\vspace{1mm}
$A_S^{(b)}$, $A_R^{(b)}$ &  min &  & & 53.4, (41.7)    \\
\vspace{1mm}
$A_S^{(c)}$, $A_R^{(c)}$  &  min &  & & 233.2, (148.3)   \\
\vspace{0.5mm}
$\beta_S$, \, $\beta_R$ & deg & $[78.1]^{+1.5}_{-1.4}$ \hspace{1.0mm} $[137.3]^{+2.2}_{-1.9}$ & [76.1], [138.1]& 
(76.5, 144.3)  \\
\vspace{0.5mm}
$\lambda_n$ & deg & $85.8^{+1.7}_{-1.9}$  &  82.8 &    & \phantom{-}37.5$^{(4)}$ \\
$R$ & & $[14.81]^{+0.37}_{-0.26}$ & [15.49] & 13.28   & [15.7]\\
\vspace{1mm}
$\Delta\sigma$ & & $[0.0029]^{+1.e-5}_{-2.e-5}$ &  [0.0029]  &  0.0030    &  [0.0029]\\
\vspace{1mm}
$e_b^{(+)}$ ($e_b$)    &  & $[0.015]^{+0.0002}_{-0.0002}$ ($0.043^{+0.024}_{-0.024}$)  & [0.015] (0.069) & 0.015   &  [0.015]  ($0.20^{+0.03}_{-0.03}$) &   \\
\vspace{1mm}
$e_c^{(+)}$  ($e_c$)  &  & $[0.044]^{+0.0005}_{-0.0004}$ ($0.067^{+0.017}_{-0.017}$) & [0.045] (0.078)& 0.043    &  [0.047]  ($0.21^{+0.03}_{-0.03}$)&  \\
\hline
\end{tabular}
\tablefoot{\tiny 
\phantom{.}\\
(1): $\nu_R/\nu_S$, $A_S$ / Fourier A uses the frequency ratio and superamplitudes only to
estimate masses;\\
(2): Petigura / Photodyn uses system parameters from the photodynamical analysis of \citet{Petigura2020};\\
(3): Estimates using mass-radius relation $m_p/M_E=(R_p/0.56\,R_E)^{1/0.67}$ \citep{Mueller2024};\\
(4): \citet{Petigura2020} value minus $180^{\rm o}$ to account for the fact that  the authors assumed $\Omega_b(t_{\rm ref})=0$;\\
$[${\it Square brackets}$]$ show the quantity calculated analytically from elements and masses using the formalism of Mardling (2025a);\\
$\Delta m_{b,c}/m_{b,c}=$ relative difference in masses compared to median value;\\
$\overline P_S$ (defined from ephemeris periods such that $\bar P_S^{-1}= n\bar P_b^{-1}-(n+1)\bar P_c^{-1}$) listed in the last column for comparison with the Fourier superperiod;\\
{\textit{\textbf {Boldface entries}}}: highlighted for comparison;\\
Amplitudes and phases in brackets were not used directly in mass estimates.  \\
Last two rows: Entries in brackets are the median and 68.3\% CI eccentricities for comparison with the fixed-point values $e_b^{(+)}$ and  $e_b^{(+)}$.\\
The reference mean densities are $\rho_{\rm Saturn}=0.69\,{\rm g\,cm}^{-3}$, $\rho_{\rm Uranus}=1.15\,{\rm g\,cm}^{-3}$.
}
\end{table*}
together with $R$, $\Delta\sigma$, $e_b^{(+)}$ and $e_c^{(+)}$.
The planet mass estimates listed in the column labeled ``Fourier'' were obtained by solving
\rn{PR}, \rn{ebc} and \rn{ASbc} for $\overline m_b$, $\overline m_c$ and $\Delta\sigma$, yielding
values that are 2.5\% and 0.4\% lower than the median values obtained from the photodynamical analysis
for planets b and c, respectively, and are well inside our error bars.
We also list our median and MAP values for the planet masses (taken from Table~\ref{table:results}) and those of \citet{Petigura2020}. 

The accuracy of the Fourier mass estimates can be attributed to a good estimate of $\Delta\sigma$ via the resonance parameter $R$ (which is vital because the superamplitudes go like $m_{b,c}/\Delta\sigma$); if we had used $\Delta S=\bar P_c/\bar P_b-(n+1)/n$ as an estimate for $\Delta\sigma$ in \rn{ASbc}, the planet masses would both have been underestimated by about 30\%.

\subsection{Transit-timing variations: Eccentricity constraints}\label{section:TTVs_ecc}

The high eccentricities of K2-19\,b and c reported by \citet{Petigura2020} (both about 0.2) have inspired much 
discussion about the formation of the system,
including the influence of additional planets \citep[e.g.,][]{Petit2020,Laune2022,Choksi2023,Goldberg2023}
and stochastic stirring by turbulent eddies \citep{Goldberg2023}.
Based on seven additional years of transit data, however, our posterior is significantly better constrained,
with median and 68.3\% CI values at $e_b=0.043\pm0.024$ and $e_c=0.067\pm 0.017$.    
While substantially lower than 0.2, these values still imply some free eccentricity,
with values lower than 0.005 unlikely at a confidence of 99\%. 
Panel (f) of Figure~\ref{figure:eccs-masses} again shows the posterior coverage of 
the $e_b-e_c$ plane (see the figure caption for the meaning of the colors and the discussion below). 
The shape of this posterior projection
is a result of a third integral (in addition to the total energy and angular momentum), which exists at first order
in eccentricity for systems that are close to a first-order commensurability. Because of this integral,
the free components of the Runge-Lenz vectors for each planet orbit are approximately constant,\footnote{The
median and 68.3\% CI values of their magnitudes are \hbox{$e_b^{({\rm free})}=0.046\pm0.023$} and 
\hbox{$e_c^{({\rm free})}=0.037\pm 0.019$}, and are related via $e_b^{({\rm free})}/e_c^{({\rm free})}=\mv_2/|\mv_1|$ (C2).} and as a result, their contribution to the TTVs is limited because their variation in time is second order in 
eccentricity. This leads to the degeneracy
that is visible in the figure and is confined to the upper half of the allowed region (the latter being the entire colored part) as a result of second-order effects and of the sensitivity to the average transit duration time (see the next section).

\begin{figure}
  \centering
\includegraphics[width=0.5\textwidth]{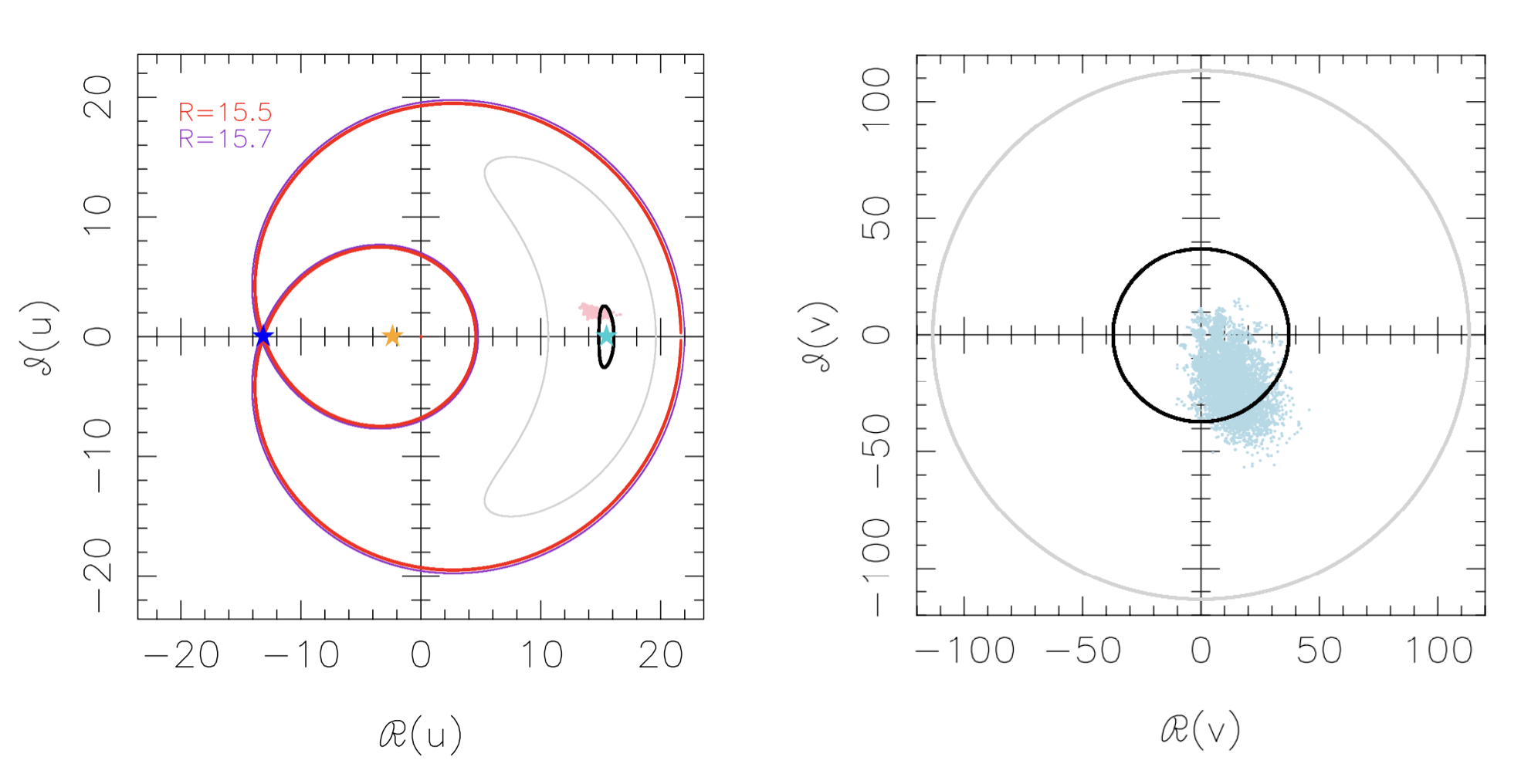}
  \caption{Phase space in which K2-19\,b,c reside. 
{\it Left panel}: Our posterior values of the real and imaginary parts of $u$ ({\lightpink 7000 pink points} using Eq.~\rn{u}); 
a complete resonant cycle starting from the MAP value of $u$ (black curve);
the associated primary fixed point for which $u=R$ ({\cyan cyan star}) and the secondary and hyperbolic fixed points
({\orange orange} and {\blue blue stars}, respectively, these two only exist when the phase space is resonant);
the associated separatrix ({\red red curve});
a complete resonant cycle using the system parameters of \citet{Petigura2020} for the initial value of $u$ 
({\grey gray curve});
and the associated separatrix, which is almost identical to ours ({\purple purple curve}).
{\it Right panel}: 
Our posterior values of the real and imaginary parts of $\upsilon$ ({\lightblue 7000 light blue points} using Eq.~\rn{v}); 
a complete supercycle starting from the MAP value of $\upsilon$ (black curve)
and the system parameters of \citet{Petigura2020} ({\grey gray curve}).
Each panel uses different scales.
  } 
  \label{figure:poincare}
\end{figure}

The colors in Figure~\ref{figure:eccs-masses}(f) 
are associated with the libration and/or circulation behavior of the standard resonance angles
$\phi_1=n\lambda_b-(n+1)\lambda_c+\varpi_b$
and
$\phi_2=n\lambda_b-(n+1)\lambda_c+\varpi_c$
and the difference in the longitudes of periastron $\varpi_b-\varpi_c$; this can be derived 
from the new definitions of forced
and free eccentricity given in C2 together with the complex functions\footnote{ $u$ and $\upsilon$ are the equivalent of $Y_1$ and $Y_2$ in \citet{Petit2020} with different scalings.}
\be
u=-3\left(\frac{\sigma}{\Delta\sigma}\right)^2\left(\overline m_b+ \alpha\, \overline m_c\right)
\left(\mv_1\,z_b+\mv_2\,z_c\right)
\label{u}
\ee
and
\be
\upsilon=-3\left(\frac{\sigma}{\Delta\sigma}\right)^2\left(\overline m_b+ \alpha\, \overline m_c\right)
\left(g^{-1}\mv_2\,z_b-g\mv_1\,z_c\right),
\label{v}
\ee
posterior values that are plotted in Figure~\ref{figure:poincare}.
%
Here, $g=(m_c/m_b)^{1/2}\alpha^{-1/4}$,
$z_1=e_b\,{\rm e}^{i\phi_1}$
and
$z_2=e_c\,{\rm e}^{i\phi_2}$. We note from \rn{PR} that $u=R$ at the primary fixed point (at which
$\phi_1=0$ and $\phi_2=\pi$). We also note that the true resonance angle $\psi$, defined such that
$u=|u|{\rm e}^{i\psi}$, librates with a small amplitude for the whole posterior.
Finally, we note that the third integral mentioned above is simply $\upsilon\upsilon^*={\rm constant}$
(so that the radius of the black curve in the right panel of Figure~\ref{figure:poincare} is constant).
The details are provided in C2.

\subsection{The transit duration: Further eccentricity constraints}\label{section:TDVs}

\begin{figure}
  \centering
  \includegraphics[width=0.49\textwidth]{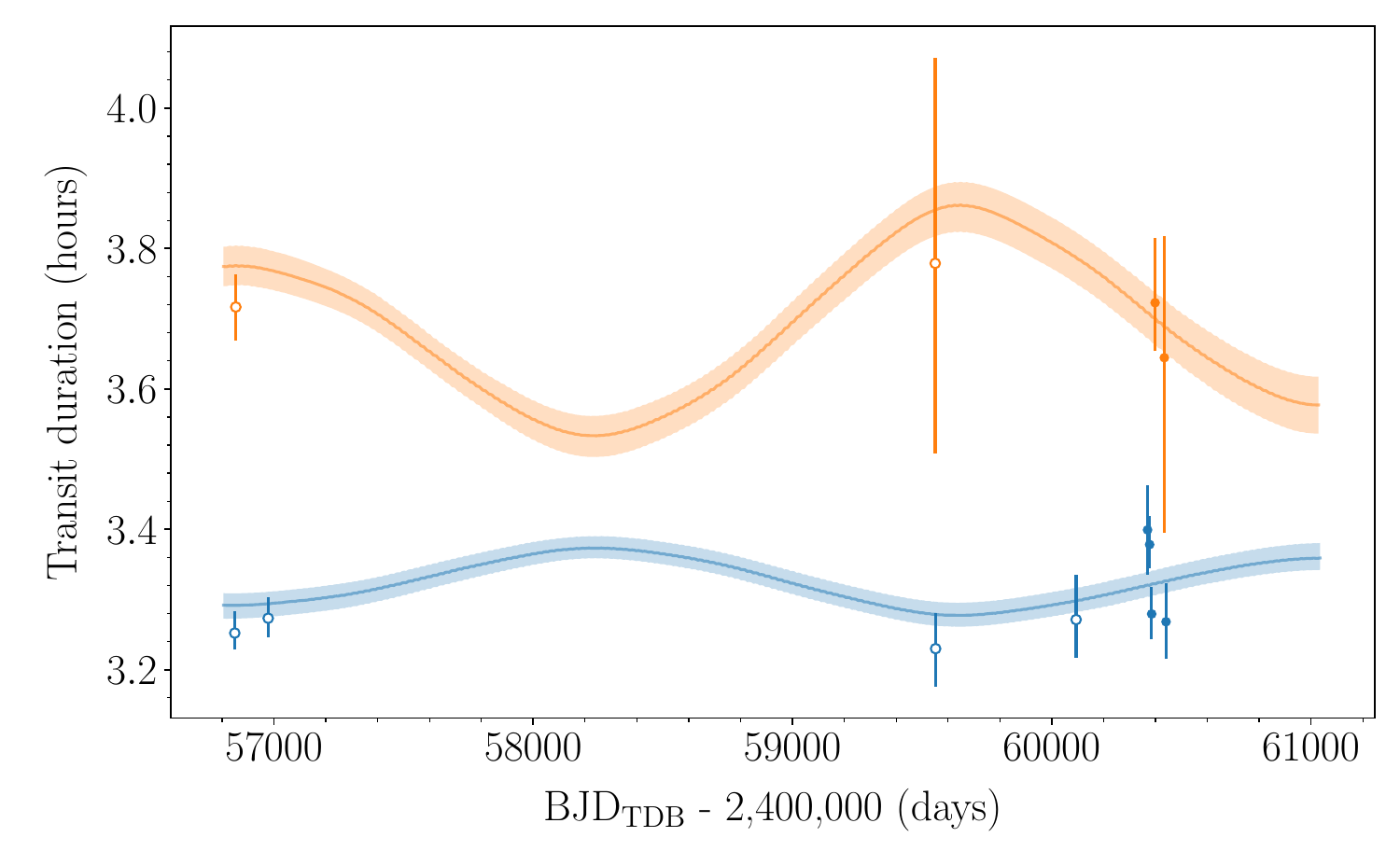}
  \caption{Posterior of the transit duration of K2-19\,b (blue band) and K2-19\,c (orange band). We used 1000 random draws from the posterior distribution to estimate the median transit duration and its uncertainty (68.3\% CI). The error bars represent the transit durations we measured for individual transits (filled dots) or groups of transits (open dots placed at the average timing of the transits) using the procedure described in Sect.~\ref{section:transit_times}.} \label{figure:TDVs}
\end{figure}

The transit duration was computed as the difference between the first and fourth contact using the sky-projected planet-star separation. The transit durations of K2-19\,b and c are shown in Fig.~\ref{figure:TDVs} 
and exhibit transit duration variations (TDVs). 
The transit duration of K2-19\,b varies by about 6~minutes peak to peak over the 10-year baseline, and
that of K2-19\,c varies by about 20~minutes. This is
3\% and 11\%, respectively, of the median transit durations. As we show in C2
the variation period is the superperiod, and the amplitudes are proportional to the primary fixed-point eccentricities
$e_b^{(+)}$ and $e_c^{(+)}$ and not the true eccentricities, as is commonly assumed. 
The {\it average} transit durations of planets b and c do provide constraints via the
quantities $e_b\sin\omega_b$ and $e_c\sin\omega_c$, however, which further confines the $e_b-e_c$ posterior projection
in the upper half of the colored region in panel (f) of Figure~\ref{figure:eccs-masses}.

\subsection{Comparison with previous works}\label{section:comparison}

The first photodynamical analysis of K2-19 with a time span of 285~days was presented by \citet{Barros2015}. They found $e_{\rm b} = 0.119^{+0.082}_{-0.035}$ and $e_{\rm c} = 0.095^{+0.073}_{-0.035}$ at 2456813~BJD$_{\rm TDB}$. These values are different by $1.8\,\sigma$ ($e_{\rm b}$) and $0.7\,\sigma$ ($e_{\rm c}$) from our results. Their mass ratios $m_b/m_\star = (13.4^{+4.3}_{-2.9})\times10^{-5}$, $m_c/m_\star = (5.5^{+2.0}_{-1.3})\times10^{-5}$, and $m_c/m_b = 0.42 \pm 0.12$ agree with our results with differences of $1.0\,\sigma$, $1.3\,\sigma$, and $0.5\,\sigma$, respectively. 

The next photodynamical analysis of K2-19 was performed by \citet{Petigura2020} with a time span of 3.3~years. They obtained osculating $e_{\rm b} = 0.20 \pm 0.03$, $e_{\rm c} = 0.21 \pm 0.03$, and well-aligned apsides ($\omega_{\rm c} - \omega_{\rm b} = 2 \pm 2\degree$). We did not find the reference time for the orbital parameters in \citet{Petigura2020}. Assuming that their reference time is close to ours, the difference from our result is $4.1\,\sigma$ for both planets, which cannot be explained by orbital evolution if the reference time was significantly different from ours. Despite the differences in eccentricities, their masses of K2-19\,b and c agree with our results (for the reasons explained in Sect.~\ref{section:TTVs}).
Based on the difference between the transit times predicted by \citet{Petigura2020} and those observed (Fig.~\ref{figure:PetiguraTTVs}), which are up to $6.7\,\sigma$ and 14~hours, we suspect an issue in their analysis. As shown in Sect.~\ref{section:analysis_TTVs}, K2-19\,d and e cannot have caused this discrepancy, and neither can the shorter time span of the observations. Perhaps the discrepant individual timing they found for the LCOGT partial transit ($3.5\,\sigma$ off based on our results) has driven their solution. As shown in Fig.~6 of \citet{Petigura2020}, their model already struggled to fit this transit timing. When \citet{Petit2020} enforced a low-eccentricity solution, the model for the LCOGT timing fell at the same value as we found, which corresponds to $12.4 \pm 0.9$~minutes in the residuals of Fig.~7 in \citet{Petit2020}.

We briefly compare our results with the analyses that only used radial velocities from \citet{Dai2016} and \citet{Nespral2017}. \citet{Dai2016} found radial velocity amplitudes of $K_{\rm b} = 9.6^{+1.8}_{-1.6}$~\ms\ and $K_{\rm c} = 7.5 \pm 2.1$~\ms, which differ by $0.6\,\sigma$ and $2.0\,\sigma$, respectively, from our results. On the other hand, \citet{Nespral2017} found $K_{\rm b} = 18.8 \pm 2.4$~\ms, which differs by $3.4\,\sigma$ from our results, and did not significantly detect K2-19\,c. 

\subsection{Internal structure}\label{section:internal_structure}

Internal structure retrieval is especially degenerate for gas giants. \citet{Petigura2020} used the models of \citet{Lopez2014} and found that K2-19\,b has a hydrogen-helium envelope mass fraction of $f_{\rm env} = 44 \pm 3$\% by mass, which challenges the core-accretion theory. Runaway accretion begins at about $f_{\rm env} \approx 50$\%, and fine-tuning would be required for the disk to disappear at just the moment when K2-19\,b approached the runaway phase.

We followed \citet{CastroGonzalez2024b} and investigated the internal structure of K2-19\,b using \gastli \citep{Acuna2021,Acuna2024}, which is a one-dimensional coupled interior-atmosphere model for warm ($T_{\rm eq} < 1000$~K) gas giants (between 17~\MEarth and 6~\Mjup). The interior model is composed of two layers, that is, the core and the envelope. The core is assumed to be a 1:1 mixture of silicate rock and water, whereas the envelope is a mixture of H/He and water. The model is parameterized with two values: the core mass fraction (CMF) and the mass fraction of water (representative of metals) in the envelope ($Z_{\rm env}$).  

We generated a grid of \gastli models for $T_{\rm eq}=857$~K, C/O=0.55 (solar value), masses from 24 to 38~\MEarth with increments of 0.25~\MEarth, a CMF from 0 to 0.99 with increments of 0.1, log(Fe/H) from -2 to 2.4 with increments of 0.5, and $T_{\rm int}$ from 50 to 250 K with increments of 25~K. For the retrieval, we used the planetary mass and radius and the stellar age as observable parameters. We sampled from the posterior distribution with \emcee \citep{Goodman2010, emcee}. The interior parameter priors and posteriors are listed in Table~\ref{table:interior_retrieval}. The one- and two-dimensional projections of the posterior sample \citep{corner} of selected system parameters are shown in Fig.~\ref{figure:interior_retrieval}. The envelope mass fraction \citep[$f_{\rm env}$ for][]{Lopez2014} is $0.283^{+0.38}_{-0.049}$. The total water (metals) mass fraction is $Z_{\rm planet} = 0.729^{+0.041}_{-0.14}$. We compared our results with those by \citet{Thorngren2016}. We computed the stellar metal mass fraction as $Z_{\star} = 0.014 \times 10^{\rm [Fe/H]_\star}$ to infer the heavy-element enrichment relative to the host star: $Z_{\rm planet}/Z_{\star} = 48.1^{+4.1}_{-8.8}$. This value is compatible with formation via core accretion for a planet mass of $0.0968 \pm 0.0039$~\Mjup \citep[see Fig.~11 of][]{Thorngren2016}.

\begin{table}[]
\tiny
\centering
\renewcommand{\arraystretch}{1.18}
\setlength{\tabcolsep}{3pt}
\caption{Interior structure retrieval of K2-19~b with \gastli.}
\label{table:interior_retrieval}
\begin{tabular}{lcc}
\hline \hline
Parameter               & Prior & Posterior median \\
& & and 68.3\% CI                  \\ \hline
Core mass fraction, CMF                      & $U$(0.01, 0.99) & $0.718^{+0.049}_{-0.38}$ \\
Atmospheric metallicity, $\log$(Fe/H)        & $U$(-2, 2.4)    & $0.4^{+1.1}_{-1.6}$ \\
Envelope water mass fraction, $Z_{\rm env}$  &                 & $0.054^{+0.31}_{-0.049}$ \\
Total water mass fraction, $Z_{\rm planet}$  &                 & $0.729^{+0.041}_{-0.14}$ \\
Internal temperature, $T_{\rm int}$ ($K$)    & $U$(50, 200)    & $98.8^{+8.6}_{-13}$ \\
\hline  
\end{tabular}
\tablefoot{$U$(l,u): Uniform distribution prior in the range [l, u].}
\end{table}

\section{Discussion}\label{section:discussion}

\begin{figure*}
    \includegraphics[width=0.67\columnwidth]{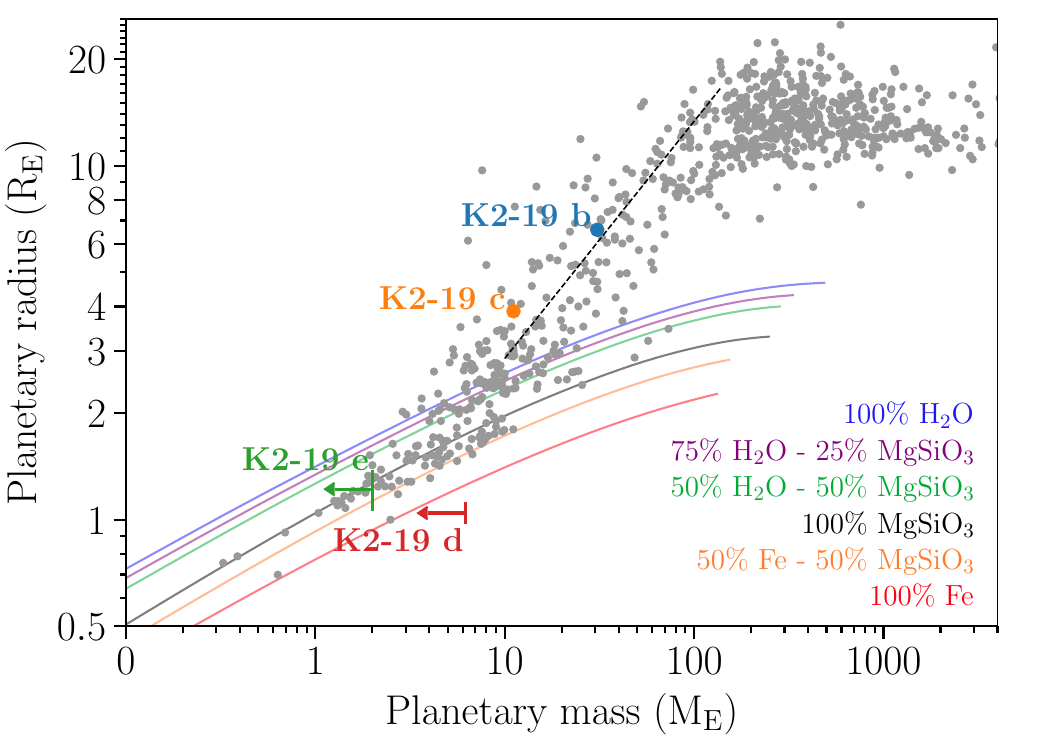}
    \includegraphics[width=0.67\columnwidth]{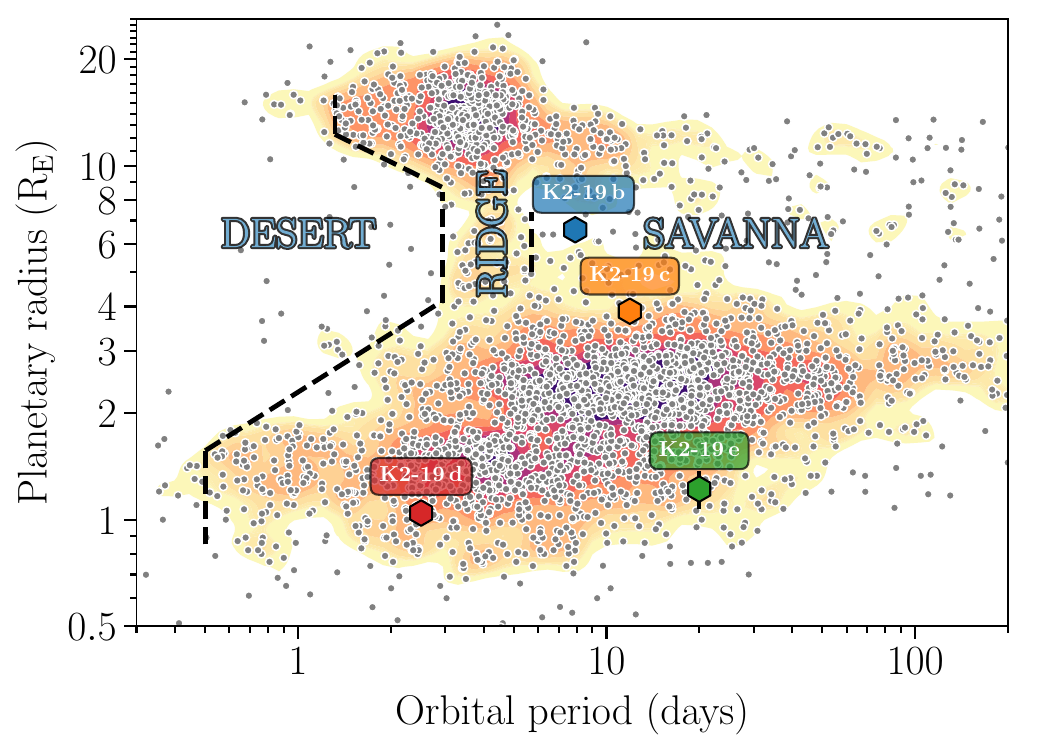}
    \includegraphics[width=0.67\columnwidth]{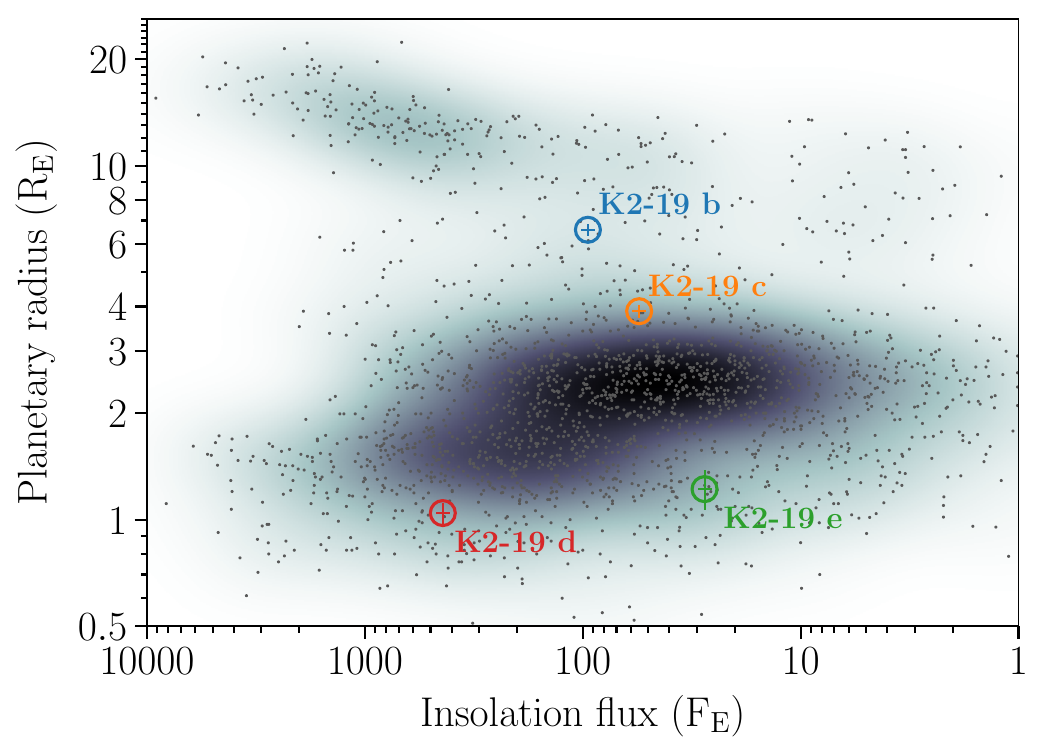}
    \caption{Comparison of the K2-19 planets with known exoplanets. {\it Left}: Mass--radius diagram of known exoplanets. The gray dots show the transiting planets listed in the NASA Exoplanet Archive \citep[\url{https://exoplanetarchive.ipac.caltech.edu/},][]{Akeson2013} with planetary radius and mass uncertainties below 20\%. The dashed black line shows the sub-Neptune mass–radius relation from \citet{Parc2024}. The solid lines represent theoretical models for different compositions \citep{Zeng2013}. {\it Center}: Idem for planetary radii vs. orbital period (restricted to planets with planetary radius uncertainties below 20\%), with exo-Neptunian boundaries from \citet{CastroGonzalez2024a}. {\it Right}: Idem for planetary radii vs. insolation flux. A Gaussian kernel density estimate is shown in different color intensities.}
    \label{figure:context}
\end{figure*}

Figure~\ref{figure:context} illustrates that K2-19\,b and c lie within the Neptune savanna \citep{Bourrier2023}, which is a moderately populated region in the period–radius diagram that is found at orbital periods longer than those that define the Neptune desert \citep[e.g.,][]{Lecavelier2007,Mazeh2016} and the Neptunian ridge \citep{CastroGonzalez2024a}. Both planets exhibit low densities ($0.590 \pm 0.053~\mathrm{g\;cm^{-3}}$ and $1.046 \pm 0.097~\mathrm{g\;cm^{-3}}$ for K2-19\,b and K2-19\,c, respectively), as anticipated for the planets in the savanna \citep{CastroGonzalez2024b}. K2-19\,c satisfies the selection criteria of the sample defined by \citet{Leleu2024}, which supports their finding that resonant sub-Neptunes tend to be puffier.

The 3:2 mean-motion resonance is one of the most common configurations observed in known multiplanetary systems \citep{Dai2024}. This resonance is thought to be particularly robust, to often act as the first capture point for migrating planets, and to provide long-term dynamical stability. The updated orbital parameters of the K2-19 system are broadly consistent with a disk-driven migration scenario. The nonzero free eccentricity suggests, however, that an additional excitation mechanism must have operated during or after migration. The system is part of the ATREIDES spin-orbit angle survey \citep{Bourrier2025}, which will bring further constraints on the 3D orbital architecture of the planets and their dynamical origin.

The precision on the planet-to-star mass ratio is 0.8\% for planet~b and 1.5\% for planet~c, and the masses have a relative precision of 4.2\% for planet~b and 4.0\% for planet~c, which are dominated by the 3.8\% uncertainty on the stellar mass. 
The two well-separated periodicities in the TTVs (super and resonant) enables a robust determination of the planetary masses independent of the eccentricities. In contrast, constraints on the free eccentricities are weak because an inherent degeneracy is associated with low-eccentricity two-planet systems, and the libration and/or circulation behavior of the usual resonance angles and difference in the longitudes of periastron is similarly degenerate. Thus, while future transit observations aimed at refining the system architecture should prioritize epochs with the largest predicted timing uncertainties, these observations are unlikely to constrain the free eccentricities much more than they are at present.

Although faint (J=11.6~mag), K2-19\,b has a transmission spectroscopy metric \citep[TSM;][]{Kempton2018} of $64.8 \pm 4.5$, which is among the ten most favorable planets of this size (6-7~\REarth) with an equilibrium temperature below 1000~K. The {\it James Webb} Space Telescope \citep{Gardner2006} might be able to measure the metallicity of the atmosphere of K2-19\,b and place an additional constraint on the planet interior.

\citet{Petigura2020} used photodynamical modeling for K2 data, but only transit timing for the other observations. As we showed in Sect.~\ref{section:analysis_TTVs}, the likely cause of their high retrieved eccentricities is a single timing from a partial transit near the end of the transit-timing dataset that was miscalculated ($3.5\,\sigma$, or $12.4\pm3.6$~minutes, off). This highlights the susceptibility of the results to single outliers when the available data are barely enough to constrain the solution. In this case, using a photodynamical model for all observations might have enabled a better error propagation and avoided an overestimation of the eccentricities in the past.

\section*{Data availability}

Light curves, samples from the posterior, and transit times forecast up to the year 2035 can be found at \url{https://zenodo.org/records/17084030}.

\begin{acknowledgements}

We are grateful to the ESO/La Silla staff for their continuous support of ExTrA. We acknowledge funding from the European Research Council under the ERC Grant Agreement n. 337591-ExTrA. 

CHEOPS is an ESA mission in partnership with Switzerland with important contributions to the payload and the ground segment from Austria, Belgium, France, Germany, Hungary, Italy, Portugal, Spain, Sweden, and the United Kingdom. The CHEOPS Consortium would like to gratefully acknowledge the support received by all the agencies, offices, universities, and industries involved. Their flexibility and willingness to explore new approaches were essential to the success of this mission. CHEOPS data analyzed in this article will be made available in the CHEOPS mission archive (\url{https://cheops.unige.ch/archive_browser/}).

This paper includes data collected by the TESS mission. Funding for the TESS mission is provided by the NASA's Science Mission Directorate. We acknowledge the use of public TESS data from pipelines at the TESS Science Office and at the TESS Science Processing Operations Center. Resources supporting this work were provided by the NASA High-End Computing (HEC) Program through the NASA Advanced Supercomputing (NAS) Division at Ames Research Center for the production of the SPOC data products. This paper includes data collected by the TESS mission that are publicly available from the Mikulski Archive for Space Telescopes (MAST).

This paper includes data collected by the Kepler mission and obtained from the MAST data archive at the Space Telescope Science Institute (STScI). Funding for the Kepler mission is provided by the NASA Science Mission Directorate. STScI is operated by the Association of Universities for Research in Astronomy, Inc., under NASA contract NAS 5–26555.

This work makes use of observations from the Las Cumbres Observatory global telescope network.

We thank the Swiss National Science Foundation (SNSF) and the Geneva University for their continuous support to our planet search programs. This work has been carried out within the framework of the National Centre of Competence in Research PlanetS supported by the Swiss National Science Foundation under grants 51NF40\_182901 and 51NF40\_205606. The authors acknowledge the financial support of the SNSF. 

This publication makes use of data products from the Two Micron All Sky Survey, which is a joint project of the University of Massachusetts and the Infrared Processing and Analysis Center/California Institute of Technology, funded by the National Aeronautics and Space Administration and the National Science Foundation.

This publication makes use of data products from the Wide-field Infrared Survey Explorer, which is a joint project of the University of California, Los Angeles, and the Jet Propulsion Laboratory/California Institute of Technology, funded by the National Aeronautics and Space Administration.

This work has made use of data from the European Space Agency (ESA) mission
{\it Gaia} (\url{https://www.cosmos.esa.int/gaia}), processed by the {\it Gaia}
Data Processing and Analysis Consortium (DPAC,
\url{https://www.cosmos.esa.int/web/gaia/dpac/consortium}). Funding for the DPAC
has been provided by national institutions, in particular the institutions
participating in the {\it Gaia} Multilateral Agreement.

Simulations in this paper made use of the \rebound code which can be downloaded freely at \url{http://github.com/hannorein/rebound}. 

This research made use of \texttt{nep-des} (available in \url{https://github.com/castro-gzlz/nep-des}).

Part of these simulations have been run on the {\it Lesta} and {\it Bonsai} clusters kindly provided by the Observatoire de Gen\`eve.

This project has received funding from the European Research Council (ERC) under the European Union's Horizon 2020 research and innovation programme (project {\sc Spice Dune}, grant agreement No 947634).
M.L. acknowledges support of the Swiss National Science Foundation under grant number PCEFP2\_194576.
I.G.J. acknowledges support of the National Science and Technology Council in Taiwan under grant number NSTC 113-2112-M-007-030.
We acknowledge grants Spanish program Unidad de Excelencia María de Maeztu CEX2020-001058-M and 2021-SGR-1526 (Generalitat de Catalunya).

In addition to the Python packages mentioned in the text, we utilized \textit{NumPy} \citep{numpy}, \textit{SciPy} \citep{scipy}, \textit{matplotlib} \citep{matplotlib}, and \textit{seaborn} \citep{Waskom2021}.

\end{acknowledgements}

\bibliographystyle{aa}
\bibliography{K2-19}

\begin{appendix}

\FloatBarrier
\section{Additional figures and tables}

\begin{table}[!ht]
    \tiny
    \renewcommand{\arraystretch}{1.25}
    \setlength{\tabcolsep}{2pt}
\centering
\caption{Modeling of the SED.}\label{table:sed}
\begin{tabular}{lccc}
\hline
Parameter & & Prior & Posterior median   \\
&  & & and 68.3\% CI \\
\hline
Effective temperature, $T_{\mathrm{eff}}$ & (K)     & $N$(5350, 110)     & $5363 \pm 74$ \\
Surface gravity, \logg\                   & (cgs)   & $N$(4.42, 0.06)   & $4.527 \pm 0.029$ \\
Metallicity, $[\rm{Fe/H}]_\star$          & (dex)   & $N$(0.03, 0.02)   & $0.026 \pm 0.019$ \\
Distance                                  & (pc)    & $N$(296.9, 1.7)   & $297.2 \pm 1.7$ \\
$E_{\mathrm{(B-V)}}$                      & (mag)   & $U$(0, 3)         & $0.030^{+0.052}_{-0.023}$ \\
Jitter Gaia                               & (mag)   & $U$(0, 1)         & $0.169^{+0.24}_{-0.092}$ \\
Jitter 2MASS                              & (mag)   & $U$(0, 1)         & $0.057^{+0.090}_{-0.037}$ \\
Jitter WISE                               & (mag)   & $U$(0, 1)         & $0.042^{+0.10}_{-0.032}$ \\
Mass, $m_\star$                           & (M$_\odot$) &               & $0.885 \pm 0.034$ \\
Radius, $R_\star$                         & (R$_\odot$) &               & $0.845 \pm 0.017$ \\
Density, $\rho_\star$                     & ($\mathrm{g\;cm^{-3}}$) &   & $2.06 \pm 0.16$ \\
Isochronal age                            & (Gyr) &                     & $4.8 \pm 4.3$ \\
Luminosity                                & (L$_\odot$) &               & $0.531 \pm 0.030$ \smallskip\\
\hline
\end{tabular}
\tablefoot{$N$($\mu$,$\sigma$): Normal distribution prior with mean $\mu$, and standard deviation $\sigma$. $U$(l,u): Uniform distribution prior in the range [l, u].}
\end{table}

\begin{figure}[t]
  \centering
  \includegraphics[width=0.5\textwidth]{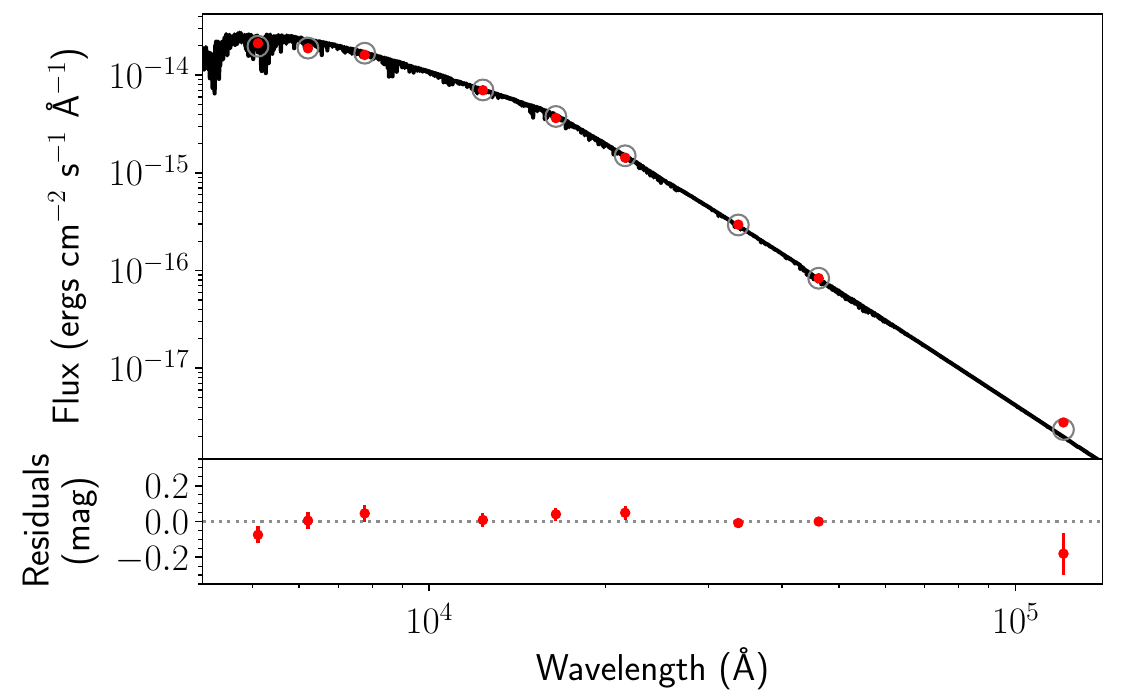}
  \caption{The SED of K2-19. {\it Top panel}: The solid line is the MAP PHOENIX/BT-Settl interpolated synthetic spectrum, red circles are the absolute photometric observations, and the open gray circles are the result of integrating the synthetic spectrum in the observed bandpasses. {\it Bottom panel}: Residuals of the MAP model (the jitter has been added quadratically to the data error bars).} \label{figure:sed}
\end{figure}

\begin{table}
\tiny
\caption{Parameters of K2-19 from the literature.}             
\label{table:literature}      
\centering          
\begin{tabular}{c c c c c c c c}
\hline\hline       
Work                        & \teff~(K) & \logg & [Fe/H]$_\star$ & [M/H]$_\star$ & $m_{\star}$ ($M_{\odot}$) & $R_{\star}$ ($R_{\odot}$) & d (pc) \\ 
\hline                    
\citet{Montet2015} & $5519^{+49}_{-82}$ & & $-0.27 \pm 0.10$ & & $0.84 \pm 0.04$ & $0.81^{+0.09}_{-0.05}$ &  $291^{+33}_{-20}$ \\
\cite{Armstrong2015} & $5230 \pm 420$ & $4.39 \pm 0.79$ & $0.38 \pm 0.23$ & & $0.92 \pm 0.14$ & $1.03 \pm 0.20$  \\
\citet{Barros2015} & $5390 \pm 180$ & $4.42 \pm 0.34$ & $0.19 \pm 0.12$ & & $0.918^{+0.086}_{-0.070}$ & $0.926^{+0.19}_{-0.069}$ & \\ 
\citet{Narita2015} & $5345 \pm 17$ & $4.394 \pm 0.050$ & $0.07 \pm 0.03$ & & $0.902 \pm 0.011$ & $0.914 \pm 0.027$ \\
\citet{Huber2016} & $5340 \pm 160$ & $4.520 \pm 0.040$ & $0.019^{+0.12}_{-0.080}$ & & $0.901^{+0.055}_{-0.031}$ & $0.857^{+0.069}_{-0.054}$ & $296^{+32}_{-29}$ \\ 
\cite{Vanderburg2016} & $5391 \pm 50$ & $4.60 \pm 0.10$ & & $-0.03 \pm 0.08$ & \\ 
\citet{Sinukoff2016} & $5430 \pm 60$ & $4.63 \pm 0.07$ & $0.10 \pm 0.04$ & & $0.93 \pm 0.05$ & $0.86 \pm 0.04$ \\ 
\citet{Brewer2016} & $5322 \pm 25$ & $4.510 \pm 0.028$ & & $0.05 \pm 0.01$ \\ 
\citet{Nespral2017} & $5250 \pm 70$ & $4.50 \pm 0.10$ & $0.10 \pm 0.05$ & & $0.918 \pm 0.064$ & $0.88 \pm 0.11$ \\
\citet{Sousa2018} & $5355 \pm 35$ & $4.46 \pm 0.06$ & $0.05 \pm 0.02$ & & $0.86 \pm 0.06$ \\ 
\citet{Stassun2018} & $5320 \pm 100$ & $4.553 \pm 0.077$ & & $0.05 \pm 0.10$ & $0.92 \pm 0.11$ & $0.840 \pm 0.043$  & $289.8 \pm 5.6$ \\ 
\citet{Teske2019} & $5418 \pm 21$ & $4.54 \pm 0.04$ & $0.03 \pm 0.02$ \\ 
\citet{Hardegree-Ullman2020} & $5481 \pm 31$ & $4.726 \pm 0.051$ & $0.216 \pm 0.029$ & & $1.22^{+0.17}_{-0.15}$ & $0.793^{+0.021}_{-0.020}$ & $289.8^{+5.7}_{-5.5}$ \\
\citet{Petigura2020} & $5320 \pm 100$ & $4.51 \pm 0.08$ & $0.06 \pm 0.05$ & & $0.88 \pm 0.03$ & $0.82 \pm 0.03$  \\ 
\citet{Sousa2021} & $5346 \pm 27$ & $4.42 \pm 0.06$ & $0.03 \pm 0.02$ & & $0.846 \pm 0.006$ & $0.864 \pm 0.028$ & 299.3 \\ 
\citet{Magrini2022} & $5285 \pm 82$ & $4.53 \pm 0.15$ & $0.09 \pm 0.09$ & & $0.89^{+0.07}_{-0.11}$ & \\ 
\citet{Su2022} & $5378 \pm 82$ & $4.712 \pm 0.049$ & $0.146 \pm 0.029$ \\
\citet{Xiong2022} & $5360 \pm 110$ & $4.50 \pm 0.16$ & $0.055 \pm 0.054$ \\ 
\citet{Andrae2023} & $5348^{+16}_{-20}$ & $4.509^{+0.013}_{-0.015}$ & & $0.026^{+0.015}_{-0.018}$ & & & $295.7^{+1.7}_{-1.6}$ \\ 
\citet{Wang2023} & $5247 \pm 38$ & $4.511 \pm 0.027$ & $0.032 \pm 0.055$ \\ 
\citet{Zink2023} & $5420 \pm 100$ & $4.55^{+0.01}_{-0.02}$ & $0.14 \pm 0.06$ & & $0.91 \pm 0.02$ & $0.84 \pm 0.01$ & \\ 

This work (prior) & & $4.531 \pm 0.044$ & & & $0.885 \pm 0.034$ & $0.845 \pm 0.039$ & $296.9 \pm 1.7$ \\
This work (posterior) & & $4.556 \pm 0.024$ & & & $0.892 \pm 0.034$ & $0.824 \pm 0.024$ &  \\

\hline                  
\end{tabular}
\end{table}

\newpage
\clearpage
\begin{figure*}
  \hspace{-2cm}\includegraphics[width=1.2\textwidth]{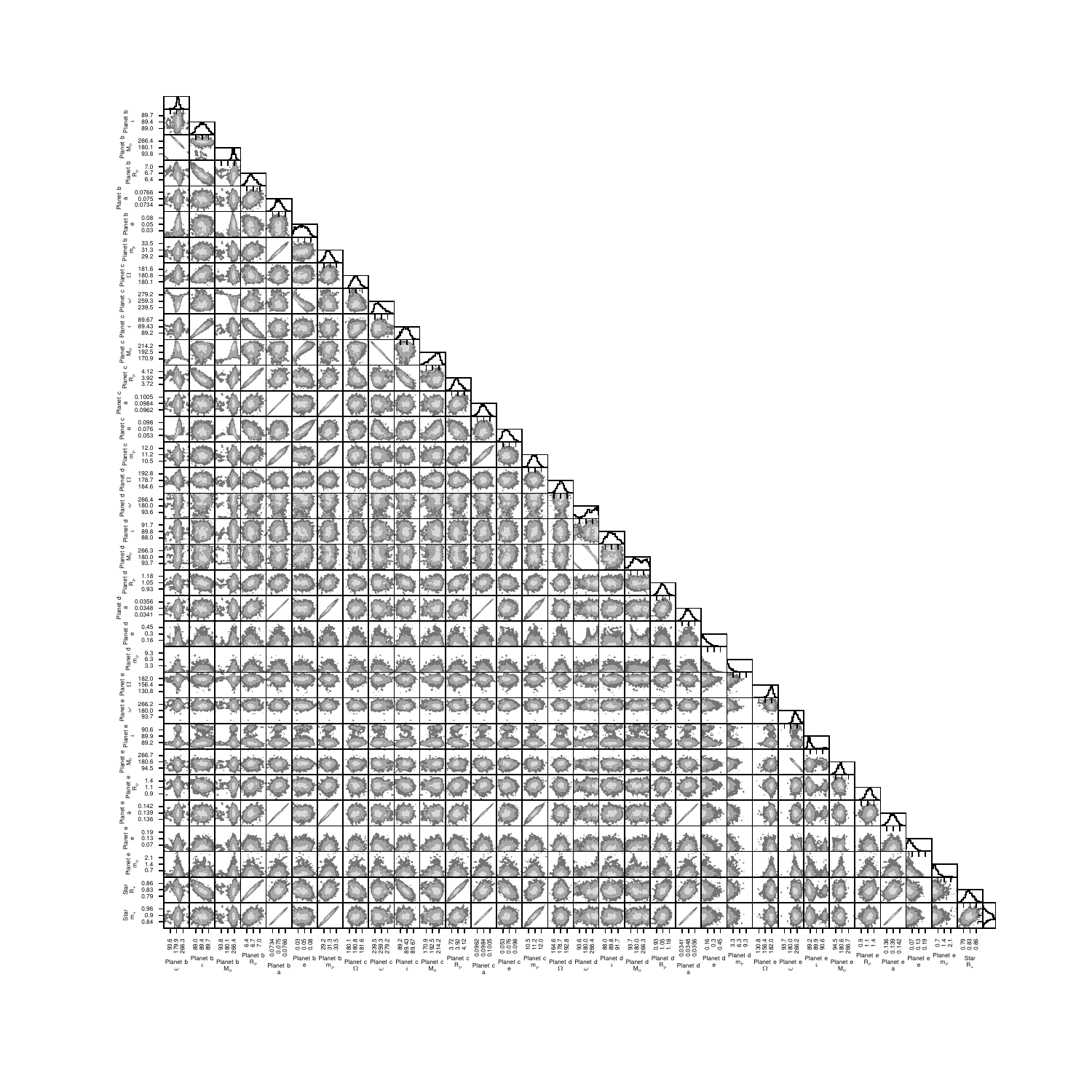}
  \vspace{-2cm}\caption{Two-parameter joint posterior distributions for the most relevant model parameters from the photodynamical modeling (Sect.~\ref{section:photodynamical}). The 39.3, 86.5, and 98.9\% two-variable joint confidence regions are denoted by three different gray levels; in the case of a Gaussian posterior, these regions project on to the one-dimensional 1, 2, and $3\,\sigma$ intervals. The histogram of the marginal distribution for each parameter is shown at the top of each column, except for the parameter on the last line, which is shown at the end of the line. Units are the same as in Table~\ref{table:results}.} \label{figure:pyramid}
\end{figure*}

\newpage
\clearpage
\begin{figure*}
  \centering
  \includegraphics[width=1\textwidth]{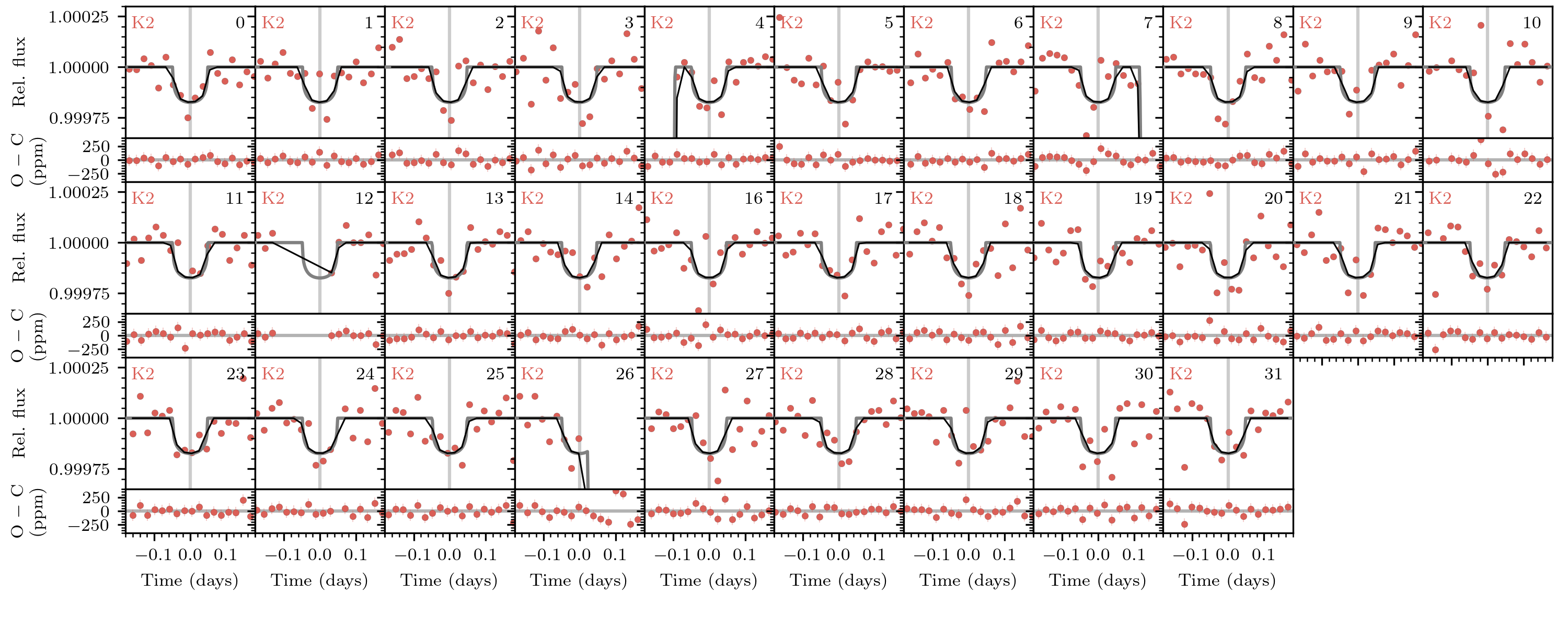}
  \caption{Noise-model-corrected transits of K2-19\,d (dots with the same color code as in Fig.~\ref{figure:phot}), the MAP model (black line), and the continuous oversampled MAP model (gray line) from the photodynamical modeling. Each panel is centered at the linear ephemeris (indicated by the vertical gray lines). Each panel is labeled with the telescopes that observed and the epoch number. The residuals after subtracting the MAP model are shown in the lower part of each panel.} \label{figure:transit1}
\end{figure*}
\begin{figure*}
  \centering
  \includegraphics[width=1\textwidth]{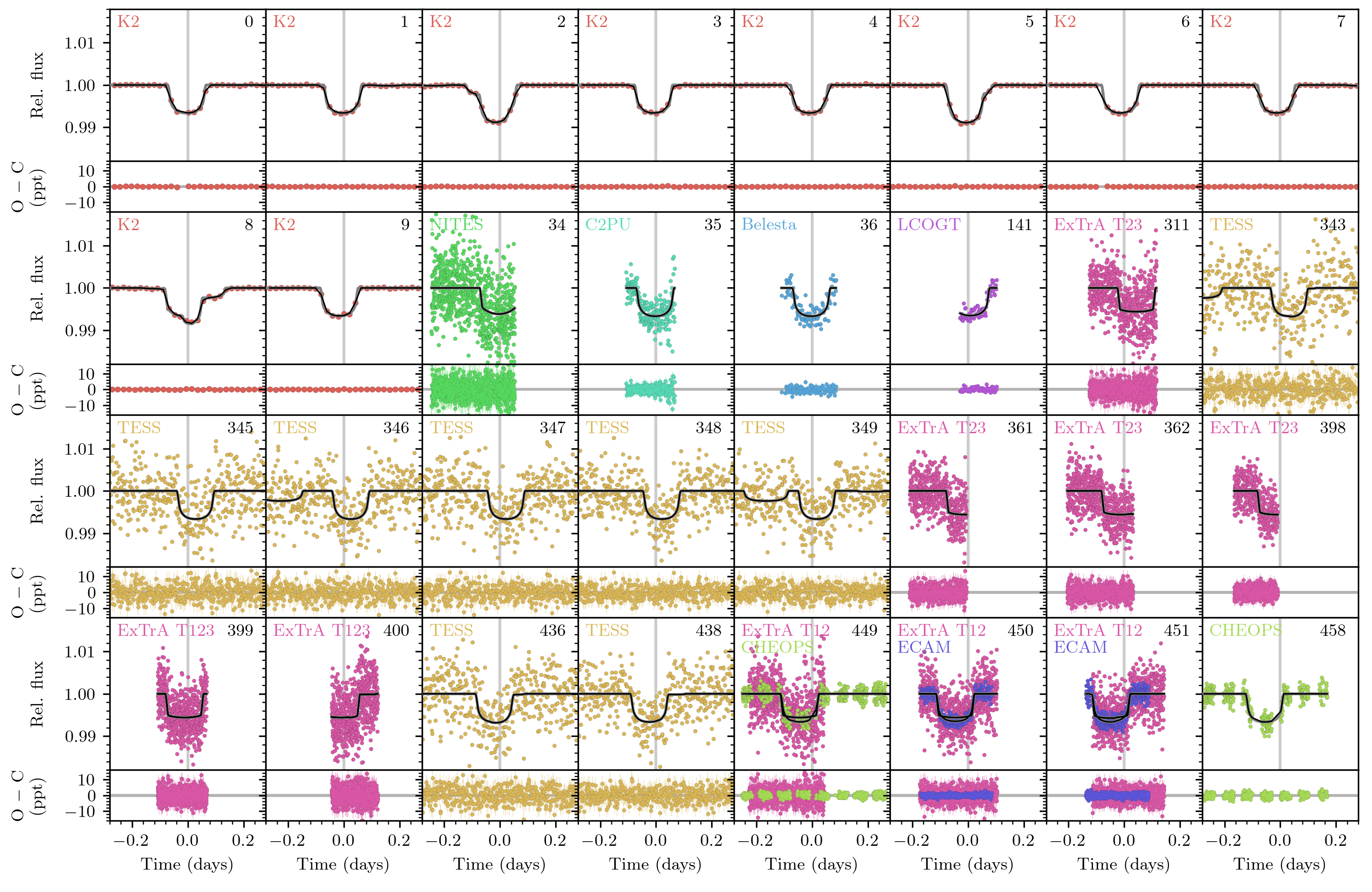}  
  \caption{Same as Fig.~\ref{figure:transit1}, but for K2-19\,b.} \label{figure:transit2}
\end{figure*}
\begin{figure*}
  \centering
  \includegraphics[width=1\textwidth]{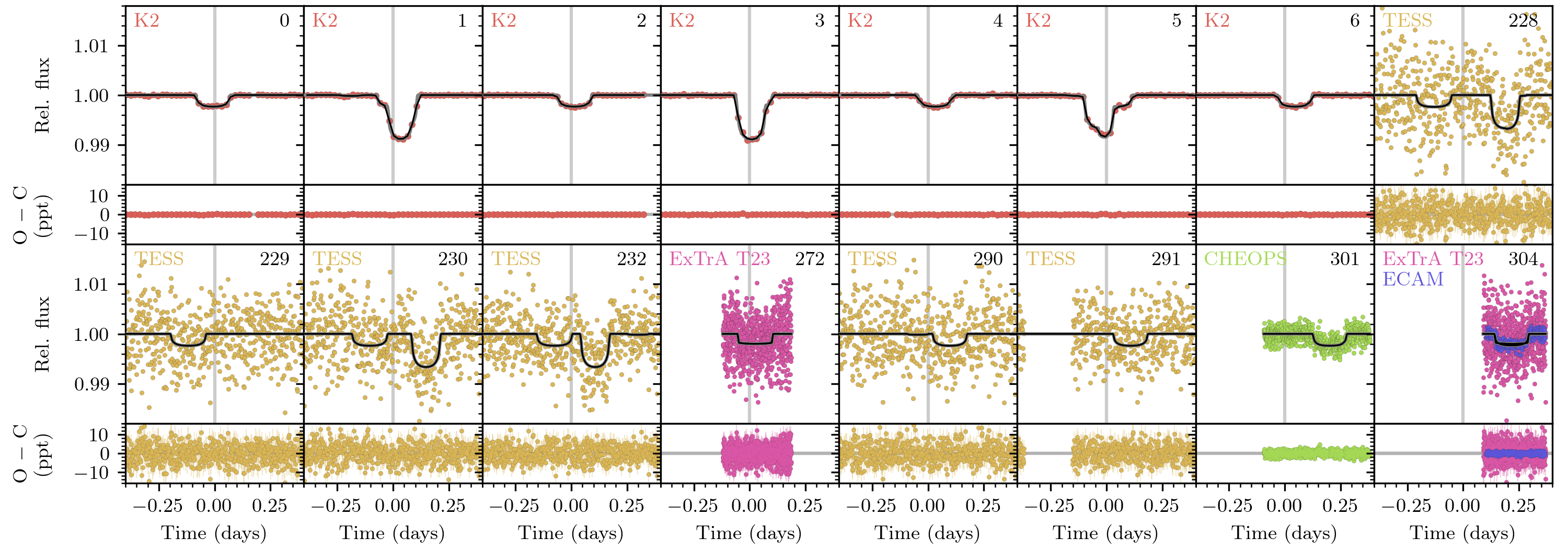}  
  \caption{Same as Fig.~\ref{figure:transit1}, but for K2-19\,c.} \label{figure:transit3}
\end{figure*}
\begin{figure*}
  \centering
  \includegraphics[width=1\textwidth]{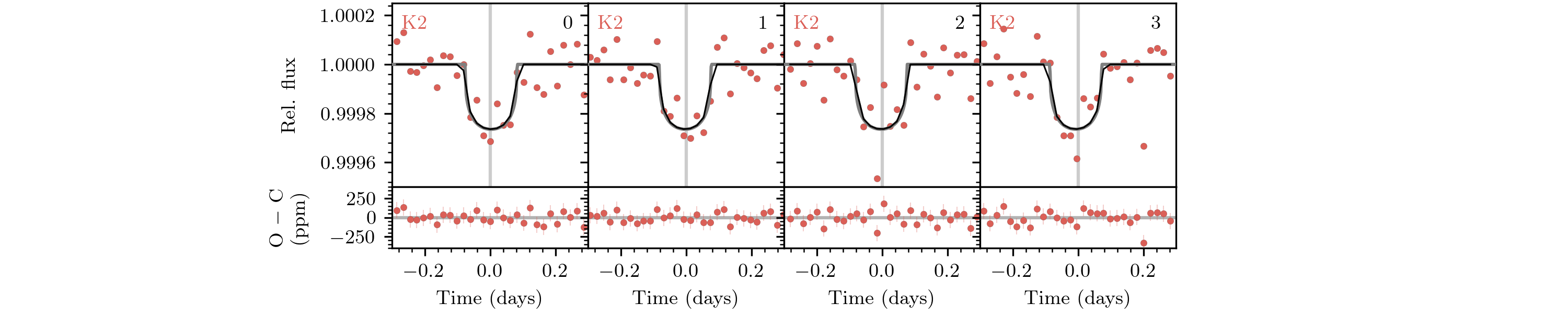}  
  \caption{Same as Fig.~\ref{figure:transit1}, but for K2-19\,e.} \label{figure:transit4}
\end{figure*}

\begin{figure*}
\centering
  \includegraphics[width=0.45\textwidth]{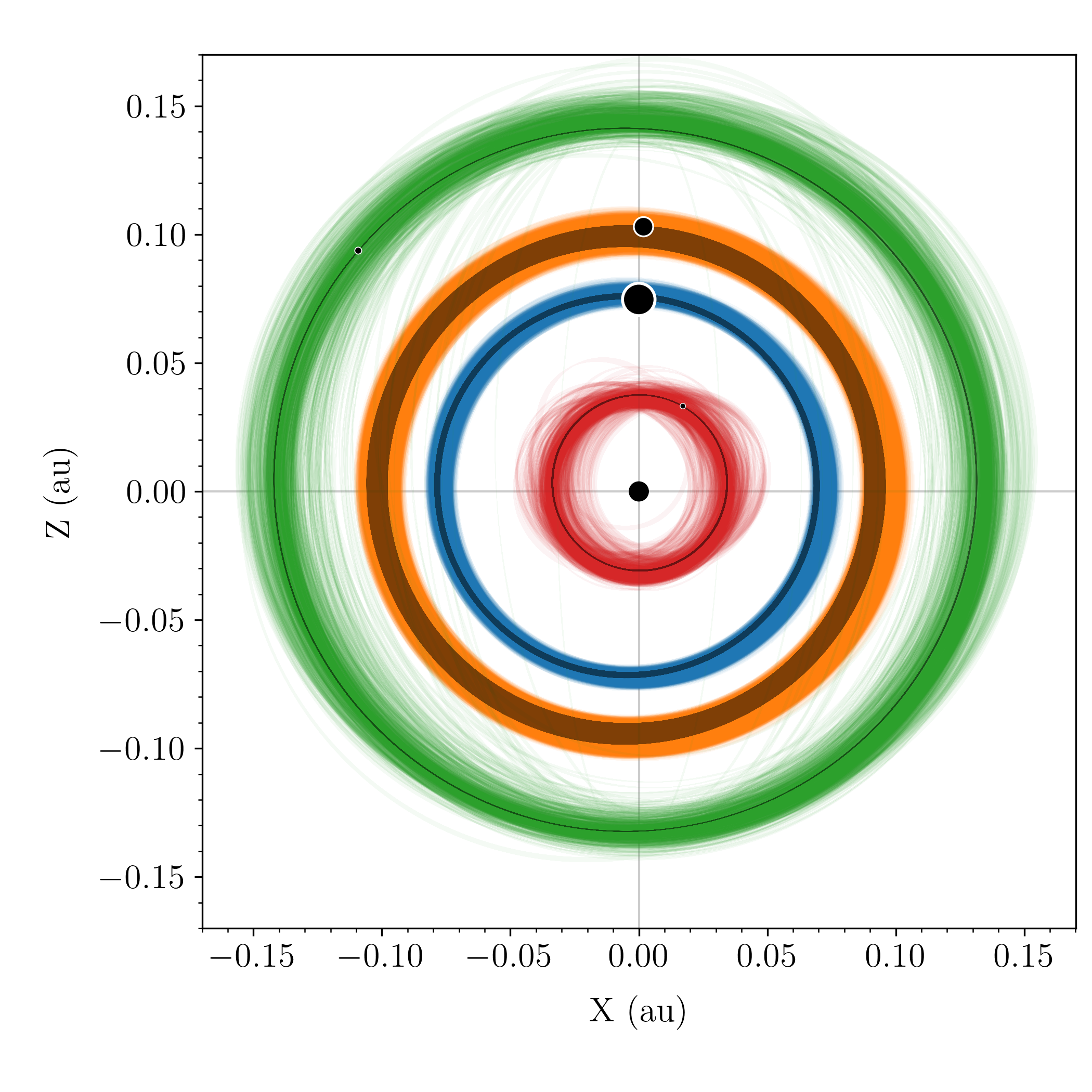}
  \hspace{0.2cm}\includegraphics[width=0.45\textwidth]{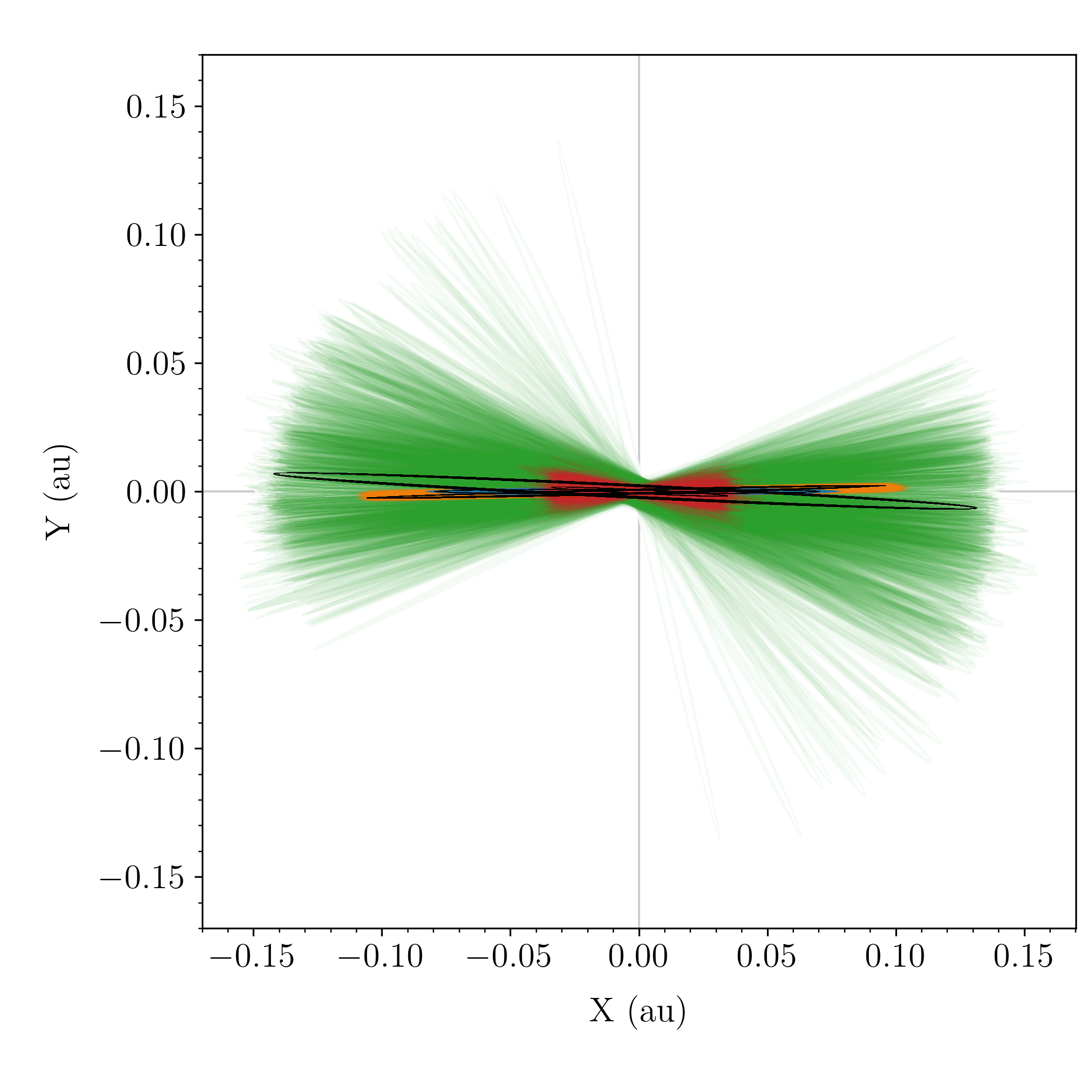}
  \caption{Orbital projections for the time-span covered by the observations. The origin is the system barycenter, and the orbits are projected in the X-Z plane (left panel, system top view, movement is clockwise, the positive Z-axis points towards the observer) and in the sky plane seen by the observer (X-Y, right panel). A thousand random orbits were drawn from the posterior samples, and the MAP is shown as a gray orbit. In the left panel, the black circles mark the position of the star (size to scale) and the planets (enlarged by a factor of 20) at $t_{\mathrm{ref}}$ for the MAP.} \label{figure:orbits}
\end{figure*}

\begin{figure*}
  \includegraphics[width=1\textwidth]{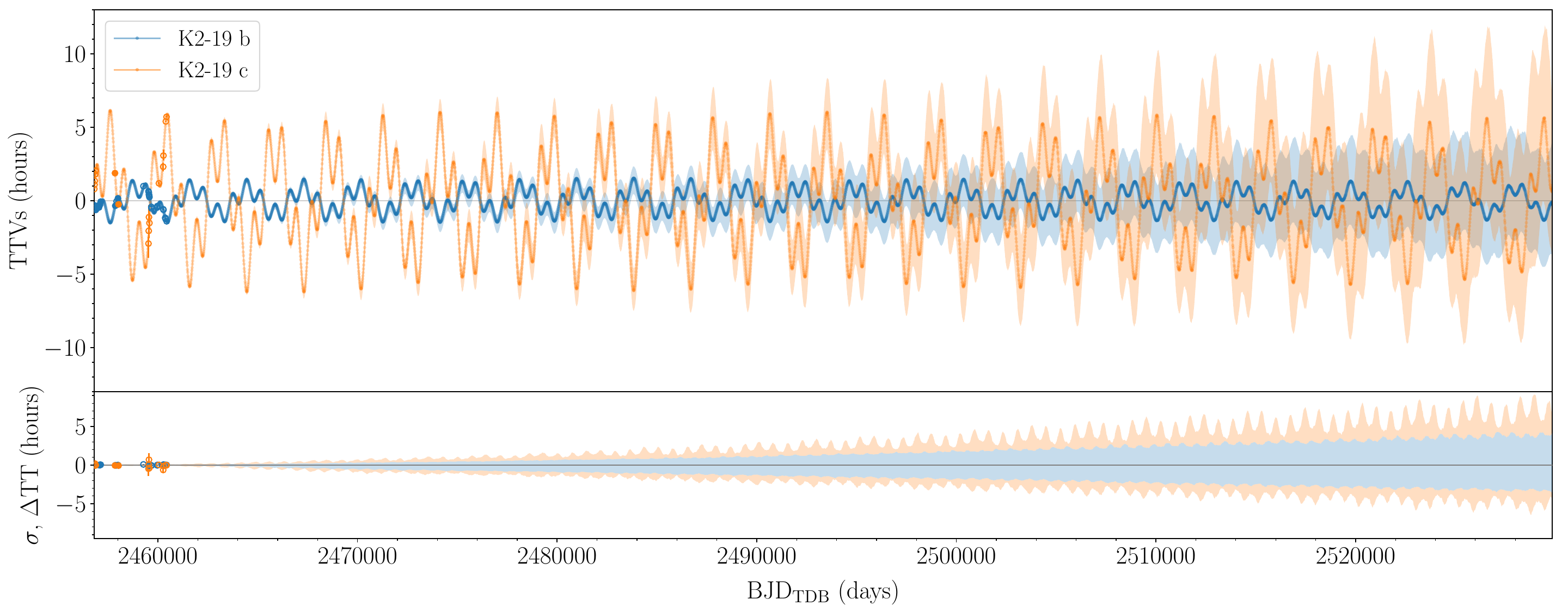}
  \caption{Same as Fig.~\ref{figure:TTVs}, but for 200~years from the start of the K2 observations. The linear ephemerides are 2456813.396578 + 7.92088840\;$\times$\;Epoch (BJD$_{\rm TDB}$), 2456817.239433 + 11.89838594\;$\times$\;Epoch (BJD$_{\rm TDB}$) for planet b and c, respectively} \label{figure:TTVs200yr}
\end{figure*}

\begin{figure*}
  \includegraphics[width=1\textwidth]{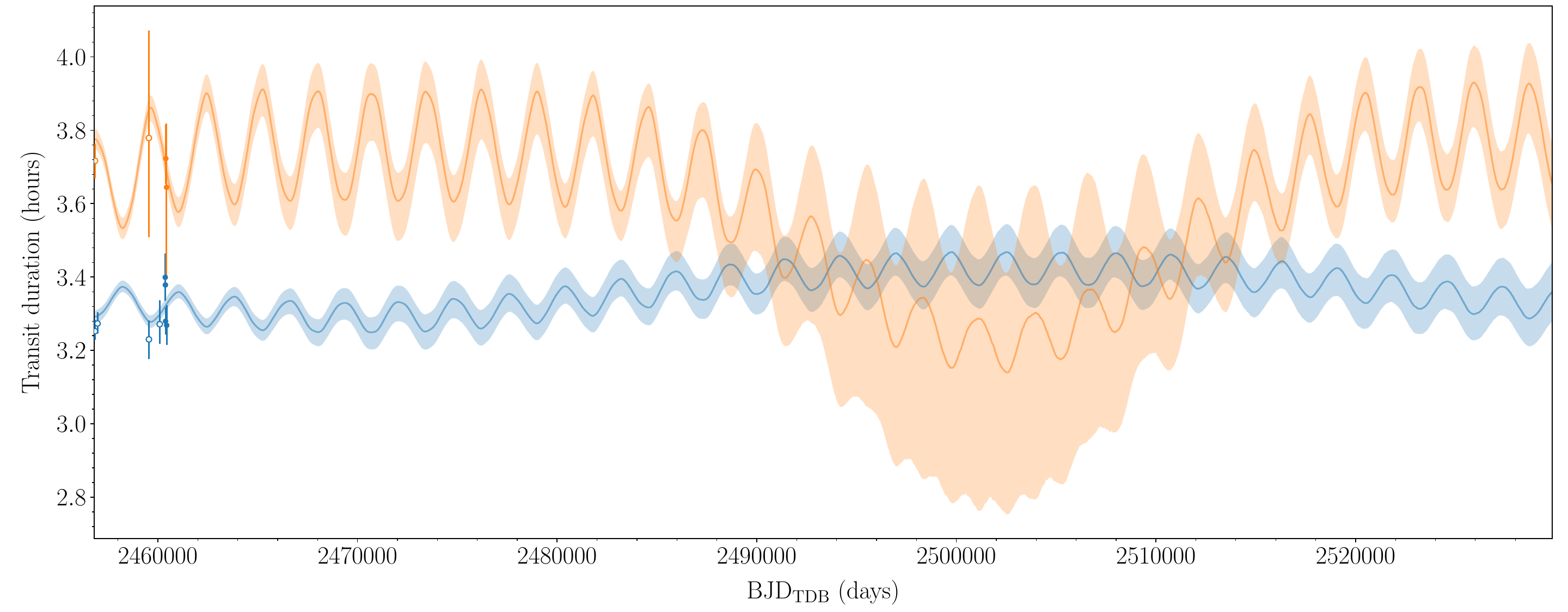}
  \caption{Same as Fig.~\ref{figure:TDVs}, but for 200~years from the start of the K2 observations.} \label{figure:TDVs200yr}
\end{figure*}

\begin{figure*}

\includegraphics[width=0.24\textwidth]{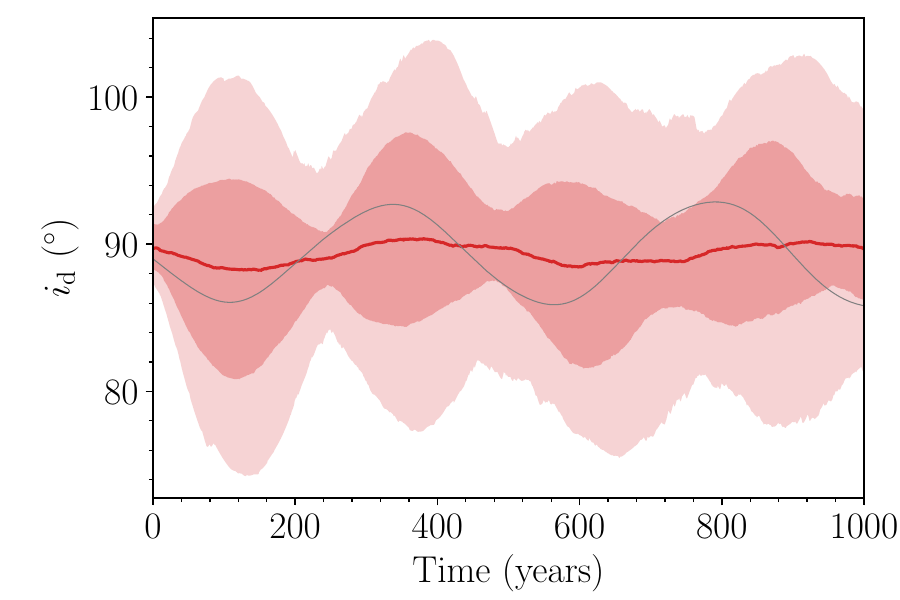}
\includegraphics[width=0.24\textwidth]{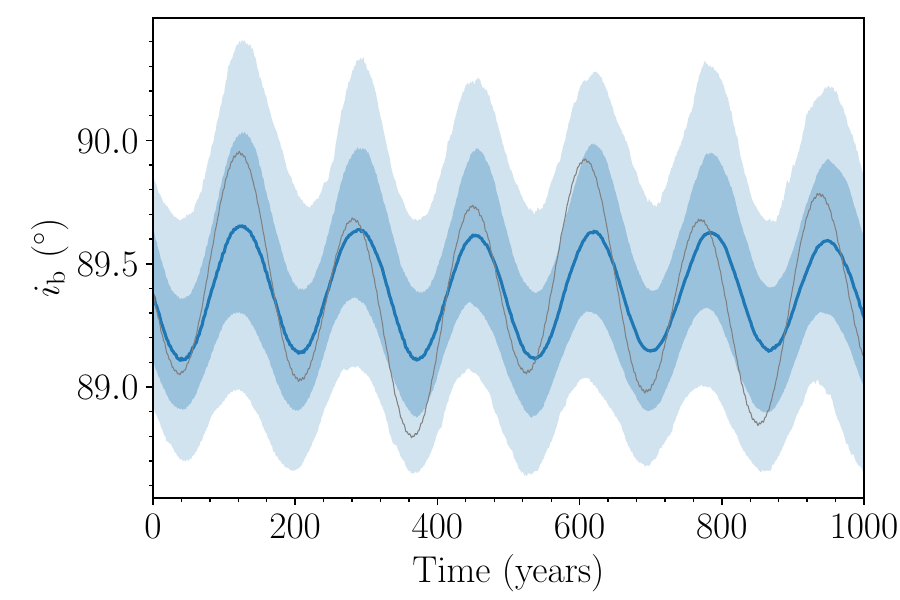}
\includegraphics[width=0.24\textwidth]{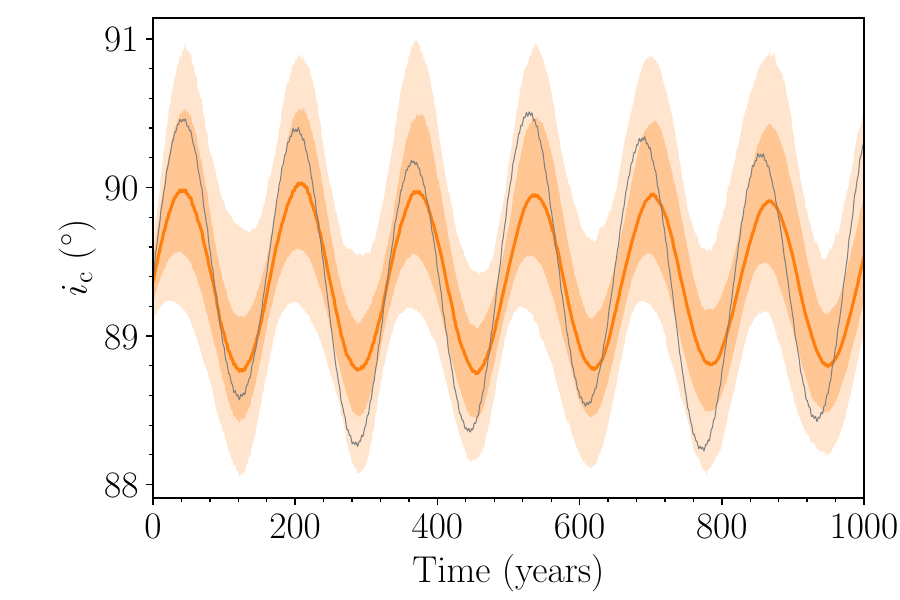}
\includegraphics[width=0.24\textwidth]{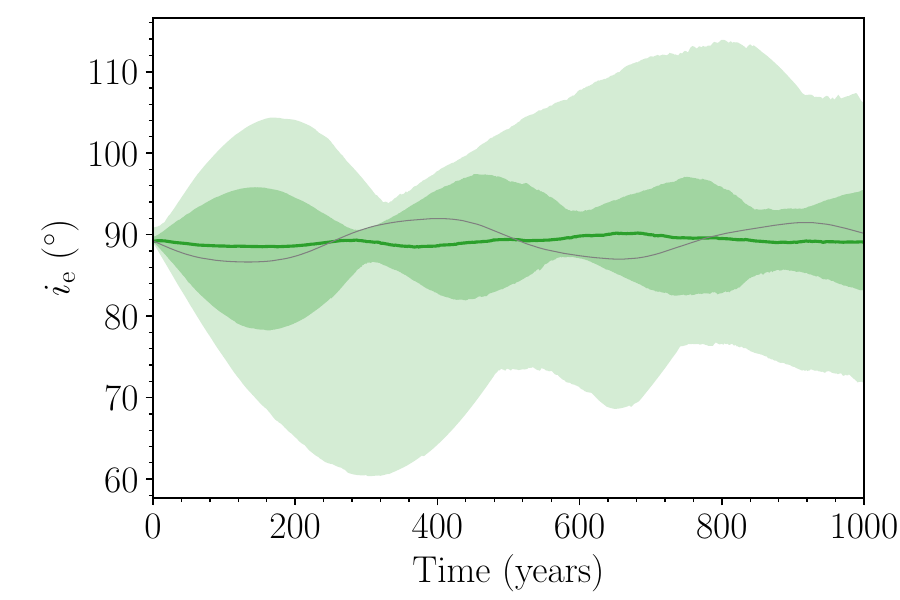}

\includegraphics[width=0.24\textwidth]{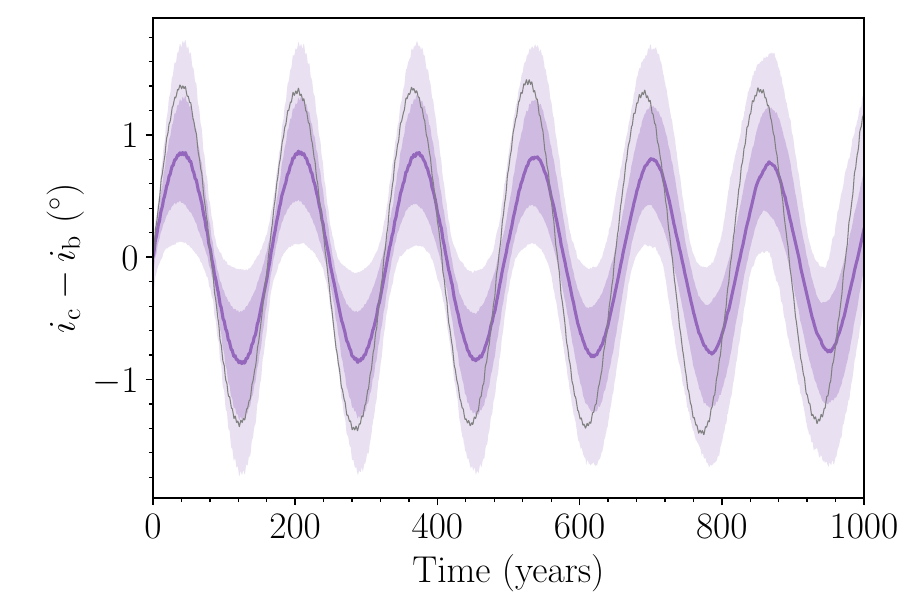}
\includegraphics[width=0.24\textwidth]{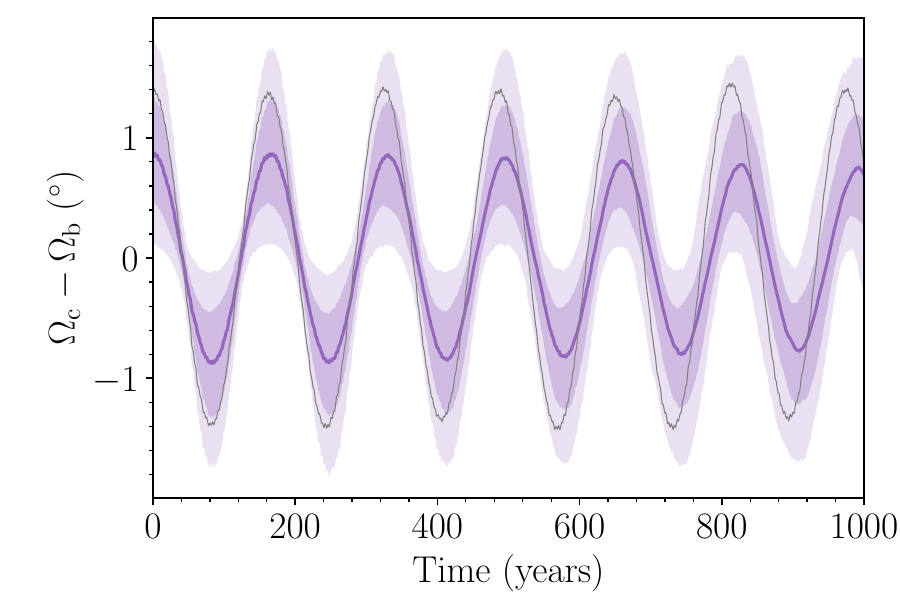}
\includegraphics[width=0.24\textwidth]{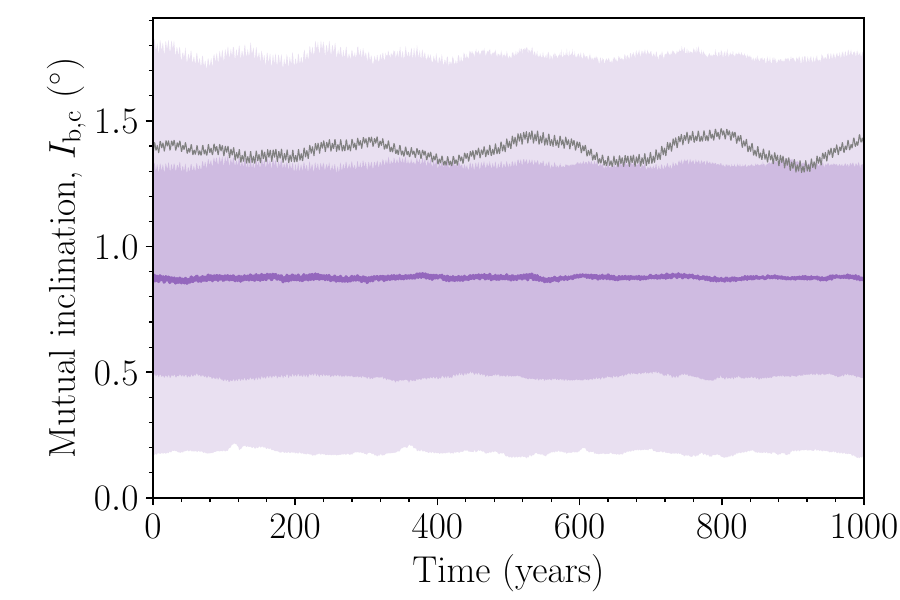}

\caption{Orbital inclination (Jacobi) evolution of planets d (first column), b (second column), c (third column), and e (fourth column) from 1000 random samples of the posterior. The 68.3\% and 95.4\% Bayesian CIs are plotted in different intensities. The solid color curve marks the median of the posterior distribution. The solid gray curves correspond to the parameters of the MAP values. The last line shows some relations of orbital parameters for planets b and c.}
\label{figure:LongTermEvolution}
\end{figure*}

\begin{figure*}[!ht]
  \includegraphics[width=1\textwidth]{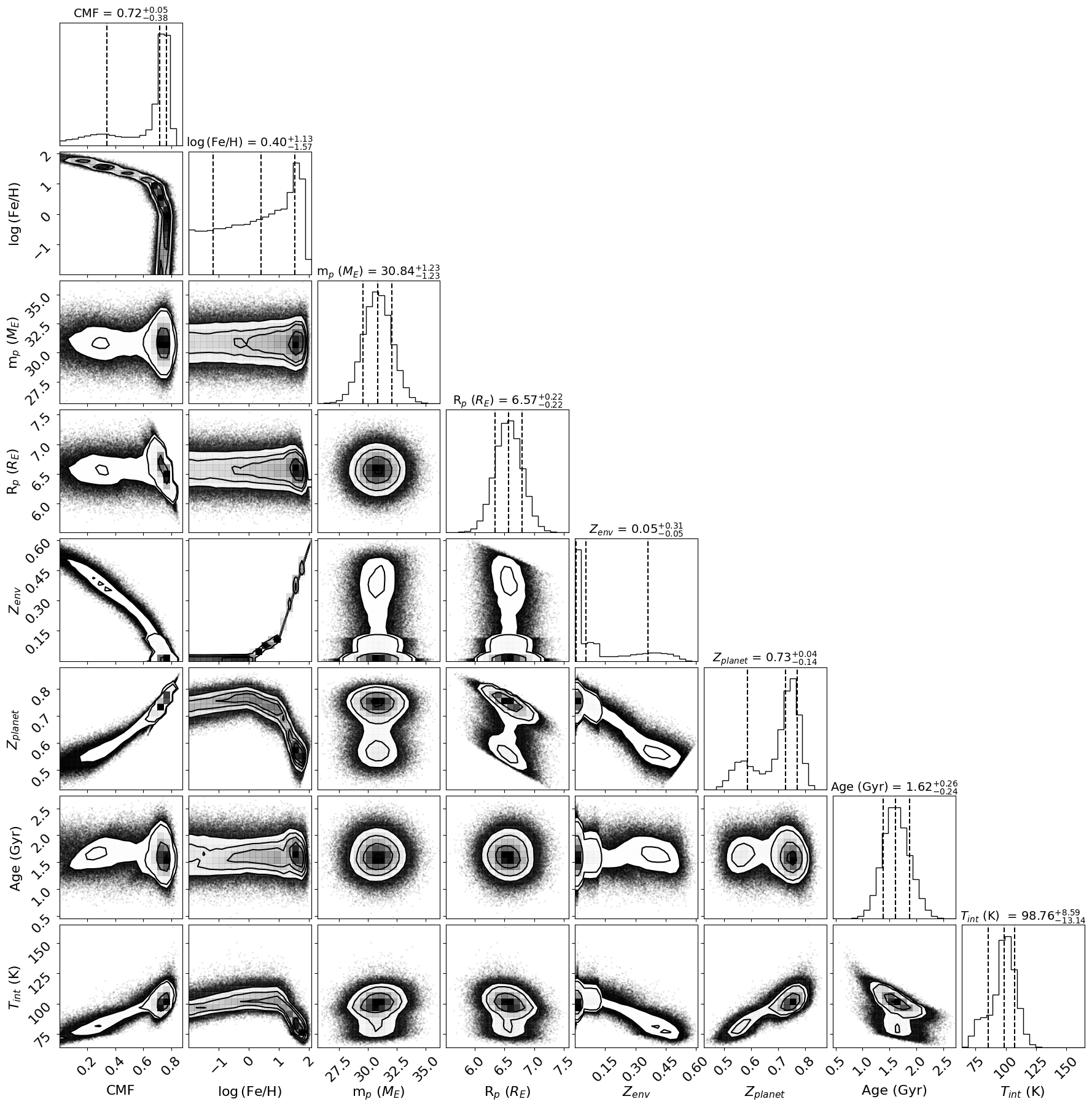}
  \caption{Two-parameter joint posterior distributions for the most relevant model parameters from the interior retrieval (Sect.~\ref{section:internal_structure}). The 1, 2, and $3\,\sigma$ confidence regions are denoted black contours. The histogram of the marginal distribution for each parameter is shown at the top of each column, except for the parameter on the last line, which is shown at the end of the line.} \label{figure:interior_retrieval}
\end{figure*}

\FloatBarrier

\newpage
\clearpage

\section{Determination of transit times}\label{section:transit_times}

Transit times of multiple transits observations can be obtained by fitting them simultaneously while assuming constant transit parameters. This approach can obtain more precise values than when each transit is analyzed individually, be less sensitive to stellar activity or instrumental systematics, and it is the safest option if there are partial transits among the observations. Depending on the photometric precision, the transit timing of partial transits presents a degeneracy with the transit duration. This degeneracy is broken if the the signal of limb darkening is detected.
However, when the transits present TDVs, assuming one transit shape for all transits can bias the derived transit timings. 
To obtain the transit timings in K2-19, we found a compromise by fitting together groups of transits observed close in time. But we also fit individual transit times for the most precise observations. Simultaneous observations of the same transit were analyzed together.
For planet b, we analyzed epochs [311-362] as a group and epochs [398-438] as another group. Epochs 449, 450, 451, and 458 were analyzed individually. For planet c, epochs [228-232] and [290-291] were analyzed as groups, while epochs 272, 301, and 304 were analyzed individually. The partial transit of planet~b at epoch 141 (observed with LCOGT), for which the timing presented in \citet{Petigura2020} represents an outlier, was analyzed assuming the transit parameters of K2 observations. The timings of planets~d, only detected in K2 observations, were analyzed simultaneously. And the same for planet~e candidate (Sect.~\ref{section:nuance}).
We modeled the datasets with \juliet \citep{Espinoza2019,Kreidberg2015,Foreman-Mackey2017,Speagle2020} using the same noise model as in Sect.~\ref{section:photodynamical}.
Table~\ref{table:transit_times} presents literature and the new transit timings we computed. These timings, derived independently of the assumptions used in the photodynamical modeling (Sect.~\ref{section:photodynamical}), were utilized to validate its results by comparing them with the posterior TTVs (Fig.~\ref{figure:TTVs}).

\begin{table}[b]
\centering
    \scriptsize
    \renewcommand{\arraystretch}{0.90}
    \setlength{\tabcolsep}{2pt}
\caption{Transit times of the observations.}\label{table:transit_times}
\begin{tabular}{rlll}
\hline
Epoch & Posterior median  & Telescope & Source \\
      & and 68.3\% CI [BJD$_{\mathrm{TDB}}$]  & Instrument  & \\
\hline
\emph{\bf Planet~d}\\ 
0	& $2456811.426_{-0.016}^{+0.015}$ & K2\hspace{1.7cm} & This work\hspace{1.5cm} \\
1	& $2456813.945_{-0.023}^{+0.031}$ & K2 & This work \\
2	& $2456816.433_{-0.014}^{+0.018}$ & K2 & This work \\
3	& $2456818.965_{-0.019}^{+0.023}$ & K2 & This work \\
4	& $2456821.468_{-0.021}^{+0.017}$ & K2 & This work \\
5	& $2456823.990_{-0.014}^{+0.019}$ & K2 & This work \\
6	& $2456826.490_{-0.018}^{+0.014}$ & K2 & This work \\
7	& $2456828.954_{-0.011}^{+0.013}$ & K2 & This work \\
8	& $2456831.486_{-0.015}^{+0.013}$ & K2 & This work \\
9	& $2456834.007_{-0.011}^{+0.012}$ & K2 & This work \\
10	& $2456836.5443_{-0.0084}^{+0.0093}$ & K2 & This work \\
11	& $2456839.025_{-0.016}^{+0.012}$ & K2 & This work \\
12	& $2456841.514_{-0.051}^{+0.033}$ & K2 & This work \\
13	& $2456844.041_{-0.016}^{+0.015}$ & K2 & This work \\
14	& $2456846.582_{-0.018}^{+0.020}$ & K2 & This work \\
16	& $2456851.523_{-0.016}^{+0.026}$ & K2 & This work \\
17	& $2456854.069_{-0.013}^{+0.014}$ & K2 & This work \\
18	& $2456856.571_{-0.016}^{+0.019}$ & K2 & This work \\
19	& $2456859.082_{-0.015}^{+0.014}$ & K2 & This work \\
20	& $2456861.6036_{-0.0088}^{+0.019}$ & K2 & This work \\
21	& $2456864.105_{-0.019}^{+0.012}$ & K2 & This work \\
22	& $2456866.610_{-0.022}^{+0.024}$ & K2 & This work \\
23	& $2456869.124_{-0.022}^{+0.042}$ & K2 & This work \\
24	& $2456871.641_{-0.018}^{+0.014}$ & K2 & This work \\
25	& $2456874.144_{-0.029}^{+0.018}$ & K2 & This work \\
26	& $2456876.651_{-0.016}^{+0.038}$ & K2 & This work \\
27	& $2456879.163_{-0.014}^{+0.028}$ & K2 & This work \\
28	& $2456881.677_{-0.022}^{+0.018}$ & K2 & This work \\
29	& $2456884.195_{-0.080}^{+0.032}$ & K2 & This work \\
30	& $2456886.677_{-0.025}^{+0.022}$ & K2 & This work \\
31	& $2456889.193_{-0.025}^{+0.028}$ & K2 & This work \smallskip\\
\hline
\end{tabular}
\end{table}

\begin{table}
\centering
    \scriptsize
    \renewcommand{\arraystretch}{1.35}
    \setlength{\tabcolsep}{2pt}
\begin{tabular}{rlll}

\hline
Epoch & Posterior median  & Telescope & Source \\
      & and 68.3\% CI [BJD$_{\mathrm{TDB}}$]  & Instrument  & \\
\hline

\emph{\bf Planet~b}\\
0 & $2456813.38390 \pm 0.00056$ & K2 & \citet{Barros2015} \\ 
1 & $2456821.30410 \pm 0.00088$ & K2 & \citet{Barros2015} \\ 
2 & $2456829.22194 \pm 0.00025$ & K2 & \citet{Barros2015} \\ 
3 & $2456837.13900 \pm 0.00064$ & K2 & \citet{Barros2015} \\ 
4 & $2456845.06156 \pm 0.00053$ & K2 & \citet{Barros2015} \\ 
5 & $2456852.9794 \pm 0.0028$ & K2 & \citet{Barros2015} \\ 
6 & $2456860.90005 \pm 0.00079$ & K2 & \citet{Barros2015} \\ 
7 & $2456868.82080 \pm 0.00052$ & K2 & \citet{Barros2015} \\ 
8 & $2456876.73856 \pm 0.00064$ & K2 & \citet{Barros2015} \\ 
9 & $2456884.65834 \pm 0.00080$ & K2 & \citet{Barros2015} \\ 
30 & $2457051.00413^{+0.00218}_{-0.00225}$ & FLWO 1.2~m & \citet{Narita2015} \\
34 & $2457082.69550^{+0.00140}_{-0.00116}$ & TRAPPIST-South & \citet{Narita2015} \\ 
34 & $2457082.6892 \pm 0.0019$ & NITES & \citet{Barros2015} \\ 
35 & $2457090.6157 \pm 0.0014$ & 1-m C2PU Omicron & \citet{Barros2015} \\
36 & $2457098.5368 \pm 0.0022$ & Belesta 82-cm & \citet{Barros2015} \\ 
41 & $2457138.15047^{+0.00145}_{-0.00176}$ & OAO 1.88~m / MuSCAT & \citet{Narita2015} \\
133 & $2457866.8604 \pm 0.0009$ & Spitzer & \citet{Petigura2020} \\
141 & $2457930.2403^{+0.0015}_{-0.0013}$ & LCOGT-SAAO 1~m & This work\\ 
150 & $2458001.5368 \pm 0.0014$ & Spitzer & \citet{Petigura2020} \\
311 & $2459276.8341^{+0.0019}_{-0.0020}$ & ExTrA T23 & This work \\ 
343 & $2459530.2882^{+0.0029}_{-0.0032}$ & TESS s45 & This work \\ 
345 & $2459546.1269 \pm 0.0023$ & TESS s45 & This work \\ 
346 & $2459554.0421^{+0.0021}_{-0.0023}$ & TESS s46 & This work \\ 
347 & $2459561.9601 \pm 0.0023$ & TESS s46 & This work \\ 
348 & $2459569.8789^{+0.0033}_{-0.0029}$ & TESS s46 & This work \\ 
349 & $2459577.7967^{+0.0024}_{-0.0025}$ & TESS s46 & This work \\ 
361 & $2459672.8198^{+0.0016}_{-0.0017}$ & ExTrA T23 & This work \\ 
362 & $2459680.7365^{+0.0022}_{-0.0023}$ & ExTrA T23 & This work \\ 
398 & $2459965.8909^{+0.0021}_{-0.0019}$ & ExTrA T23 & This work \\ 
399 & $2459973.8158^{+0.0014}_{-0.0017}$ & ExTrA T123 & This work \\ 
400 & $2459981.7360 \pm 0.0019$ & ExTrA T123 & This work \\ 
436 & $2460266.8799^{+0.0033}_{-0.0032}$ & TESS s72 & This work \\ 
438 & $2460282.7201^{+0.0021}_{-0.0022}$ & TESS s72 & This work \\ 
449 & $2460369.8272 \pm 0.0012$ & CHEOPS, ExTrA T12 & This work \\ 
450 & $2460377.74640^{+0.00057}_{-0.00055}$ & Euler / ECAM, ExTrA T12 & This work \\ 
451 & $2460385.66556^{+0.00070}_{-0.00072}$ & Euler / ECAM, ExTrA T12 & This work \\ 
458 & $2460441.10519^{+0.00064}_{-0.00066}$ & CHEOPS & This work \smallskip\\ 

\emph{\bf Planet~c}\\
0 & $2456817.2730 \pm 0.0015$ & K2 & \citet{Barros2015} \\ 
1 & $2456829.1835 \pm 0.0013$ & K2 & \citet{Barros2015} \\ 
2 & $2456841.0922 \pm 0.0014$ & K2 & \citet{Barros2015} \\ 
3 & $2456853.0091 \pm 0.0077$ & K2 & \citet{Barros2015} \\ 
4 & $2456864.9083 \pm 0.0019$ & K2 & \citet{Barros2015} \\ 
5 & $2456876.8150 \pm 0.0014$ & K2 & \citet{Barros2015} \\ 
6 & $2456888.7161 \pm 0.0013$ & K2 & \citet{Barros2015} \\ 
87 & $2457852.4774 \pm 0.0074$ & Spitzer & \citet{Petigura2020} \\
102 & $2458030.8645 \pm 0.0059$ & Spitzer & \citet{Petigura2020} \\
228 & $2459529.950^{+0.019}_{-0.040}$ & TESS s45 & This work \\ 
229 & $2459541.884^{+0.015}_{-0.022}$ & TESS s45 & This work \\ 
230 & $2459553.822^{+0.034}_{-0.017}$ & TESS s46 & This work \\ 
232 & $2459577.602^{+0.027}_{-0.018}$ & TESS s46 & This work \\ 
272 & $2460053.6499^{+0.0038}_{-0.0028}$ & ExTrA T23 & This work \\ 
290 & $2460267.868^{+0.011}_{-0.013}$ & TESS s72 & This work \\ 
291 & $2460279.798^{+0.017}_{-0.014}$ & TESS s72 & This work \\ 
301 & $2460398.8787^{+0.0016}_{-0.0017}$ & CHEOPS & This work \\ 
304 & $2460434.5866^{+0.0023}_{-0.0022} $ & Euler / ECAM, ExTrA T23 & This work \smallskip\\ 

\emph{\bf Candidate planet~e} \\
0 &	$2456812.250^{+0.019}_{-0.014}$	& K2 & This work\\ 
1 &	$2456832.220^{+0.016}_{-0.011}$ & K2 & This work\\ 
2 &	$2456852.185^{+0.011}_{-0.013}$	& K2 & This work\\ 
3 &	$2456872.143^{+0.013}_{-0.016}$	& K2 & This work \smallskip\\ 
\hline
\end{tabular}
\end{table}

\newpage
\clearpage

\section{TTVs analysis}\label{section:analysis_TTVs}

In this section, we analyze only the transit times of planets~b and c (listed in Table~\ref{table:transit_times}). We exclude the partial transit observed by LCOGT, which was analyzed assuming the transit parameters of K2 observations. However, there is a significant time span between both observations. According to the photodynamical modeling, the difference between transits durations is about 6~minutes, so the determination can be biased. For epoch 34 of planet~b, for which two timings can be found in the literature, we keep the one by \citet{Narita2015} for the analysis in this section because it is more precise. 

To investigate the effect of the coverage of the TTVs signal, we separate the timings into seven datasets: 2014 (only K2 data), 2015 \citep[time span comparable with the photodynamical modeling of][]{Barros2015}, 2017 \citep[this one is equivalent to the modeling of][in terms of time span]{Petigura2020}, 2021, 2022, 2023, and 2024. For each dataset, only transit times up to the end of the named year were considered.

\begin{figure}
  \centering
  \includegraphics[width=0.42\textwidth]{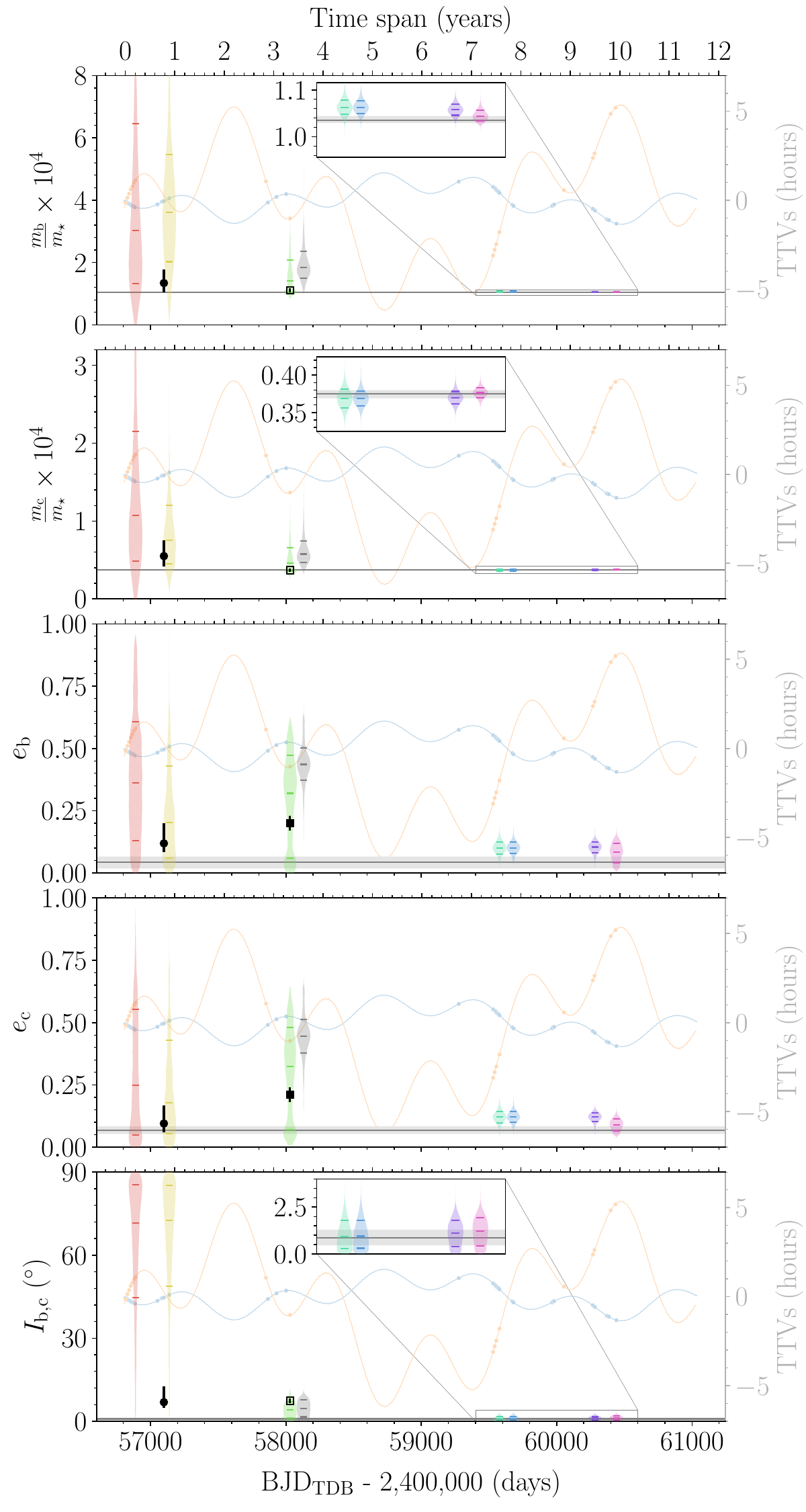}
  \caption{Comparison of the marginal posterior distribution (violins) of selected parameters (orbital parameters are at $t_{\mathrm{ref}}$) for the analysis of the transit times with different time spans. Violins are positioned at the time of the last observation considered in the analysis and horizontal ticks inside them marks the median and 68.3\% CI of the marginal posterior distribution. The gray violins corresponds to the 2017 dataset using the LCOGT timing from \citet{Petigura2020} and are offset to the right for clarity. The horizontal line and gray band represent the median and 68.3\% CI of the marginal posterior distribution from the photodynamical analysis of Sect.~\ref{section:photodynamical}. The insets provide a zoomed view of the last four transit timing analyses. Results from the photodynamical analysis of \citet{Barros2015} are shown with black circle error bars, and those of \citet{Petigura2020} with black square error bars (open squares correspond to derived values assuming normal distributions for the quoted parameters, and thus do not account for correlations between them). In the background, the median posterior TTVs for planets~b (light blue) and c (light orange) from the photodynamical analysis of Sect.~\ref{section:photodynamical} are shown, and dots mark the observed transits.} \label{figure:TTruns_parameters}
\end{figure}

\begin{figure}
  \centering
  \includegraphics[width=0.47\textwidth]{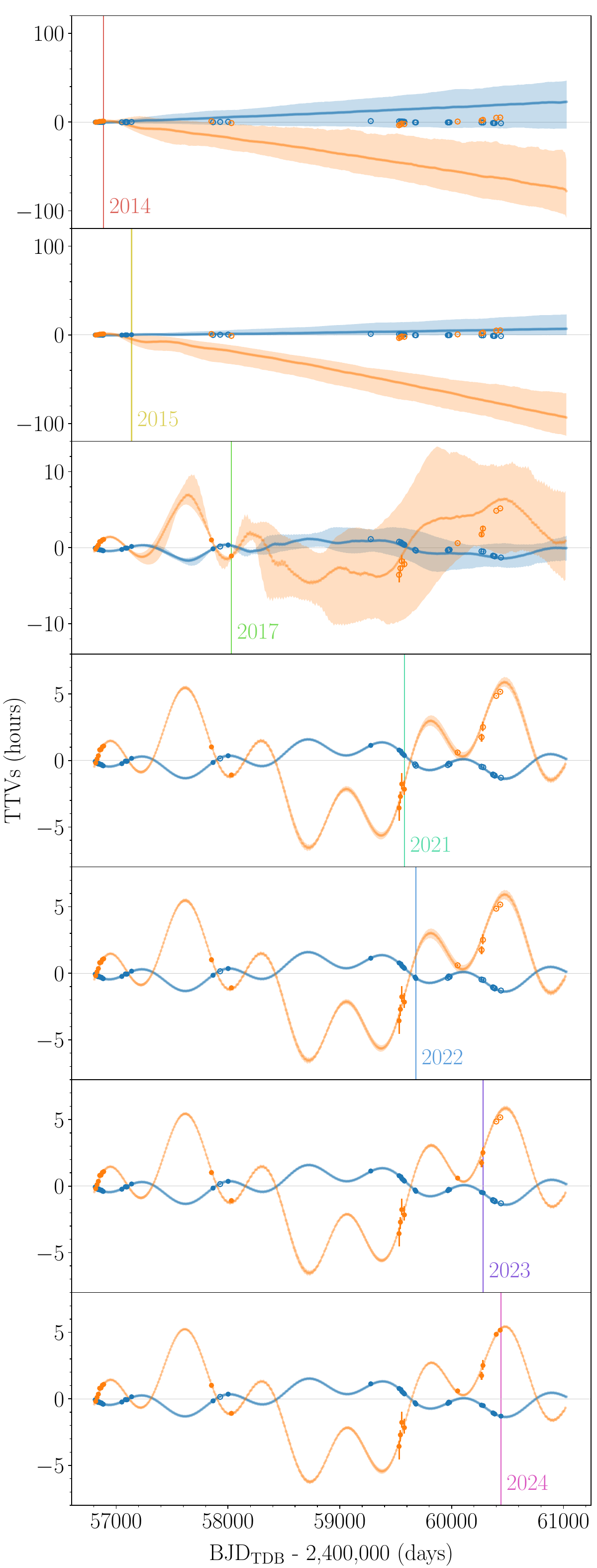}
  \caption{Same as Fig.~\ref{figure:TTVs}, but for the analysis using only transit times up to the end of the labeled year and only planets b and c in the model. The labeled vertical line (the color code is the same as in Fig.~\ref{figure:TTruns_parameters}) marks the last observation modeled. Filled circles represent the times used in the analysis of each of the panels, while open circles indicate observations that are not used.} \label{figure:TTruns_TTVs}
\end{figure}

The model only includes the star and planets~b and c. Transit times were modeled as the moments of minimum sky-projected planet–star separation, calculated using Eq.~1 from \citet{Deck2014}, with n-body integrations performed in \rebound, employing WHFast and an integration step of 0.01~days. We fixed the orbital inclination of both planets at 90\degree\ (which biases the mutual inclination) and included a jitter term \citep{Gregory2005} for the dataset of each planet. The eccentricities and argument of pericenter were parameterized using $\sqrt{e}\cos{\omega}$ and $\sqrt{e}\sin{\omega}$. We used uniform priors for all parameters. The joint posterior distribution was sampled using \emcee \citep{Goodman2010, emcee}. The marginal posterior distribution of selected parameters are shown in Fig.~\ref{figure:TTruns_parameters}, and the TTVs plots are in Fig.~\ref{figure:TTruns_TTVs}.

\begin{figure}
  \centering
  \includegraphics[width=0.465\textwidth]{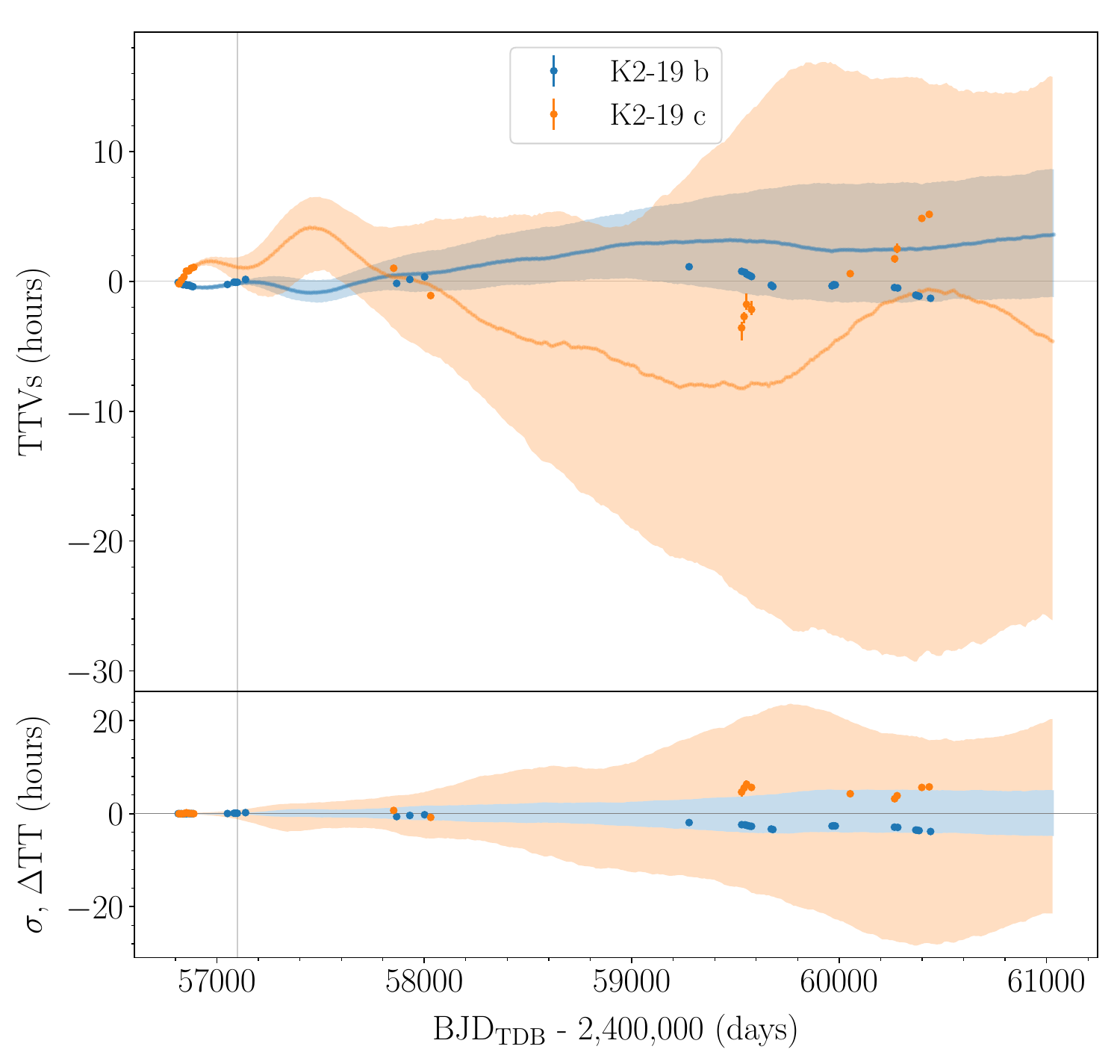}
  \caption{Same as Fig.~\ref{figure:TTVs} but for \citet{Barros2015} TTVs posterior (blue and orange bands) compared with observations (error bars, Table~\ref{table:transit_times}). The vertical gray line marks the last observation modeled.}\label{figure:BarrosTTVs}
\end{figure}

\citet{Nascimbeni2024} asserts that the period modulation of the TTV needs to be fully mapped to avoid degeneracy in the dynamical retrieval. When comparing the 2015 dataset to \citet{Barros2015}, the photodynamical modeling appears to be less impacted by this degeneracy than the TTV analysis. The distinction between modeling just transit times and employing photodynamical modeling is more noticeable in 2015, which has less coverage of the TTV signal (see Fig.~\ref{figure:BarrosTTVs} and the 2015 panel in Fig.~\ref{figure:TTruns_TTVs}), than in 2024 (see Fig.~\ref{figure:TTVs} and the 2024 panel in Fig.~\ref{figure:TTruns_TTVs}). The posteriors from the photodynamical modeling (Sect.~\ref{section:photodynamical}) are slightly more precise than those from the transit-time analysis of the 2024 dataset for the shared parameters, although both agree within $1\,\sigma$. With the exception of the 2015 dataset (and 2024, which uses all available data), the remaining datasets predict future observed transits within $1.6\,\sigma$ for planet~b and within $2.6\,\sigma$ for planet~c (Fig.~\ref{figure:TTruns_TTVs}). 

\begin{figure}
  \centering
  \includegraphics[width=0.465\textwidth]{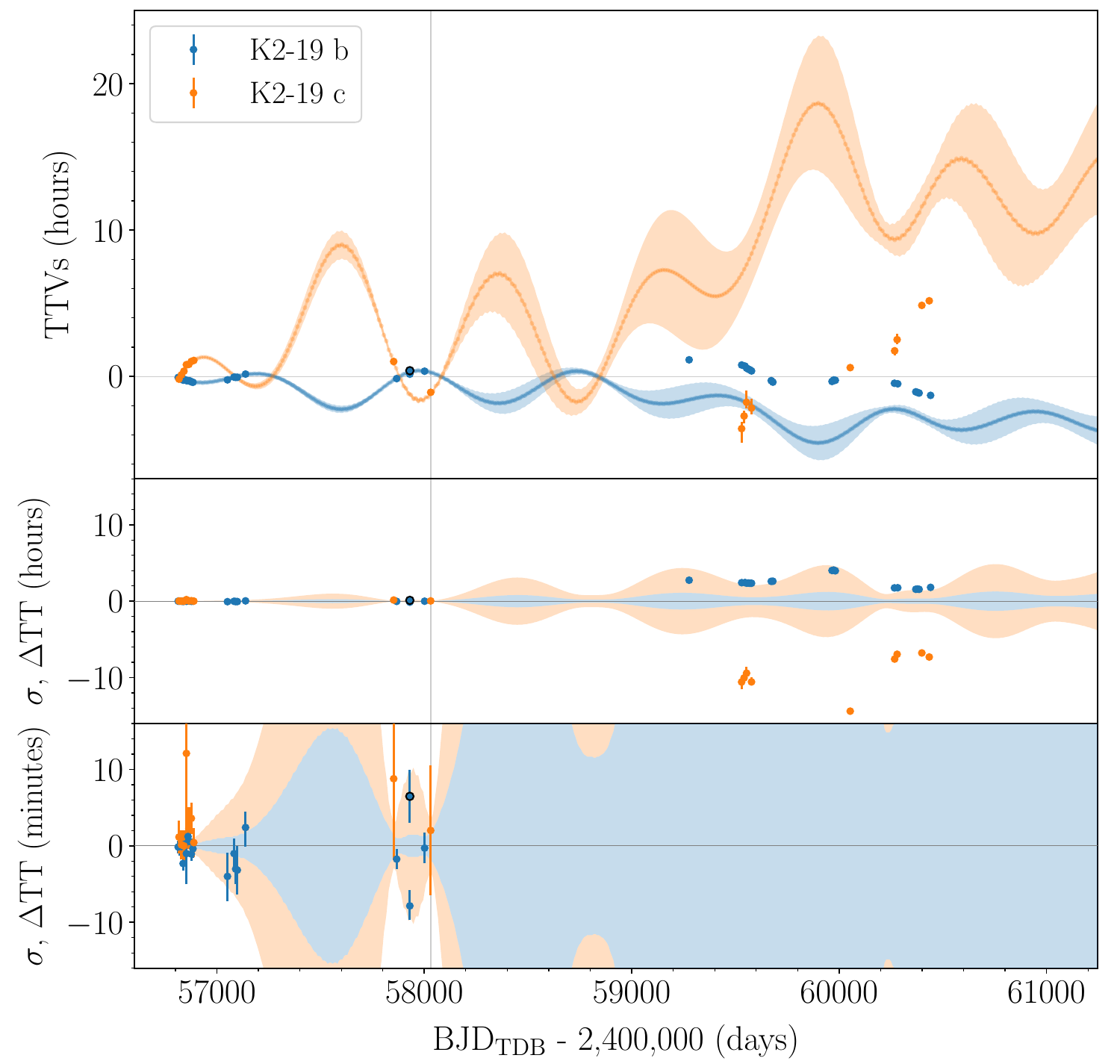}
  \caption{Same as Fig.~\ref{figure:BarrosTTVs} but for \citet{Petigura2020}. Their 141~epoch of planet~b timing is the error bar encircled in black.} \label{figure:PetiguraTTVs}
\end{figure}

\begin{figure}
  \centering
  \includegraphics[width=0.47\textwidth]{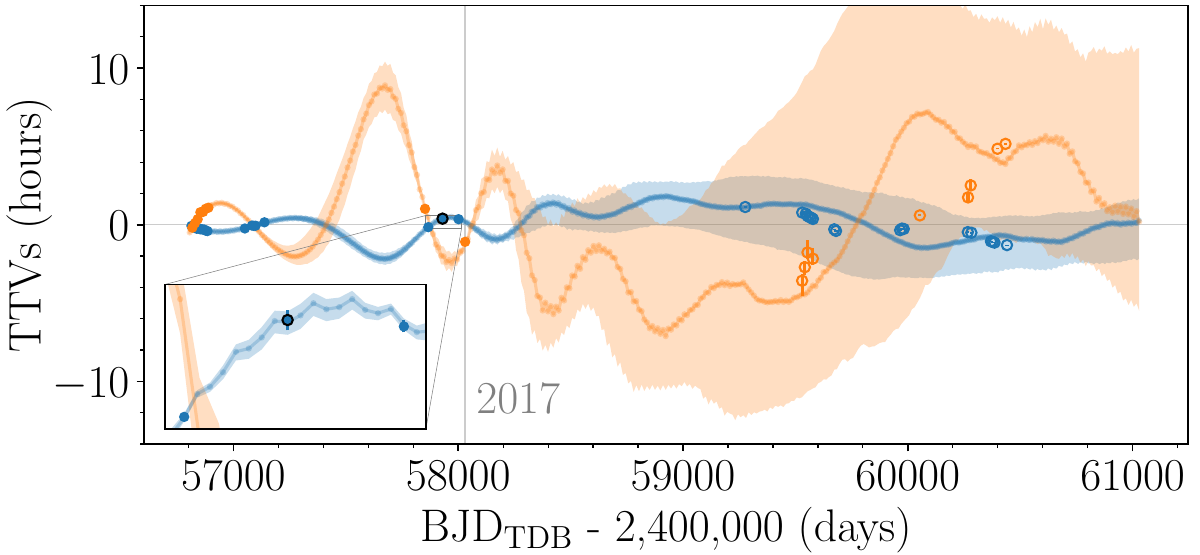}
  \caption{Same as Fig.~\ref{figure:TTruns_TTVs} for the 2017 dataset with the 141~epoch of planet~b timing from \citet{Petigura2020} observed with LCOGT (error bar encircled in black).} \label{figure:TTVs2017LCO}
\end{figure}

We repeated the analysis of the 2017 dataset, including the discrepant LCOGT transit timing from \citet{Petigura2020}, to investigate if it could be responsible for their retrieved higher eccentricities and poor transit timing forecasts (Fig.~\ref{figure:PetiguraTTVs}). The TTV model is able to fit the discrepant timing and produces transit timing forecasts similar to those of the 2017 dataset without the LCOGT timing (see Figs.~\ref{figure:TTruns_TTVs} and \ref{figure:TTVs2017LCO}). However, the posterior eccentricities exhibit notable differences (see Fig.~\ref{figure:TTruns_parameters}). The posteriors for the dataset that includes the LCOGT transit timing from \citet{Petigura2020} are narrower and centered around a higher eccentricity. Therefore, this discrepant timing could be at least partially responsible for the higher eccentricity value found by these authors. The reasoning is as follows: The \citet{Petigura2020} analysis used a photodynamical modeling for the K2 data and modeled only the transit timings for the other datasets. The discrepant LCOGT transit timing from \citet{Petigura2020} pushes to moderate eccentricity, but the photodynamical modeling, which has an additional constrain on the eccentricity from the transit shape, pushes to lower eccentricities. In the end, their posterior converges to an intermediate value between these two. If this is correct, performing photodynamical modeling of all datasets would have avoided this problem.

\clearpage

\end{appendix}
\end{document}